\documentclass[sigconf,screen,nonacm]{acmart}

\usepackage{algorithm}
\usepackage[noend]{algpseudocode}
\usepackage{mathtools}
\usepackage{mathpartir}
\usepackage{wrapfig}
\usepackage{picins}
\usepackage{bussproofs}
\usepackage{cleveref}
\usepackage{thmtools}

\newcommand{\err}{\mathrm{err}}
\newcommand{\init}{\mathrm{init}}

\newcommand{\loc}{\lambda}
\newcommand{\val}{\mu}
\newcommand{\sval}{\texttt{\textmu}}
\newcommand{\sloc}{\texttt{l}}
\newcommand{\src}{\mathsf{src}}
\newcommand{\tgt}{\mathsf{tgt}}
\newcommand{\Lang}{\mathcal{L}}
\newcommand{\Global}{\mathsf{Global}}
\newcommand{\setcomp}[1]{\overline{{#1}}}

\DeclarePairedDelimiter{\tp}{\langle}{\rangle}
\DeclarePairedDelimiter{\sem}{\llbracket}{\rrbracket}
\newcommand{\nodes}{\mathsf{nodes}}
\newcommand{\Node}{\mathsf{Node}}
\newcommand{\Data}{\mathsf{Data}}
\newcommand{\Loc}{\mathsf{Loc}}
\newcommand{\htri}[3]{\left\{ #1 \right\}\; \allowbreak {#2}\; \allowbreak \left\{ #3 \right\}}
\newcommand{\Assertions}{\textsc{Assertions}}
\newcommand{\QFAshcroft}{\textsc{QF-Ashcroft}}
\newcommand{\Ashcroft}{\textsc{Ashcroft}}
\newcommand{\VNode}{\cV_\Node}
\newcommand{\VData}{\cV_\Data}
\newcommand{\VLoc}{\cV_\Loc}
\newcommand{\substructures}{\mathsf{sub}}
\newcommand{\lstate}{\mathsf{LocalState}}
\newcommand{\ldata}{\mathsf{LocalData}}
\newcommand{\gstate}{\mathsf{GlobalState}}
\newcommand{\id}[1]{%
  \ensuremath{\mathit{\mathcode`\-=`\-\relax#1}}}

\DeclareMathOperator{\concat}{+\kern -0.4em+}
\newcommand{\rw}{\mathsf{rw}}

\newcommand{\ti}{\mathtt{i}}

\newcommand{\ar}{\mathsf{ar}}
\newcommand{\start}{\mathsf{start}}

\newcommand{\sC}{\mathscr{C}}
\newcommand{\sF}{\mathscr{F}}
\newcommand{\atom}{\mathsf{atom}}

\newcommand{\topo}{{topology}}
\newcommand{\topos}{{topologies}}

\newcommand{\Topos}{{Topologies}}
\newcommand{\tpy}{\mathbb{T}}

\newcommand{\nbh}{{neighbourhood}}

\newcommand{\nb}{N}

\newcommand{\prog}{\mathbf{P}}

\newcommand{\locs}{\mathbf{L}}

\newcommand{\iv}[2][\sval]{#1(#2)}

\newcommand{\lmt}[1]{#1_{\infty}}
\newcommand{\cls}[1]{{\sf CL}(#1)}
\newcommand{\clsk}[2]{{\sf CL}(#1,#2)}

\let\textAA\AA
\renewcommand{\AA}{\ifmmode\mathbb{A}\else\textAA\fi}
\newcommand{\BB}{\mathbb{B}}

\newcommand{\DD}{\mathbb{D}}

\newcommand{\LL}{\mathbb{L}}

\newcommand{\NN}{\mathbb{N}}

\let\textSS\SS
\renewcommand{\SS}{\ifmmode\mathbb{S}\else\textSS\fi}
\newcommand{\TT}{\mathbb{T}}

\newcommand{\cC}{\mathcal{C}}

\newcommand{\cF}{\mathcal{F}}

\newcommand{\cH}{\mathcal{H}}
\newcommand{\cI}{\mathcal{I}}
\newcommand{\cJ}{\mathcal{J}}

\newcommand{\cP}{\mathcal{P}}
\newcommand{\cQ}{\mathcal{Q}}

\newcommand{\cS}{\mathcal{S}}
\newcommand{\cT}{\mathcal{T}}

\newcommand{\cV}{\mathcal{V}}

\theoremstyle{definition}

\declaretheorem[name=Example,numberwithin=section]{example}
\declaretheorem[name=Definition,numberwithin=section]{definition}

\declaretheorem[name=Remark]{remark}
\declaretheorem[name=Theorem,numberwithin=section]{theorem}
\declaretheorem[name=Lemma,sibling=theorem]{lemma}
\declaretheorem[name=Proposition,sibling=theorem]{proposition}
\declaretheorem[name=Corollary,sibling=theorem]{corollary}
\declaretheorem[name=Claim,sibling=theorem]{claim}

\begin{document}
\title{Symmetric Proofs of Parameterized Programs}

\author{Ruotong Cheng}
\orcid{0009-0004-1857-7251}
\email{chengrt@cs.toronto.edu}
\affiliation{%
  \institution{University of Toronto}
  \city{Toronto}
  \country{Canada}
}

\author{Azadeh Farzan}
\orcid{0000-0001-9005-2653}
\email{azadeh@cs.toronto.edu}
\affiliation{%
  \institution{University of Toronto}
  \city{Toronto}
  \country{Canada}
}

\begin{abstract}
We investigate the problem of safety verification of infinite-state parameterized programs that are formed based on a rich class of {\em topologies}. We introduce a new proof system, called {\em parametric proof spaces}, which exploits the underlying {\em symmetry} in such programs. This is a {\em local} notion of symmetry which enables the proof system to reuse proof arguments for {\em isomorphic neighbourhoods} in program topologies. We prove a sophisticated relative completeness result for the proof system with respect to a class of universally quantified invariants. We also investigate the problem of algorithmic construction of these proofs. We present a construction, inspired by classic results in model theory, where an infinitary {\em limit} program can be soundly and completely verified in place of the parameterized family, under some conditions. Furthermore, we demonstrate how these proofs can be constructed and checked against these programs without the need for axiomatization of the underlying topology for proofs or the programs. Finally, we present conditions under which our algorithm becomes a decision procedure. 
\end{abstract}

\maketitle

\section{Introduction}\label{sec:intro}
This paper focuses on the problem of algorithmic safety verification of infinite-state parameterized programs, which appear in many contexts, from device drivers and file systems to distributed programs to highly parallel GPU code. Classically, a {\em parameterized program} $\mathcal{P}$ is an infinite family of programs $\{P_n\}_{n\in \NN_+}$ where $P_n$ stands for the member of family with $n$ threads. Our focus is on a subclass of safety properties whose violations can be encoded as the violation of a (thread) local assertion. 
A parameterized program is safe if every member of the infinite family is safe. This new dimension of infinity (in addition to program data domains) adds a new level of challenge to the verification problem. 
The problem is generally undecidable~\cite{AptK1986}, but since these programs are broadly used, having robust semi-decision procedures is highly desirable.

The technical challenge of verifying parameterized programs can alternatively be viewed as the discovery of global inductive invariants that prove their safety. These inductive invariants are, in general, arbitrarily quantified statements, with quantifier alternation \cite{PadonLSS2017}, which makes even checking the inductiveness of a candidate invariant tricky, let alone the automated discovery of one. Some success has been found in automated discovery of invariants that are purely universally quantified \cite{FarzanKP2015,HoenickeMP2017}, where the target programs belong to a limited class based on replicated identical threads communicating via shared memory.

Our target is a rich class of infinite-state parameterized programs over a rich class of underlying {\em topologies}. Consider, for example, threads on a ring topology, where an instance with four threads is illustrated in \Cref{fig:ring}. A thread is associated with each circle node, which is instantiated from a template, and a shared variable (over an infinite data domain) is associated with each rectangle node.
\piccaption{Ring\label{fig:ring}}\parpic[r]{\includegraphics[width=0.27\columnwidth]{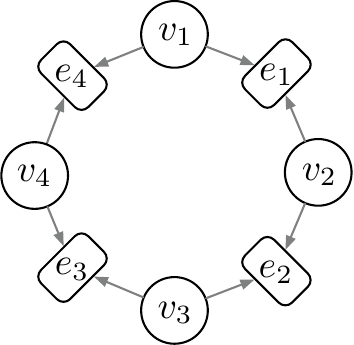}
}
\noindent This type of example lies outside the program models of \cite{EmersonN1995,FarzanKP2015,HoenickeMP2017}. Beyond the two dimensions of infinity of the program family and the data domain, there is an additional actor, the topology, that has to be formally modelled and used in the verification process.

We propose a new proof system, called a {\em parametric proof space}, which can be instantiated for different underlying topologies as parameters. The key insight of this work is that exploiting a generic notion of {\em symmetry}, based on the inherent symmetry present in the underlying topology family, is the key to the definition of this proof system. Remarkably, the symmetry captures all the information required from the topology family for the proof system without the need for a further axiomatization of the topologies.

Parametric proof spaces can be viewed as a generalization of \emph{proof spaces} introduced in \cite{FarzanKP2015}, which work only for a fixed topology family called the {\em star} topology (see \Cref{ex:star}): a collection of identical replicated threads sharing a bounded number of global variables as their means of communication. A proof space is a set of valid Hoare triples which is closed under {\em sequencing}, {\em conjunction}, and {\em symmetry}; that is, it is a theory of this deductive system. At the high level, the operations are defined as follows, where $\sigma$/$\sigma_1$/$\sigma_2$ are sequences of program commands:
\begin{itemize}
\item {\em Sequencing:}  if $\{P_1\}\;\sigma_1\;\{P_2\}$ and $\{P'_2\}\;\sigma_2\;\{P_3\}$ are valid triples, then so is $\{P_1\}\;\sigma_1;\sigma_2\;\{P_3\}$, where $P_2 \vdash P'_2$ and $\vdash$ is a {\em combinatorial entailment}\footnote{Without interpreting atomic predicates, checking if conjuncts that appear in $P'_2$ are a subset of the ones that appear in $P_2$}.
\item {\em  Conjunction:} if $\{P_1\}\;\sigma\;\{Q_1\}$ and $\{P_2\}\;\sigma\;\{Q_2\}$ are valid, then so is $\{P_1 \land P_2 \}\;\sigma\;\{Q_1 \land Q_2\}$.
\item {\em Symmetry:}  if $\{P\}\;\sigma\;\{Q\}$ is valid, then any triple $\{P'\}\;\sigma'\;\{Q'\}$ that can be obtained by consistently renaming thread identifiers in all three components of the triple is also valid.
\end{itemize}
A proof space $\cH$ is called finitely generated if it can be generated from a finite set of triples. 
$\cH$ proves a set of runs $\mathcal{R}$ correct with respect to a specification ${\text{\sf Pre}}/{\text{\sf Post}}$ if for each run $\sigma$ in $\mathcal{R}$, the Hoare triple $\{{\text{\sf Pre}}\}\;\sigma\;\{{\text{\sf Post}}\}$ belongs to $\cH$; that is, there is a derivation of $\{{\text{\sf Pre}}\}\;\sigma\;\{{\text{\sf Post}}\}$ using the sequencing, conjunction, and symmetry rules, from the axiom triples.

Parametric proof spaces retain the conjunction and sequencing rules, and propose a new {\em parametric symmetry rule}, to replace the symmetry rule of proof spaces, which is unsound for topologies other than the star topology. 
One can remedy this unsoundness by using isomorphisms of the underlying topologies to define the symmetry rule instead. However, this notion of symmetry is too restrictive and does not grant the proof system enough generalization power. 

Recall the ring example in Figure \ref{fig:ring}.
Suppose that each shared variable (on a rectangle node) is a Boolean $t$, representing a token. 
Consider a command $\sigma$ that passes a true token from right to left: \vspace{-3pt}
\[\texttt{assume t(r); t(r) := false; t(l) := true} \vspace{-3pt}\] 
where $r$ and $l$ refer to the rectangle nodes on the right and left, 
and the Hoare triple $\{\neg t(e_1) \}\; \tp{\sigma : v_3}\; \{ \neg t(e_1) \}$, which is valid because the node $v_3$ executing $\sigma$ should not affect the value of the boolean variable at $e_1$. It is natural to expect a symmetry rule to imply the validity of the Hoare triple $\{\neg t(e_4) \}\; \tp{\sigma : v_3}\; \{ \neg t(e_4) \}$, which is {\em valid for the same reason}. Yet, no isomorphism on a ring would fix $v_3$ and map $e_1$ to $e_4$. 
The situation becomes even more dire if we want the same type of valid Hoare triples in a ring of size 5 instead. There is no isomorphism mapping nodes in a ring of size 4 to those in a ring of size 5.

Our new symmetry rule uses a {\em local} notion of symmetry based on the concept of isomorphic {\em neighbourhoods} in the underlying structure of the program. We show that the new rule {\em soundly} (\Cref{thm:s-param-sound}) produces a new valid Hoare triple for a new neighbourhood based on an existing one for a symmetric neighbourhood. The use of {\em local symmetry} is significant, and we make this claim precise through a formal {\em relative completeness theorem}. Intuitively, we show that whenever the program can be proved safe using a type of universally quantified invariant, then it can be proved correct using a finitely generated parametric proof space (\Cref{thm:rc-informal}).

We define these universally quantified invariants as a generalization of Ashcroft invariants \cite{Ashcroft1975} in Section \ref{sec:gai}. The existing Ashcroft invariants in the literature \cite{FarzanKP2015,HoenickeMP2017} then become a special case of these for the star topology. We show that if a generalized Ashcroft invariant exists, then there is a way of extracting valid Hoare triples (\Cref{lem:ashcroft-ext,lem:fin-gen-ext}). The tricky part, however, is to show that finitely many such Hoare triples are sufficient to generate a parametric proof space that proves the entire parameterized program correct. This requires the family of topologies to satisfy a condition related to the width $k$ of the invariant (i.e., the number of universal quantifiers). Intuitively, this says that the topology family (1) includes enough small instances such that (2) in any instance of any size, any neighbourhood of size roughly $k$ can be covered by one of these small instances. Condition (1) provides the finite basis, and condition (2) guarantees the sufficiency of this basis (\Cref{thm:relative-completeness}).
Intuitively, when the topology family satisfies this condition, one can check the generalized Ashcroft invariant against the parameterized program through finitely many queries in the theory of the program data; these queries and the finite basis of the parametric proof space are closely related.

Certain topologies, like rings, do not satisfy this condition. Fortunately, one can {\em close} most of them {\em downwardly} by adding the missing substructures (e.g. pieces of rings) to the family, while preserving the safety of the parameterized program. We formalize such a construction in Section \ref{sec:subprograms}, which broadens the applicability of our relative completeness result.

As a proof system, algorithmic deduction for proof spaces involves arguing that they prove every member of a family of parameterized programs correct. To have effective algorithms, one has to deal with the dimensions of infinity present in both parametric proof spaces and families of parameterized programs. 
To ensure finite representability of parametric proof spaces for algorithms, we need additional conditions on the very general notion of symmetry. 
A key advantage of our methodology is that we do not need to axiomatize the topologies, although they play a central role in parametric proof spaces. We use the equivalence classes of our local symmetry equivalence relation instead, which can be represented through quantifier-free formulas based on {\em concrete} (tuples of) nodes from the topology.

To address the problem related to the infinitely many members in the family, we introduce {\em limit} programs. A limit program is a single infinitary program that (1) in some formal sense simulates behaviours of all members of the parameterized family, and (2) does not include additional behaviour that can change the safety verdict. Both conditions can be achieved as a consequence of the implicit property of the underlying topologies, that smaller instances of the parameterized family can be \emph{embedded} in larger ones. We propose a choice for {\em limit} programs in Section \ref{sec:limit-programs} that satisfies (1) and (2), and give model-theoretic conditions, formalizing the embedding property, under which the limit exists, can be constructed, and is finitely representable.
More specifically, under certain conditions, we can construct an infinitary topology that embeds the topologies for all members of the parameterized family without redundancy, and as such, the corresponding program can simulate the behaviours of all family members and does not introduce extra behaviours. 

Once we have a finitely representable parametric proof space and a finitely representable limit program, the following {\em operational} proof rule can be implemented as a verification algorithm:\vspace{-8pt}

\begin{equation}
  \AxiomC{$\mathcal{L}(P_{\sf limit}) \subseteq \mathcal{L}(\cH) $, \ $\cH$ is correct by construction}\label{rule:OR}\tag{OR}
  \UnaryInfC{$\forall i \in \mathcal{I}: P_i$ is safe}
  \DisplayProof \vspace{-2pt}
\end{equation}
where $\mathcal{L}(-)$ is used to denote both the behaviours of the limit program and the behaviours proved correct by proof space $\cH$. 

To effectively represent the parametric proof spaces and the limit program,  we generalize predicate automata (introduced in \cite{FarzanKP2015}) so that they can be defined parametrically on a choice of topology family. We then show that {\em parametric predicate automata} preserve the closure properties of predicate automata, which reduces the inclusion check from rule \ref{rule:OR} to an emptiness check on the language of a single automaton (\Cref{thm:PAL-closedness}). 

We present an algorithm for this emptiness check in the style of backward reachability in well-structured transition systems, which relies on the existence of a {\em well-quasi-order} for termination.  We give conditions on the structure that guarantee this termination and yield a decision procedure for emptiness (\Cref{thm:A-PA-emp}). When emptiness is decidable, the operation proof rule \ref{rule:OR} can be implemented as a refinement-style algorithm for the verification of parameterized programs. While the parametric proof space is not large enough to prove the program correct, the emptiness check yields counterexamples. These counterexamples are finite traces and can be verified using a standard verification scheme. The resulting concrete Hoare triples can then be added to the set of generators of the parametric proof space. This is the main reason why we formally define proof spaces using a set of concrete generators.

Finally, we turn our attention to the question of the decidability of safety verification of Boolean parameterized programs on arbitrary topologies \cite{AminofKRS+2018}. First, we present that parametric proof spaces are {\em complete} for these programs on any topology family (\Cref{thm:bool-c}). If, additionally, the conditions of algorithmic decidability are satisfied by the underlying topologies, our algorithmic framework yields a decision procedure (\Cref{cor:bool-d}), which is new for parameterized Boolean programs over this specific class of topology families (see Sections \ref{sec:example} and \ref{sec:related-work}).

\section{An Example Decidable Topology}\label{sec:example}
\begin{figure}[h]
\begin{center}
\includegraphics[scale=0.22]{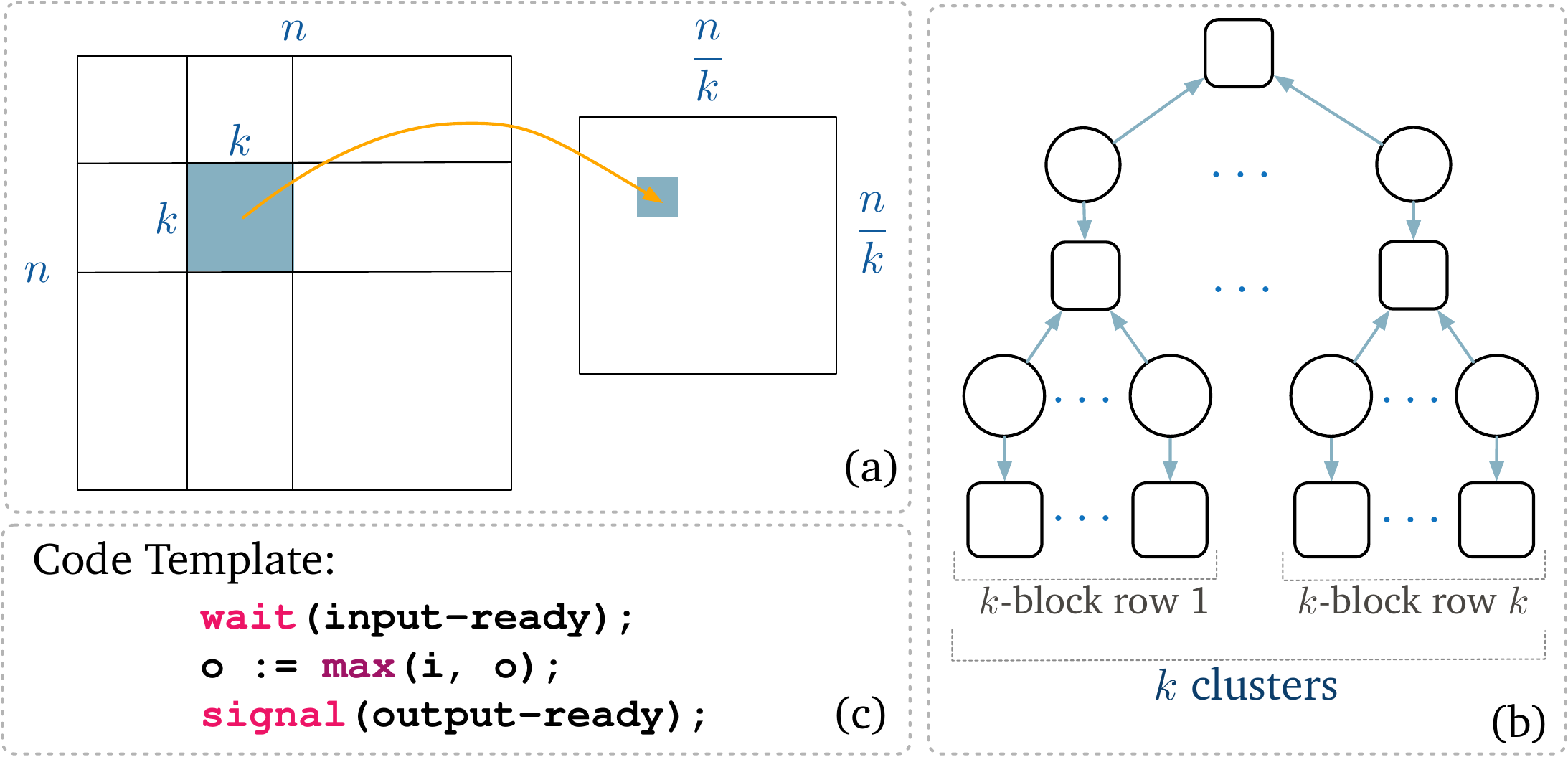}\vspace{-10pt}
\caption{Two-Phase Convolution (a,c). The topology (b).}\vspace{-10pt}
\label{fig:example}
\end{center}
\end{figure}

To ground our contributions, we use an example from parallel computing to present a class of parameterized programs over a new family of topologies that admit decidable proofs through parametric proof spaces, which are not covered by previous work \cite{AminofKRS+2018,FarzanKP2015}. Figure \ref{fig:example}(a) illustrates an instance of a {\em convolution with separable filters}, from parallel image processing. Under certain conditions, a filter with $O(m^2)$ operations can be executed in $O(2m)$ steps as {\em a sequence of two separate filters}. Consider the $n^2$ grid as an image in greyscale and the goal of reducing the image quality by replacing each $k^2$ partition with its maximum value. This can be done by first convolving each row and then convolving each column, due to the associativity of the operation $\mathtt{max}$. The underlying topology of this parallel program is a {\em forest} of trees of the form illustrated in Figure \ref{fig:example}(b). Each tree corresponds to the computation performed on a distinct $k \times k$ square. Each square node in the tree represents an array cell, and each circle node runs the code template illustrated in Figure \ref{fig:example}(c). Intuitively, the squares in the bottom row (from left to right) are the cells of one $k \times k$ block enumerated row-wise, and the squares in the middle row are the cells of the resulting array after the first convolution phase, enumerated column-wise. The root is the cell of the output array recording the result of the convolution. 

First, a {\em limit program} exists for the program family (with parameters $k$ and $n$), that satisfies the conditions of \Cref{thm:prog-PAL} and it can be constructed as a Fra\"{i}ss\'{e} limit (see \Cref{prop:fraisse}). The underlying (limit) topology consists of $\omega$ (i.e., countably infinite) copies of trees of depth 5, where each tree has a branching factor $\omega$ replacing $k$. Incidentally, this holds for any family of fixed-depth, arbitrary-branching trees (see \Cref{app:example} for a proof sketch).

Furthermore, for any \emph{monadic} parametric proof space for this program, the state space admits a well-quasi-order through {\em Kruskal's tree theorem}. This implies that checking any monadic parametric proof space candidate is {\em decidable} (see \Cref{lem:wsts}).  

\def\upd{\mathtt{upd}}

When the array data is finite, for example, if we have boundedly many shades of grey in the greyscale image, then the results in this paper culminate in a new decision procedure for verifying this program: We prove that there exists a universal monadic parametric proof space that proves the program correct (see \Cref{thm:bool-c}) and checking this proof is decidable (see \Cref{cor:bool-d}).

\section{Background} \label{sec:background}

We recall relevant basic concepts from model theory. A \emph{vocabulary} consists of predicate symbols and function symbols, and we assume a global map $\ar$ from symbols to their arity. We treat equality as a logical symbol. A vocabulary without function symbols is \emph{relational}. A \emph{$\cV$-structure} $\AA$ is defined as an \emph{underlying set} with an \emph{interpretation} of the symbols in $\cV$. We abuse notation and use $\AA$ to also refer to the underlying set of $\AA$.

The vocabulary $\cV[\AA]$ is the expansion of the vocabulary $\cV$ with one fresh constant symbol for every element of $\AA$. When a $\cV$-structure $\AA$ is viewed as a $\cV[\AA]$-structure, the constant symbol for every $a \in \AA$ is interpreted as $a$.

Let $\AA, \BB$ be two $\cV$-structures. An \emph{embedding} $f : \AA \hookrightarrow \BB$ is an injection that preserves and reflects all predicates and functions; when such an embedding exists, we say $\AA$ is an \emph{embedded substructure} of $\BB$. We say $\AA$ is a \emph{substructure} of $\BB$ if the underlying set of $\AA$ is a subset of that of $\BB$, and the inclusion map is an embedding from $\AA$ into $\BB$. An \emph{isomorphism} is a surjective embedding. An \emph{automorphism} of $\AA$ is an isomorphism from $\AA$ to itself.

The \emph{substructure generated} by a subset $A$ of the underlying set of $\AA$ is the smallest substructure containing $A$ that is closed under applying functions from $\cV$. We say $\AA$ is \emph{homogeneous} if any isomorphism between two finitely generated substructures extends to a full automorphism.

\section{Parameterized Concurrent Programs}\label{sec:programs}

The goal of this section is to formally define a parameterized program, which is a family of programs defined over an underlying family of topologies. In Section \ref{sec:prog-topo}, we formally define topologies as (parametric) logical structures, and then formalize the syntax and semantics of programs. Then, in Section \ref{sec:param-prog-topo}, we formally define parameterized programs and a notion of {\em symmetry} for them that is later used in Section \ref{sec:param-prog-topo} to define parametric proof spaces.  

\subsection{Programs over \Topos} \label{sec:prog-topo}

Our programs, denoted as $\prog = \tp{P, \sem{-}}$, comprise a syntactic part $P$ and a semantic part $\sem{-}$. The syntax is intuitively defined over program (communication) topology, which is formally defined as a structure $\tpy$, over some vocabulary $\VNode$, that has a sufficient amount of information to precisely define the set of {\em syntactic runs} of the program; i.e. a sequence of program commands executed by the program without any meaning attached to them. For convenience, we refer to this structure as the \topo. In turn, $\sem{-}$ is defined over a \emph{data domain} $\DD$, which is a structure over some vocabulary $\VData$, that captures the type of data that is stored in program variables and is manipulated by the program commands.

Formally, a \topo~ is a (possibly) infinite logical structure $\tpy$, whose underlying elements are the \emph{nodes} in the network\footnote{If all function symbols in $\VNode$ have arity $1$, when it comes to the underlying sets, the class of substructures of $\tpy$ is indeed a topology on $\tpy$, rendering it a topological space. Nevertheless, we choose to model the communication topology as a logical structure due to the ease of integration into a proof system.}. We refer to the substructure generated by a set of nodes $V$ as the \emph{\nbh} of $V$ denoted by $\nb(V)$. We occasionally abuse notation and consider a tuple of nodes $\vec{v}$ to be a set and hence let $\nb(\vec{v})$ be the neighbourhood of the set nodes in the tuple $\vec{v}$.

To build on a \topo~ to get a complete program syntax, we further assign a variable\footnote{To encode several local variables, this variable can have a structure type.} and a process to each node in $\tpy$. 
We then assume a finite set of {\em commands} $\Sigma$. Each command $\sigma \in \Sigma$ becomes a program command by being indexed with a node $v$ from $\tpy$: $\tp{\sigma: v}$. 

A global state of the program is the valuation of each variable and program location for each running process. For the latter, we fix a (finite) set of program locations $\locs$, with a distinct location $l_\init$ for the initial location. Then, we write $\gstate(\TT) := (\TT \to \locs) \times (\TT \to \DD)$. To formalize $\sem{-}$, we also define $\ldata(v) := \nb(v) \to \DD$ for every node $v$.

A \emph{concurrent program over a {\topo}~ $\tpy$} is formally a pair $\prog = \tp{P, \sem{-}}$, where $P : \tpy \to 2^\Sigma$ gives the syntactic runs of the program and  $\sem{-} : \prod_{\tp{\sigma : v} \in \Sigma \times \tpy} \ldata(v) \times \ldata(v)$ describes the semantics of (indexed) commands; we enforce further conditions on them when it comes to parameterized programs (\Cref{def:pp}).

\paragraph{\bfseries Syntactic Runs of The Program} The program effectively comprises a set of nodes, each running some code in parallel. Here, we formally define the set of syntactic runs, which is intuitively the set of interleaved (syntactic) runs from each node. 

Let functions $\src, \tgt : \Sigma \to \locs$ map a command to its \emph{source} and \emph{target}. Intuitively, these functions distinguish different occurrences of the same command in a control flow graph. Combined with a map $P : \tpy \to 2^\Sigma$, they induce a transition relation between location maps $\tpy \to \locs$, intuitively a product control flow graph: For any pair of location maps $\loc, \loc'$, we define that $\loc \xrightarrow{\sigma : v} \loc'$ if $\sigma$ is in $P(v)$, the location $\loc(v)$ is the source of $\sigma$, and $\loc' = \loc \oplus \{ v \mapsto \tgt(\sigma) \}$.

We can then define the set of syntactic runs of $\prog$, more specifically based on its syntactic component $P$, as the set of all runs $\tp{\sigma_1 : a_1} \cdots \tp{\sigma_n : a_n} \in (\Sigma \times \tpy)^*$ such that there exist location maps $\loc_0, \ldots, \loc_n : \tpy \to \locs$ satisfying $\loc_0 \equiv l_\init$ and $\loc_{i-1} \xrightarrow{\sigma_i : a_i} \loc_i$ for all $i$. This set of syntactic runs is denoted by $\Lang(P)$.

\paragraph{\bfseries Semantics} The semantics $\sem{-} : \prod_{\tp{\sigma, v} \in \Sigma \times \tpy} \lstate(v) \times \lstate(v)$ describes how an indexed command $\tp{\sigma : v}$ changes the valuation on the neighbourhood $\nb(v)$. For any larger neighbourhood $V \supset \nb(v)$, extend the semantics of $\tp{\sigma : v}$ to valuations on $V$ by declaring that the command does not access (i.e. read or write) anything outside $\nb(v)$; formally, set $\Global_V\sem{\sigma : v}$ to
\[
 \left\{ (\mu, \mu') \in (V \to \DD)^2 : (\mu, \mu')|_{\nb(v)} \in \sem{\sigma : v},\, \mu|_{\setcomp{\nb(v)}} = \mu'|_{\setcomp{\nb(v)}} \right\},
\]
where the notation $(\mu, \mu')|_X$ abbreviates $(\mu|_X, \mu'|_X)$. Via relation composition, we lift $\Global_V\sem{-}$ from individual commands to any run $\rho \in (\Sigma \times \tpy)^*$ whose nodes are contained in $V$.

\paragraph{\bfseries Transition Relation} 

The syntax and semantics of a program $\prog = \tp{P, \sem{-}}$ over topology $\TT$ induce a transition relation over global states $\tp{\loc, \val} \in \gstate(\TT)$: For any indexed command $\tp{\sigma : v}$, we say $\tp{\loc, \val} \xrightarrow{\sigma : v} \tp{\loc', \val'}$ if $\loc \xrightarrow{\sigma : v} \loc'$ and $(\val, \val') \in \Global_\TT\sem{\sigma : v}$. We say $\tp{\loc, \val} \to \tp{\loc', \val'}$ if $\tp{\loc, \val} \xrightarrow{\sigma : v} \tp{\loc', \val'}$ for some $\tp{\sigma : v}$.

\paragraph{\bfseries Safety Specifications}

In this paper, we focus on a class of safety specifications where violation of safety is witnessed by violation of a (local) assertion in one thread. To formally encode the specification, we assume a distinct {\em error} location, denoted by $l_\err$, in the set of program locations $\locs$. The set of \emph{error runs} of the program, denoted by $\Lang_\err(P)$, is then the subset of the syntactic runs $\Lang(P)$ that end with a location map whose range contains $l_\err$.

A run $\rho \in \Lang(P)$ is \emph{infeasible} if $\Global_\tpy\sem{\rho}$ does not contain any pair $(\mu, \mu')$; that is, no state at an error location is reachable, starting from any initial state. We say the program $\tp{P, \sem{-}}$ is \emph{safe} if every error run is infeasible; equivalently, for any location maps $\loc_\init \equiv l_\init$ and $\loc_\err$ whose range contains $l_\err$, for any valuations $\mu$ and $\mu'$, the relation $\tp{\loc_\init,\mu} \to^* \tp{\loc_\err,\mu'}$ does not hold.

\subsection{Parameterized Programs over a Family of \Topos} \label{sec:param-prog-topo}

In the context of this paper, a \emph{parameterized program} is a family of programs whose syntax and semantics are {\em symmetric} in a formally definable sense. This notion of symmetry is defined using \emph{local isomorphisms} and their pullbacks.

Let $\tpy$ and $\tpy'$ be two structures over the same vocabulary. Given tuples $\vec{a} \in \tpy^d$ and $\vec{b} \in \tpy'^d$, an isomorphism $\beta : \nb(\vec{a}) \to \nb(\vec{b})$ is called {\em local} if $\beta(\vec{a}) = \vec{b}$ ($\beta$ applies to $\vec{a}$ component-wise). A local isomorphism $\beta$ from $\nb(\vec{a})$ to $\nb(\vec{b})$ is uniquely identifiable by the pair of node tuples $\vec{a}$ and $\vec{b}$:
\begin{proposition} \label{prop:li-unique}
    Let $\tpy$ and $\tpy'$ be two structures over the same vocabulary. For any $\vec{a} \in \TT^d$ and $\vec{b} \in \TT'^{d}$, there is at most one isomorphism from $\nb(\vec{a})$ to $\nb(\vec{b})$ that maps $\vec{a}$ to $\vec{b}$.
\end{proposition}
\noindent Hence, we refer to it as the local isomorphism \emph{from $\vec{a}$ to $\vec{b}$} for short,  which induces an (equivalence) relation on tuples over $\TT$:

\begin{definition}[Local Symmetry $\simeq$]\label{def:ls}
    Let $\tpy$ and $\tpy'$ be two structures over the same vocabulary. For any node tuple $\vec{a}$ over $\tpy$ and $\vec{b}$ over $\tpy'$, we say $\vec{a}$ and $\vec{b}$ are \emph{locally symmetric}, denoted as $\vec{a} \simeq \vec{b}$, if they have the same length and there is a local isomorphism from $\vec{a}$ to $\vec{b}$.
\end{definition}

\begin{example} \label{ex:ring-ls}
    The ring topology of Figure \ref{fig:ring} can be modelled with the vocabulary $\VNode = \tp{\{ d, p\}, \{ l, r \}}$, where $d$ and $p$ are unary predicate symbol indicating whether the node is data (rectangle) or a process (circle), and $l$ and $r$ are unary function symbols representing the left and right node of a process node; for a data node $e$, we set $l(e) = r(e) = e$. We have $(e_1, v_3, e_1) \simeq (e_4, v_3, e_4)$ but $(e_1, v_3, e_1) \not\simeq (e_2, v_3, e_2)$.
\end{example}

First, we observe that $\simeq$-equivalence tuples satisfy the same set of quantifier-free formulas, and vice versa: 
\begin{proposition} \label{prop:local-symm-qf}
    Let $\TT$ and $\TT'$ be structures over the vocabulary $\cV$. For any $\vec{a} \in \TT^d$ and $\vec{b} \in {\TT'}^d$, we have $\vec{a} \simeq \vec{b}$ if and only if for every quantifier-free $\cV$-formula $\varphi[x_1, \ldots, x_d]$, $\TT \models \varphi(\vec{a}) \iff \TT' \models \varphi(\vec{b})$.
\end{proposition}

If, additionally, $\TT^d/{\simeq}$ is finite for all dimensions $d$, then every equivalence class can be defined by a unique quantifier-free formula: \vspace{-10pt}
\begin{proposition} \label{prop:local-symm-fin}
  For any $d \in \NN_+$, if the quotient $\TT^d/{\simeq}$ is finite, then every equivalence class of $\simeq$ is definable by a quantifier-free $\cV$-formula.
\end{proposition}

The local isomorphism $\beta : \nb(\vec{a}) \to \nb(\vec{b})$ induces a map $\beta^* : (\nb(\vec{b}) \to \DD) \to (\nb(\vec{a}) \to \DD)$ between valuations by $\mu \mapsto \mu \circ \beta$. We lift $\beta^*$ to (sets of) tuples of valuations in the natural way.

Intuitively, a {\em parameterized program} is a family of programs defined over structures over the same vocabulary, where any pair of locally symmetric nodes run the same code with the same semantics for individual commands. Formally:

\begin{definition}[Parameterized Program] \label{def:pp}
Given a vocabulary $\VNode$ for \topos{} and a data domain $\DD$ over a vocabulary $\VData$, a parameterized program is a family of programs where:
\begin{itemize}
\item Each program $\tp{P, \sem{-}}$ in the family is defined over a \topo~ over the vocabulary $\VNode$, the data domain $\DD$, and the same set of commands $\Sigma$ and locations $\locs$.
\item For every pair of programs $\tp{P, \sem{-}}$ over $\tpy$ and $\tp{P', \sem{-}'}$ over $\tpy'$ in the family, for any $a \in \tpy$ and $a' \in \tpy'$, if $a \simeq a'$ via $\beta$, i.e., $a' = \beta(a)$, then $P(a) = P'(a')$ and $\sem{\sigma : a} = \beta^* \sem{\sigma : a'}$ for every $\sigma \in \Sigma$.
\end{itemize}
\end{definition}
Note that the second bullet point also applies to a pair of the same program. We say a \emph{program} $\prog$ is \emph{locally symmetric} if $\{\prog\}$ satisfies \Cref{def:pp}. Clearly, every member of a parameterized program is locally symmetric.

We say a parameterized program is {\em safe} if every program in the family is safe. We write the parameterized program as an indexed set $\{\prog_i\}_{i\in\cI}$, where $\cI$ is possibly infinite.

\parpic[r]{\includegraphics[width=0.2\columnwidth]{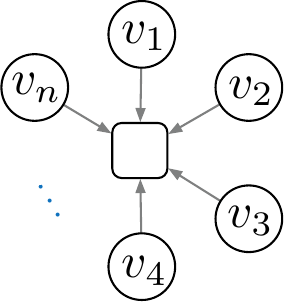}
}\vspace{-7pt}\begin{example}\label{ex:star}
    Let us see how the most classic instance of parameterized programs, namely shared memory concurrent programs in which the threads are indistinguishable  \cite{FarzanKP2015}, is a special case of Definition \ref{def:pp}. The topology is illustrated on the right. We use a constant symbol $\mathsf{g}$ to model the global (shared) variables, hosted on the square node in the middle. The $n$ circle nodes host the $n$ threads. Formally, the vocabulary is $\VNode = \tp{\varnothing, \{ \mathsf{g} \}}$, and the $\VNode$-structure $\TT_n$ corresponding to a program with $n$ threads has the universe $\{ -1, 0, \ldots, n-1 \}$, where $-1$ is the interpretation of $\mathsf{g}$ and the natural numbers model the $n$ threads.
\end{example}

Our definition of parameterized programs is intentionally very general. Notably, the parameterized family does not have to contain programs over every instance of a type of communication topology. For instance, given a parameterized program $\cP$ described by Example \ref{ex:star}, if we remove the program over $\TT_{42}$ from $\cP$, the result is still a parameterized program. In Section \ref{sec:subprograms}, we take advantage of the flexibility afforded by this generality.

We add restrictions to the family when they are required for proving specific results. For instance, the relative completeness theorem requires restrictions for the topologies (Theorem \ref{thm:relative-completeness} and Proposition \ref{prop:topo-sub}).

\section{Parametric Proof Spaces} \label{sec:proofs}
We outline in Section \ref{sec:intro} how a {\em parametric proof space} is a proof space that is parametric on a notion of local symmetry from the underlying topology of the program. We formally defined what {\em local symmetry} is for an arbitrary structure in Definition \ref{def:ls}. As the first step of defining a symmetry rule, we need to determine what the appropriate underlying structure for our parametric proof space is, so that we can use its local symmetry. 


Recall that our parameterized family of programs looks like $\{ \tp{P_i, \sem{-}_i} \}_{i\in\cI}$, where each $\tp{P_i, \sem{-}_i}$ is over the topology $\TT_i$. The idea, however, is for the proof space to provide a proof of correctness for every member of this family. Hence, we take a (generally infinite) structure $\TT$, over the common vocabulary $\VNode$, of which every $\TT_i$ is an embedded substructure. We call $\TT$ a \emph{universal structure}.

When the vocabulary has no constant symbol\footnote{When modelling a topology as a logical structure, it is always possible to replace constant symbols with unary function symbols interpreted as constant functions.}, such a structure can simply be defined over the disjoint union of the universe of each $\TT_i$; another way of producing universal structures is presented in Section \ref{sec:limit-programs}. We identify each $\TT_i$ with one of the substructures of $\TT$ that is isomorphic to $\TT_i$.

Recall that we define parameterized programs via the local symmetry relation $\simeq$. In Section \ref{sec:nrule}, we present the new symmetry rule based on the same local symmetry relation, on the universal structure $\TT$. Before that, we start by defining what the valid Hoare triples are in such a universal structure.

\subsection{Hoare Triples in the Universal Structure} \label{sec:htri}

The semantics of the indexed commands can be extended to the elements in the universal structure $\TT$: For any $\tp{\sigma : v} \in \Sigma \times \TT$, for any $i$ and $u \in \TT_i$ with a local isomorphism $\beta$ from $v$ to $u$
\begin{equation*}
\sem{\sigma : v} = 
\begin{cases}
\varnothing & \text {if no such $i$ and $u$ exist} \\
\beta^*\sem{\sigma : u}_i & \text{otherwise}
\end{cases}
\end{equation*}
The definition of universal structures allows nodes in $\TT$ that do not correspond to any node in any $\TT_i$, which is why the first case may happen.

We consider Hoare triples whose pre- and post-conditions are assertions about variables indexed by nodes. In Section \ref{sec:prog-topo}, we have assumed without loss of generality that there is a unique variable on each node $u$; we denote that variable by $\sval(u)$, where $\sval$ is a formal symbol of arity 1. Using $\VData$-formulas $\phi[x_1, \ldots, x_n]$, we construct a set of formulas, denoted as $\Assertions$, of the form $\phi[\iv{u_1}, \ldots, \iv{u_n}]$ where $u_1, \ldots, u_n$ are nodes in $\TT$. For an assertion $\varphi$, let $\nodes(\varphi)$ be the set of nodes appearing in $\varphi$.

Given $\DD$ and $\TT$, the truth value of an assertion $\varphi$ depends only on the interpretation of $\sval$, which is given by a valuation $\mu : U \to \DD$ where $U$ is a subset of $\TT$ containing $\nodes(\varphi)$. We define the satisfiability relation $\val \models \varphi$ in the standard way.

A Hoare triple has the form
\[ \htri{\phi[\overrightarrow{\iv{u}}] }{\tp{\sigma_1 : v_1} \cdots \tp{\sigma_n : v_n}}{\psi[\overrightarrow{\iv{w}}]}, \]
where $\overrightarrow{\iv{u}}$ abbreviates the tuple $\iv{u_1}, \dots, \iv{u_n}$. Let $V$ be the neighbourhood $\nb(\vec{u} \vec{v} \vec{w})$. The Hoare triple above is \emph{valid} if for any valuations $\val, \val' : V \to \DD$ such that $(\val, \val') \in \Global_V\sem{\tp{\sigma_1 : v_1} \cdots \tp{\sigma_n : v_n}}$, if $\val \models \phi[\overrightarrow{\iv{u}}]$, then $\val' \models \psi[\overrightarrow{\iv{w}}]$.

\subsection{The New Symmetry Rule}\label{sec:nrule}
Our symmetry rule intuitively says that if one has a valid Hoare triple for a tuple of nodes, then the same Hoare triple is valid for any other tuple that is locally symmetric to this tuple. Formally:

\[
\inferrule [S-param]
  {\htri{\phi[\overrightarrow{\iv{u}}]}{\tp{\sigma_1 : v_1} \cdots \tp{\sigma_n : v_n}}{\psi[\overrightarrow{\iv{w}}]}}
  {\htri{\phi[\overrightarrow{\iv{u'}}]}{\tp{\sigma_1 : v'_1} \cdots \tp{\sigma_n : v'_n}}{\psi[\overrightarrow{\iv{w'}}]}}
\quad
\inferrule
  {|\vec{u}| = |\vec{u}'| \\\\
   |\vec{w}| = |\vec{w}'| \\\\
    \vec{u} \vec{v} \vec{w} \simeq \vec{u}' \vec{v}' \vec{w}'}
  {}
\]

\begin{example}[Continuation of Example \ref{ex:ring-ls}] \label{ex:ring-nrule}
    We use the disjoint union $\TT_\text{rings}$ of rings of all sizes as the universal structure. For the token-passing example in Section \ref{sec:intro}, via \textsc{S-param}, we can deduce from $\htri{\neg t(e_1)}{\tp{\sigma : v_3}}{\neg t(e_1)}$ any Hoare triple $\{\neg t(e)\}\;\{\tp{\sigma : v}\}\;\{\neg t(e')\}$ such that $\TT_\text{rings} \models \chi(e, v, e')$ where $\chi(e, v, e') := (e = e' \land d(e) \land p(v) \land l(v) \neq e \land r(v) \neq e)$, which defines the $\simeq$-equivalence class of $(e_1, v_3, e_1)$ (see Proposition \ref{prop:local-symm-qf}). In particular, the Hoare triple $\{\neg t(e_4)\}\;\{\tp{\sigma : v_3}\}\;\{\neg t(e_4)\}$ can be deduced.
\end{example}

\begin{theorem}[Soundness of \textsc{S-param}] \label{thm:s-param-sound}
    The rule \textsc{S-param} is sound for any tuples $\vec{u}$, $\vec{v}$, $\vec{w}$, $\vec{u}'$, $\vec{v}'$, and $\vec{w}'$ from $\TT$.
\end{theorem}

\subsection{Parametric Proof Spaces and Their Denotation}
We are finally ready to formally define parametric proof spaces.
\begin{definition}[Parametric Proof Space] \label{def:pps}
    A \emph{$\TT$-proof space} is a set $\cH$ of Hoare triples closed under the rules \textsc{Sequencing}, \textsc{Conjunction}, and \textsc{S-param}. We say $\cH$ is \emph{valid} if all Hoare triples in $\cH$ are. A \emph{parametric proof space} is any $\TT$-proof space.
\end{definition}

Algorithmically speaking, we are interested in parametric proof spaces that are finitely generated\footnote{In \cite{FarzanKP2015}, there are additional syntactic restrictions on the Hoare triples in a basis for algorithmic purposes. For separation of concerns, we lift those restrictions here; verification algorithms are discussed in Section~\ref{sec:predicate-automata}.}.

\begin{definition}[Finitely Generated] \label{def:fg}
    A $\TT$-proof space $\cH$ is \emph{generated} by a set $H$ of Hoare triples if $\cH$ is the smallest $\TT$-proof space containing $H$. The set $H$ is called a \emph{basis} for $\cH$. A $\TT$-proof space is \emph{finitely generated} if it admits a {\em finite} basis.
\end{definition}

A proof space is a proof in the sense that it can provide a formal argument for the correctness of a set of behaviours. 

\begin{definition}[Language of a Parametric Proof Space] \label{def:lang}
    The \emph{language} of a $\TT$-proof space $\cH$, denoted by $\Lang(\cH)$, is the set of words $\rho \in (\Sigma \times \TT)^*$ such that $\htri{\top}{\rho}{\bot} \in \cH$. A $\TT$-\emph{proof space language} is a language obtained in this way.
\end{definition}

A direct observation is that any proof space language is locally symmetric in the following sense: lifting the definition of $\simeq$ from tuples over $\TT$ to words over $\Sigma \times \TT$ by declaring $\tp{\sigma_1 : v_1} \cdots \tp{\sigma_n : v_n} \simeq \tp{\sigma'_1 : v'_1} \cdots \tp{\sigma'_{n'} : v'_{n'}}$ if and only if $(\sigma_1, \ldots, \sigma_n) = (\sigma'_1, \ldots, \sigma'_{n'})$ and $(v_1, \ldots, v_n) \simeq (v'_1, \ldots, v'_n)$, we have

\begin{theorem} \label{thm:psl-sym}
    A proof space language is closed under the equivalence relation $\simeq$.
\end{theorem}

Since we have identified each $\TT_i$ with some substructure of $\TT$, the language $\Lang(P_i)$ of a program in the parameterized family can also be viewed as over the alphabet $\Sigma \times \TT$. Then, through language inclusion check, proof spaces provide machinery for establishing the safety of a parameterized program.

\begin{theorem}[Sound Safety Proof Rule] \label{thm:safety-proof}
A parameterized program $\cP = \{ \tp{P_i, \sem{-}_i} \}_{i\in \cI}$ is safe if there is a universal structure $\TT$ and a valid $\TT$-proof space $\cH$ such that $\bigcup_{i\in \cI} \Lang_\err(P_i) \subset \Lang(\cH)$.
\end{theorem}

For a $\TT$-proof space $\cH$ to be an \emph{effective} proof, it needs to be finitely representable, and the check $\bigcup_{i\in\cI} \Lang_\err(P_i) \subset \Lang(\cH)$ needs to be decidable. We discuss these two issues in Section \ref{sec:predicate-automata}.

\section{Completeness of Parametric Proof Spaces}\label{sec:completeness}

In this section, we study the expressive power of parametric proof space as a proof system. In particular, the majority of this section is dedicated to the development of a {\em relative completeness} result with respect to safety proofs based on {\em a family of universally quantified inductive invariants} for arbitrary data domains. Additionally, we present completeness results for programs with only Boolean data domains in Section \ref{sec:bool}. 

Ashcroft invariants were first defined in the context of fixed-thread programs, for which they are relatively complete (in Cook's sense) as a proof method \cite{Ashcroft1975}. They can also be used as universally quantified inductive invariants for parameterized programs on a star topology \cite{FarzanKP2015,HoenickeMP2017}, though they are {\em incomplete}.  
In \cite{FarzanKP2015}, it is shown that \emph{finitely generated} (non-parametric) proof spaces are complete relative to Ashcroft invariants.
To link the expressive power of parametric proof spaces to Ashcroft invariants, in Sections \ref{sec:gai} and \ref{sec:subprograms}, we address two questions: (1) What is a natural generalization of Ashcroft invariants for our richer class of parameterized programs? (2) Does the correspondence between proof spaces and Ashcroft invariants established in \cite{FarzanKP2015} still hold for parametric proof spaces and these invariants?

We propose a generalization of Ashcroft invariants in \Cref{sec:gai}. 
Establishing a correspondence between generalized Ashcroft invariants and \emph{finitely generated} parametric proof spaces, however, requires an extra condition on the family of topologies that brings a sense of finiteness to them. We start with an informal version of our relative completeness result below, which is made precise through this section, culminating in Theorem \ref{thm:relative-completeness}.  
\begin{theorem}[Relative Completeness, Informal] \label{thm:rc-informal}
    Let $\cP$ be a parameterized program over finite topologies $\cT$ with universal structure $\TT$. For any generalized Ashcroft invariant $\Phi$ that proves $\cP$ safe, if the topology family $\cT$ satisfies some condition $C$ (related to the number of quantifiers in $\Phi$), then there is a finitely generated $\TT$-proof space that proves $\cP$ safe.
\end{theorem}
Condition $C$ intuitively says that there are finitely many small topologies in $\cT$ such that any topology in $\cT$ can be decomposed into a union of their isomorphic copies. Moreover, the granularity of this decomposition is related to the width of the formula $\Phi$; i.e. the number of (universal) quantifiers in it. 
When $C$ does not hold for a parameterized program as is, one can amend the topologies through a simple construction that behaves like a {\em downward closure} computation,  adding all missing small topologies to the family, while preserving the safety/unsafety of the original program. In Section \ref{sec:subprograms}, we present this construction and argue that for a wide range of parameterized programs, the condition holds.

\subsection{Generalized Ashcroft Invariants}\label{sec:gai}
\emph{Generalized Ashcroft assertions} are defined as a fragment of first-order formulas over a vocabulary $\cV_\Ashcroft$ of three sorts: $\Node$, $\Data$, and $\Loc$. We first define the vocabulary and then state the syntactic restrictions that define the fragment. Finally, given a parameterized program, a \emph{generalized Ashcroft invariant} is an \emph{inductive} generalized Ashcroft assertion that proves the program safe. From now on, we generally omit the adjective ``generalized'' unless it is otherwise ambiguous.

The vocabulary $\VNode$ (similarly, $\VData$) has a single sort $\Node$ (resp. $\Data$): a $k$-ary relation symbol in $\VNode$ is typed $\Node^k$ and a $k$-ary function symbol is typed $\Node^k \to \Node$. In addition, we define a vocabulary $\cV_\Loc$ consisting of a sort $\Loc$ and one $\Loc$-typed constant symbol for each element in $\locs$.

The vocabulary $\cV_\Ashcroft$ is the disjoint union of $\VNode$, $\VData$, and $\cV_\Loc$ along with two extra function symbols $\sval : \Node \to \Data$ and $\sloc : \Node \to \Loc$. As a convention, $u$, $v$, and $w$ stand for $\VNode$-sorted variables, and $x$, $y$, and $z$ for $\VData$-sorted variables.

The fragment \QFAshcroft{} consists of first-order formulas over $\cV_\Ashcroft$ where no quantification is over $\Node$ and every free variable is of sort $\Node$. Ashcroft assertions of \emph{width} $k$, denoted by $\Ashcroft[k]$, are sentences of the form
\[ \forall u_1, \ldots, u_k.\, \varphi[u_1, \ldots, u_k], \]
where $\varphi$ is a \QFAshcroft{} formula. 
The set of Ashcroft assertions \Ashcroft{} is the union of $\Ashcroft[k]$ for all $k \in \NN$.

Recall that we always fix the $\VData$-structure $\DD$ and the set $\locs$ of control locations, which we view as a $\cV_\Loc$-structure $\LL$. Therefore, for a specific topology, i.e., $\VNode$-structure $\TT$ (not necessarily a universal structure), the symbols in an Ashcroft assertion $\Phi$ are all interpreted except $\sloc$ and $\sval$; hence, the truth value of $\Phi$ is given by a pair $\tp{\loc, \mu}$, where $\loc : \TT \to \LL$ is a location map and $\val : \TT \to \DD$ is a valuation. Given $\TT$, we define the satisfiability relation $\tp{\loc, \val} \models \Phi$ and the entailment relation under $\TT$, denoted by $\Phi \models \Phi'$, in the standard way. Although both relations are parametric on $\TT$, since $\TT$ is clear from the context, we omit it from our notation.

\begin{definition}[Generalized Ashcroft Invariant] \label{def:gai}
    For a program $\prog = \tp{P, \sem{-}}$, a \emph{generalized Ashcroft invariant} is a formula $\Phi \in \Ashcroft$ such that
    \begin{description}
        \item[Initialization] $\forall u.\, \sloc(u) = l_\init \models \Phi$.
        \item[Continuation] For any state $\tp{\loc, \val}$, if $\tp{\loc,\mu} \models \Phi$ and $\tp{\loc,\mu} \to \tp{\loc',\mu'}$, then $\tp{\loc',\mu'} \models \Phi$.
        \item[Safety] $\Phi \models \forall u.\, \sloc(u) \neq l_\err$.
    \end{description}
    We say $\Phi$ is an Ashcroft invariant for a parameterized program $\{ \prog_i \}_{i\in\cI}$ if it is an Ashcroft invariant for all $\prog_i$.
\end{definition}
By definition, the existence of an Ashcroft invariant for a parameterized program implies its safety.

\subsection{Hoare Triples from Ashcroft Invariants}

To prove Theorem \ref{thm:rc-informal}, there are two main challenges. We need to (1) obtain {\em valid} Hoare triples from a generalized Ashcroft invariant, and (2) show that a finite set of such triples generates a parametric proof space that is sufficient to prove the program safe.

Regarding (1), one may observe a syntactic similarity between \QFAshcroft{} and \Assertions{} (see Section \ref{sec:htri}), but there are two gaps:
\begin{itemize}
    \item[(i)] A formula in \Assertions{} is \emph{closed}, with nodes from the universal structure, whereas a formula in \QFAshcroft{} uses \emph{free variables} of sort $\Node$.
    \item[(ii)] \Assertions{} only reason about \emph{data}, whereas the \emph{control location} of each node is accessible through the function symbol $\sloc$ in \QFAshcroft{}.
\end{itemize}

We address the first two challenges in this section by introducing \emph{extended Hoare triples} as an intermediate object that bridges the gaps mentioned above. Given an Ashcroft invariant $\Phi$ for $\cP = \{\prog_i\}_{i\in\cI}$ over topologies $\cT = \{\TT_i\}_{i\in\cI}$, we associate every finite topology $\TT_i$ with a finite set $H_{\Phi}^{\TT_i}$ of \emph{valid} extended Hoare triples. (Lemma~\ref{lem:ashcroft-ext}) Then, we construct a finitely generated proof space \emph{sufficient} to prove the safety of $\cP$ by selecting a finite subset $\{\TT_j\}_{j\in\cJ}$ of finite topologies (Section \ref{sec:finite-basis}) and translating the extended Hoare triples $\bigcup_{j \in \cJ} H_{\Phi}^{\TT_j}$ to standard Hoare triples (Lemma~\ref{lem:fin-gen-ext}).

The Hoare triples in Section \ref{sec:htri} use assertions of the form $\phi[\overrightarrow{\iv{v}}]$, where $\phi[\vec{x}]$ is a $\VData$-formula. An \emph{extended} assertion is of the form $\psi[\overrightarrow{\iv{v}}, \overrightarrow{\sloc(v)}]$, where $\sloc$ is interpreted as a location map, and $\psi[\vec{x}, \vec{l}]$ is a $\cV_\Ashcroft$-formula. Based on these extended assertions, we define extended Hoare triples and extended parametric proof spaces. To avoid tedium, we omit their formal definitions. It is intuitively straightforward to observe why the soundness of the three derivation rules for parametric proof spaces remains: adding the $\sloc$ symbol can be thought of as introducing an auxiliary variable $\iv[\sloc]{a}$ for every node $a$ that maintains the program counter.

There is a simple connection between extended assertions and \QFAshcroft{} formulas. Intuitively, simplifying all interpreted terms and sub-formulas in $\varphi[\vec{a}]$ exactly gives an equivalent extended assertion. Formally:  
\begin{proposition} \label{prop:ext-ass}
    Given a $\VNode$-structure $\TT$, for any \QFAshcroft{} formula $\varphi[u_1, \ldots, u_n]$ and $\vec{a} \in \TT^n$, there is a unique (up to equivalence) extended assertion $\phi[\overrightarrow{\sval(b)}, \overrightarrow{\sloc(b)}]$, where $\vec{b}$ is a tuple from $\nb(\vec{a})$, such that for any local isomorphism $\beta$ from $\vec{a}$, the extended assertion $\phi[\overrightarrow{\sval(\beta(b))}, \overrightarrow{\sloc(\beta(b))}]$ is equivalent to $\varphi[\beta(\vec{a})]$.
\end{proposition}
As a slight abuse of notations, we refer to that extended assertion simply as $\varphi[\vec{a}]$. Then, over a finite topology $\TT_i \subset \TT$, an Ashcroft assertion $\Phi := \forall \vec{u}.\, \varphi[\vec{u}]$ of width $k$ can be turned into an extended assertion with the same set of models, namely $\bigwedge_{\vec{a} \in \TT_i^k} \varphi[\vec{a}]$. We can then associate $\Phi$ with a set $H_\Phi^{\TT_i}$ of extended Hoare triples defined as: For all indexed command $\tp{\sigma : a_0} \in \Sigma \times \TT_i$, tuple $\vec{a} \in \TT_i^k$, and node $b \in \TT_i$, 
\begin{gather}
    \textstyle
    \htri{\bigwedge_{a \in \TT_i} \sloc(a) = l_\init}{\epsilon}{\varphi[\vec{a}]} \label{htri:init} \tag{\sc Init}\\
    \textstyle
    \htri{\bigwedge_{\vec{a} \in \TT_i^k} \varphi[\vec{a}]}{\tp{\sigma : a_0}}{\varphi[\vec{a}]} \label{htri:cont} \tag{\sc Cont}\\
    \textstyle
    \htri{\bigwedge_{\vec{a} \in \TT_i^k} \varphi[\vec{a}]}{\epsilon}{\sloc(b) \neq l_\err} \label{htri:err} \tag{\sc Safe}
\end{gather}
They are valid as long as the Ashcroft assertion $\Phi$ is an Ashcroft invariant:
\begin{lemma} \label{lem:ashcroft-ext}
    Let $\prog$ be a program over a finite topology $\TT_i \subset \TT$. For any Ashcroft assertion $\Phi$, the extended Hoare triples in $H_{\Phi}^{\TT_i}$ are valid if $\Phi$ is an Ashcroft invariant for $\prog$.
\end{lemma}

Given an Ashcroft invariant for a parameterized program, the lemma provides a way of extracting valid extended Hoare triples from program instances over finite topologies. It does not, however, guarantee that \emph{finitely} many such extended Hoare triples are sufficient to generate a $\TT$-proof space that proves the \emph{entire} parameterized program correct. We address this in the next section. 

Before we conclude, we observe that extended Hoare triples can be translated back to standard Hoare triples. 
\begin{lemma} \label{lem:fin-gen-ext}
If there is a finitely generated extended $\TT$-proof space that proves $\cP$ safe, then there is a finitely generated standard $\TT$-proof space that proves $\cP$ safe.\end{lemma}
Note that we have not formally stated in the text of this paper what it means for an extended $\TT$-proof space to prove $\cP$ safe, which is formalized in the proof of this lemma in \Cref{app:completeness}.

\subsection{Guaranteeing a Finite Basis} \label{sec:finite-basis}

Suppose $\TT_j$ is a finite topology. Let us consider the question: How can $H_{\Phi}^{\TT_j}$, the set of Hoare triples obtained from the instantiation of the invariant $\Phi$ on $\TT_j$,  help prove the safety of a different program $\prog_i$ ($i \neq j$, and $\TT_i$ is possibly infinite) in the family? Suppose we are given the validity of extended Hoare triples for $\prog_j$ of the form \eqref{htri:cont} by \Cref{lem:ashcroft-ext}. If there is an isomorphism $\beta : \TT_j \to \TT'_i$ such that $\TT'_i$ is a substructure of $\TT_i$ containing $a_0$ and $\vec{a}$, then the valid Hoare triple from $H_{\Phi}^{\TT_j}$
\begin{equation}
    \textstyle \htri{\bigwedge_{\vec{a} \in \TT_j^k} \varphi[\beta^{-1}(\vec{a})]}{\tp{\sigma : \beta^{-1}(a_0)}}{\varphi[\beta^{-1}(\vec{a})]} \label{htri:Tj}
\end{equation}
is generalized through the symmetry rule \textsc{S-Param} to the triple 
\begin{equation}
    \textstyle \htri{\bigwedge_{\vec{a} \in {\TT_i'}^k} \varphi[\vec{a}]}{\tp{\sigma : a_0}}{\varphi[\vec{a}]} \label{htri:Ti-prime}
\end{equation}
which implies the validity of the Hoare triple of the form \eqref{htri:cont} for $\prog_i$. The derivation of the Hoare triple \eqref{htri:Ti-prime} from \eqref{htri:Tj} is enabled by the existence of the isomorphism $\beta$, which is a property of the topology family. The following definition, with the \emph{rank} set to $k+1$ ($k$ being the width of $\Phi$), captures a precise property of the topology family that guarantees this process can be done for any $\TT_i \in \cS$, $\TT_j \in \cT$, and extended Hoare triples of all three forms:
\begin{definition} \label{def:topo-basis}
    Let $k \in \NN_+$ and $\cT$ be a family of topologies over a common vocabulary. A \emph{basis of rank $k$} for $\cT$ is a subset $\cS$ of $\cT$ such that for any $\TT \in \cT$ and $\vec{a} \in \TT^k$, there is some $\SS \in \cS$ and embedding $f : \SS \hookrightarrow \TT$ such that $\vec{a} \in f(\SS)^k$.
\end{definition}

\parpic[r]{\includegraphics[width=0.53\columnwidth]{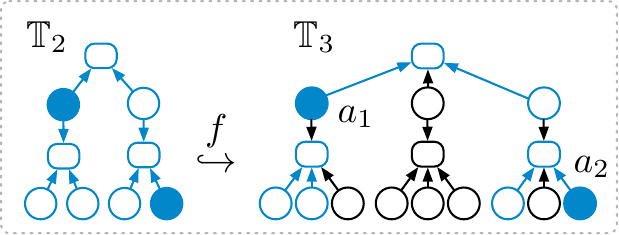}
}\vspace{-7pt}
\begin{example} \label{ex:trees-basis}
Consider a simplification of the topologies from Figure \ref{fig:example}, where every $\TT_i$ is a tree where every rectangle node has $i$ indistinguishable circle nodes as children.
    $\{\TT_1, \TT_2\}$ is a basis of rank $2$ for $\{\TT_i\}_{i\in\NN_+}$. A tuple $(a_1, a_2) \in \TT_3^2$ and an embedding $f : \TT_2 \hookrightarrow \TT_3$ whose image contains $a_1, a_2$ is illustrated.
\end{example}

\begin{remark} \label{rmk:topo-basis}
    Not every family of topologies admits a finite basis: if $\cT$ is {\em rings of all sizes}, then for any $k \in \NN_+$, the only basis of rank $k$ for $\cT$ is itself, as no ring can be embedded in another ring.
\end{remark}

We are ready to formalize our informal Theorem \ref{thm:rc-informal} fully by defining condition $C$: $\cT$ has a finite basis of rank $k+1$ consisting of finite topologies, where $k$ is the width of the Ashcroft invariant. Note that the larger $k$ is, the stricter this condition is. 
The full version of the theorem is as follows:

\begin{theorem}[Relative Completeness] \label{thm:relative-completeness}
    Let $\cP = \{\prog_i\}_{i\in\cI}$ be a parameterized program over a family $\cT = \{\TT_i\}_{i\in\cI}$ of topologies with universal structure $\TT$. If there is an Ashcroft invariant $\Phi$ of width $k$ for $\cP$, and $\cT$ has a finite basis of rank $k+1$ consisting of finite topologies, then there is a finitely generated $\TT$-proof space that proves $\cP$ safe.
\end{theorem}

\subsection{Patching Topologies for Theorem \ref{thm:relative-completeness} } \label{sec:subprograms}
As discussed in Remark \ref{rmk:topo-basis}, topologies such as rings do not admit a finite basis of any rank. 
We present a simple construction that enlarges a parameterized program $\cP$ so that the resulting program is over a topology that admits a finite basis of rank $k$. 
Central to our construction is the notion of \emph{subprograms}, which are obtained by restricting a program to a substructure of its topology:

\begin{definition}[Subprogram] \label{def:subprogram}
    Given a program $\prog = \tp{P, \sem{-}}$ over command set $\Sigma$ and topology $\TT$, for any substructure $\SS$ of $\TT$, the \emph{subprogram} of $\prog$ over $\SS$ is the program $\prog|_{\SS} = \tp{P|_\SS, \sem{-}|_{\Sigma \times \SS}}$. We also say that $\prog$ is a \emph{superprogram} of $\prog|_{\SS}$.
\end{definition}

Interestingly, the closure $\cls{\cP}$ of a parameterized program $\cP$ under subprograms (we call it a \emph{downward closure}) can be verified \emph{soundly and completely} in place of the original parameterized program when it comes to safety. This remains true if we only consider provability by parametric proof spaces.

\begin{theorem} \label{thm:subprogram-safety}
    For any parameterized programs $\cP$ and $\cP'$, if $\cP \subset \cP'$ and every program of $\cP'$ is a subprogram of some program in $\cP$, then $\cP$ is safe if and only if $\cP'$ is safe; if furthermore, $\TT$ is a universal structure for $\cP$ and $\cP'$, then any $\TT$-proof space $\cH$ proves the safety of $\cP$ if and only if it proves the safety of $\cP'$.
\end{theorem}

The second statement holds because the validity of Hoare triples in the universal structure is independent of the syntactic runs under consideration, and adding subprograms does not change the union of the languages of syntactic runs (see Theorem~\ref{thm:safety-proof}).

To establish the condition required in Theorem \ref{thm:relative-completeness}, we refine the notion of downward closures by making it parametric on a number $k \in \NN_+$. For any topology $\TT$, let $\substructures_k(\TT)$ be the set of substructures of $\TT$ generated by $k$ (not necessarily distinct) nodes. Define $\clsk{\cP}{k}$ as the parameterized program obtained from $\cP$ by adding, for every $i\in\cI$, all subprograms of $\prog_i$ over topologies in $\substructures_k(\TT_i)$. The following proposition presents a basis of rank $k$ for the topologies of $\clsk{\cP}{k}$ and provides a condition that ensures its finiteness. 

\begin{proposition} \label{prop:topo-sub}
    For any $k \in \NN_+$ and family of topologies $\cT$ over a common vocabulary $\cV$, if $\substructures_k(\TT) \subset \cT$ for any $\TT \in \cT$, then $\bigcup_{\TT\in\cT} \substructures_k(\TT)$ is a basis for $\cT$ of rank $k$. Moreover, if $\cV$ is finite and there is some $n \in \NN$ such that the cardinality of every structure in $\bigcup_{\TT\in\cT} \substructures_k(\TT)$ is bounded above by $n$, then this basis is finite up to isomorphism.
\end{proposition}

For all examples of parameterized programs $\cP$ that appear in this paper, the above-mentioned basis for $\clsk{\cP}{k}$ is finite (up to isomorphism) and consists of finite topologies. The figure below illustrates the basis of rank $2$ for the ring topology (with at least two process nodes), where we can take the bound in Proposition \ref{prop:topo-sub} to be $6$.

\begin{center}
    \includegraphics[width=\columnwidth]{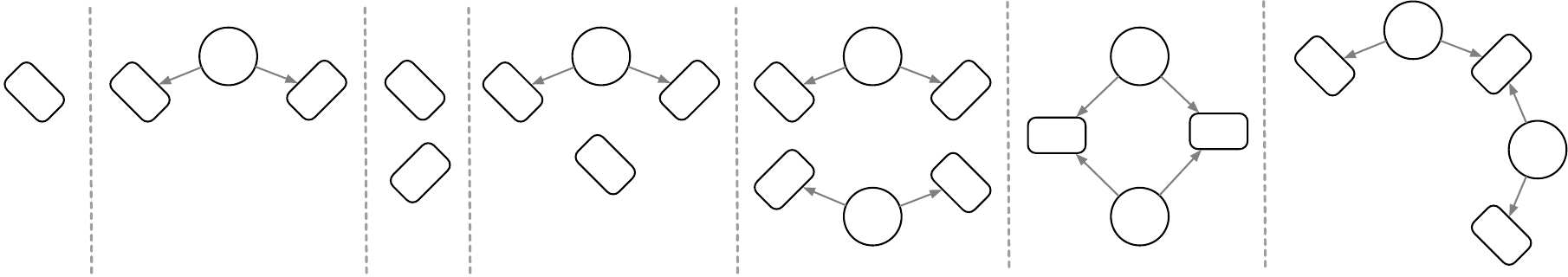}
\end{center}

\noindent In these scenarios, the relative completeness theorem simplifies to the following: for any $k \in \NN$, if $\clsk{\cP}{k+1}$ has an Ashcroft invariant of width $k$, then $\clsk{\cP}{k+1}$ and thus $\cP$ has a finitely generated $\TT$-proof space for any universal structure $\TT$ for $\cP$.

\subsection{Completeness for Parameterized Boolean Programs} \label{sec:bool}

In this section, we discuss the completeness of parametric proof spaces for parameterized Boolean programs. It is important to note that, as outlined in Section \ref{sec:prog-topo}, the verification problem becomes a {\em coverability} problem: We prove all program traces reaching an error location are infeasible. Over the star topology, it is well-known that this problem for parameterized Boolean programs is decidable (a consequence of the decidability of the coverability problem in Petri nets \cite{KarpM1969}), and in \cite[Sec. 6.3]{FarzanKP2015}, it is mentioned (and later proved in  \cite[Theorem 2]{FarzanK2025}) that (non-parametric) proof spaces are complete for the verification of such parameterized Boolean programs. Therefore, it was a natural question to ask if parametric proof spaces share this property. 

In our more general setting, a parameterized Boolean program is a parameterized program where every node has boundedly many variables, all of which are Boolean, and every indexed command only accesses (i.e. reads and writes) boundedly many variables. 

We establish a similar completeness result for a wide range of parameterized Boolean programs, which, to the best of our knowledge, also leads to the first proof of decidability of verification for this specific class. 

 Formally, our parameterized Boolean programs are those where
\begin{enumerate}
    \item the data domain $\DD$ is bit-vectors of a fixed length $n$, where $\VData$ contains one constant symbol for every one of the $2^n$ values;
    \item the (local) semantics of every indexed command is deterministic and reads and writes finitely many variables; and
    \item all variables are initialized to $0$.\footnote{Although our formal definition of safety is concerned with program runs that begin with an arbitrary valuation, it is possible to incorporate such an initialization, as we demonstrate in the proof in \Cref{app:completeness}.}
\end{enumerate}
The requirements of initialization and data determinism are for convenience and without loss of generality: Any uninitialized Boolean program with data nondeterminism can be easily converted to a Boolean program with all variables initialized to $0$ where any data nondeterminism is replaced by control-flow nondeterminism. 

The proof argument is centred around the construction of a \emph{maximal} finitely generated parameterized proof space for a given set of commands (to be used for any program). Formally: 
\begin{theorem}[Completeness for Parameterized Boolean Programs] \label{thm:bool-c}
    Any safe parameterized Boolean program $\cP$ with universal structure $\TT$ has a $\TT$-proof space that proves its safety, which only depends on $\TT$ and the set of commands and their semantics. Moreover, if $\TT^2/{\simeq}$ is finite, this proof space is finitely generated.
\end{theorem}
The proof idea is to make use of Hoare triples of the form
\[ \textstyle \htri{\bigwedge_{i=1}^k \sval(c_i) = x_i}{\tp{\sigma : a}}{\sval(b) = x_0}, \]
where $x_0, \ldots, x_k \in \DD$ and the precondition is the \emph{weakest precondition}.
When $\TT^2 / {\simeq}$ is finite, it suffices to consider finitely many pairs of nodes $(a, b)$.

Note that Theorem \ref{thm:bool-c} does not address the issue of \emph{effectiveness}, which we postpone to Corollary \ref{cor:bool-d}, after we discuss the algorithmic questions around parametric proof spaces: Under certain conditions about $\TT$, the construction in Theorem \ref{thm:bool-c} results in a decision procedure, i.e., the basis of the maximal parametric proof space is computable, and proof checking is decidable.

\section{Verification Algorithms} \label{sec:predicate-automata}
Theorem \ref{thm:safety-proof} can be interpreted as a sound proof rule for the verification of parameterized programs using parametric proof spaces. In this section, we look into how such a proof rule can be algorithmically implemented. The key is to have finite representations for the two languages, the denotation of a proof space and the set of syntactic traces of a parameterized program, and an algorithm to perform the inclusion check between them. A natural idea is to use \emph{data automata}, since the behaviours of parameterized programs and parametric proof spaces are words over an indexed alphabet of commands, where these indices come from an unbounded domain.  

First, focusing on representing the set of syntactic runs of a parameterized program, we need to overcome the challenge of finitely representing the {\em infinite union}. 
To do this, we consider the interesting case where the infinite union converges to a ``limit''. More specifically, we introduce a \emph{limit program} $\lmt{\prog}$
which is a \emph{superprogram} of \emph{all} members of $\cP$, such that $\lmt{\prog}$ can replace $\cP$ soundly and completely for safety verification.

Second, for both the program and the proof space, we need to make the automata somehow aware of the underlying topology. For this, we propose a generalization of {\em predicate automata} of \cite{FarzanKP2015} that, like parametric proof spaces, is parametric on a choice of symmetry, denoted as $\TT$-predicate automata. In effect,  the predicate automata of \cite{FarzanKP2015} become the instance where $\TT$ is $\NN$ with the empty vocabulary. We then show that under certain conditions on finite representability, both the error runs of a (limit) program over $\TT$ and the language of any $\TT$-proof space generated by a finite \emph{normalized} basis (Definition \ref{def:basis-normal}) are recognized by a $\TT$-predicate automaton. We argue that $\TT$-predicate automata enjoy the same essential closure properties as predicate automata, and hence the inclusion check can be reduced to an emptiness check. We present an algorithmic solution for the emptiness check, which cannot be a decision procedure due to the undecidability of the emptiness of $\NN$-predicate automata \cite{FarzanKP2015}. We outline conditions under which the emptiness check is decidable.

\subsection{The Limit Program} \label{sec:limit-programs}

Intuitively, for a parameterized program $\cP = \{\prog_i\}_{i\in\NN}$ (i.e., the index set $\cI$ is $\NN$), if every program $\prog_n$ can be identified with a subprogram of $\prog_m$ for every $m > n$, then one can take the ``limit'' $\lmt{\prog}$ of the sequence of programs $\prog_1, \prog_2, \ldots$. It is sound to verify $\lmt{\prog}$ in place of $\cP$, since a limit program contains all behaviours of each $\prog_i$. For completeness to hold, the limit program should not introduce extra behaviours that do not exhibit in any $\prog_i$, which requires a careful definition of ``limits''. 

We first define \emph{limits of a class of logical structures} over a common vocabulary. It is a weak notion of limit which is nevertheless sufficient for establishing soundness and completeness.

\begin{definition}[Limit of Logical Structures] \label{def:lim-struct}
    Let $\{\TT_i\}_{i\in\cI}$ be a class of logical structures over a common vocabulary $\cV$. A logical structure $\lmt{\TT}$ over $\cV$ is a \emph{limit} of $\{\TT_i\}_{i\in\cI}$ if every structure $\TT_i$ can be embedded in $\TT$, and every finitely generated substructure of $\TT$ can be embedded in some $\TT_i$.
\end{definition}

\begin{example} \label{ex:star-limit}
    For the star topology in Example \ref{ex:star}, a limit can be thought of as a star with infinitely many branches; the node for the global variables is the centre of the star, and there are infinitely many thread nodes. This is precisely the program that is verified in \cite{FarzanKP2015} in place of the parameterized family.
Another example is the limit already discussed in Section \ref{sec:example}. 
\end{example}

We observe that given an arbitrary class of structures, a limit does not necessarily exist: A limit structure first needs to be a universal structure (Section~\ref{sec:htri}), but any universal structure of the ring topology, for example, necessarily contains a disjoint copy of every ring; thus, some finitely generated substructures, such as two disjoint rings, do not embed in any structure in the class.

The well-established concept of \emph{Fra\"{i}ss\'{e} limits} from model theory (e.g., see~\cite[Sec. 6.1]{Hodges1997}) provides a systematic way to construct limits:

\begin{proposition}[Fra\"{i}ss\'{e} limit] \label{prop:fraisse}
  Let $\cT$ be a class of finite logical structures over a common vocabulary and $\cls{\cT}$ be the substructures of members of $\cT$. If $\cls{\cT}$ is a countable Fra\"{i}ss\'{e} class, then its Fra\"{i}ss\'{e} limit is a limit of $\cT$ in the sense of \Cref{def:lim-struct}.
\end{proposition}

Given a parameterized program $\cP = \{\tp{P_i, \sem{-}_i}\}_{i\in\cI}$ over a family of topologies $\cT = \{\TT_i\}_{i\in\cI}$, any limit $\lmt{\TT}$ of $\cT$ induces a limit program $\tp{\lmt{P}, \lmt{\sem{-}}}$ over $\lmt{\TT}$. Intuitively, the limit program identifies nodes across different topologies. As expected, its syntax and semantics are defined through local isomorphisms: For any node $v$ in $\lmt{\TT}$ and node $u$ in some $\TT_i$ with a local isomorphism $\beta$ from $u$ to $v$, set $\lmt{P}(v) = P_i(u)$ and $\lmt{\sem{\sigma : v}} = \beta^*\sem{\sigma : u}_i$ for all $\sigma \in \Sigma$. The well-definedness is easy to verify. To access $\lmt{P}$ and $\lmt{\sem{-}}$ in a computation, we assume there is a computable function that maps every node in $\lmt{\TT}$ to a locally symmetric node in $\cT$.

Note that the command semantics over a limit structure $\lmt{\TT}$ is defined in the same way as that of a universal structure, except that every node in $\lmt{\TT}$ is locally symmetric to some node in some $\TT_i$. With a universal structure of which $\lmt{\TT}$ can be identified with a substructure (e.g., $\lmt{\TT}$ itself), we show soundness and completeness:

\begin{theorem}[Soundness \& Completeness of a Limit Program] \label{thm:limit-safety}
  For any parameterized program $\cP$ and limit program $\lmt{\prog}$ over topology $\lmt{\TT}$ for $\cP$, we have that $\cP$ is safe if and only if $\lmt{\prog}$ is safe. Furthermore, if $\TT$ is a universal structure for $\cP$ of which $\lmt{\TT}$ can be identified with a substructure, then any $\TT$-proof space $\cH$ proves the safety of $\cP$ if and only if it proves the safety of $\lmt{\prog}$.
\end{theorem}
The intuition behind why the latter statement holds is that the languages $\Lang_\err(\lmt{P})$ and $\bigcup_{i\in\cI} \Lang_\err(P_i)$ have the same $\simeq$-closure.

\subsection{Parametric Predicate Automata}

Let $\cV$ be a vocabulary and $\TT$ be a $\cV$-structure. We still refer to the elements in $\TT$ as nodes. To perform computation over the nodes, we make two basic assumptions: both the nodes and the symbols in $\cV$ are finitely representable (i.e., encodable as bit strings), and the interpretations of the symbols are computable.

A \emph{$\TT$-predicate automaton} (PA) recognizes a language over the infinite alphabet $\Sigma \times \TT$, where $\Sigma$ is a finite set. A $\TT$-PA is equipped with a finite relational vocabulary $\cQ$ disjoint from $\cV[\TT]$, and its states, written $\sF^{\TT}_{\atom}(\cQ)$, are (uninterpreted) atomic formula of the form $q(a_1, \ldots, a_{\ar(q)})$ ($q \in \cQ$, $a_1, \ldots, a_{\ar(q)} \in \TT$). Let $\sF_+(\cV, \cQ)$ be the set of quantifier-free $(\cV \cup \cQ)$-formulas where every occurrence of $q(t_1, \ldots, t_{\ar(q)})$ ($q \in \cQ$, $t_j$ are terms) is positive. Formally, we define $\TT$-predicate automata as follows:

\begin{definition}[$\TT$-predicate automaton] \label{def:A-PA}
  A \emph{$\TT$-predicate automaton} (PA) is a 5-tuple $\tp{\cQ, \Sigma \times \TT, \delta, \varphi_\start, F}$ where $\cQ$ is a finite relational vocabulary, $\Sigma \times \TT$ is an alphabet, $\varphi_\start \in \sF_+(\cV, \cQ)$ is an initial formula without free variables, $F \subset \cQ$ is a set of accepting predicate symbols, and $\delta : \cQ \times \Sigma \to \sF_+(\cV, \cQ)$ is a \emph{syntactic transition function} such that for any $q \in \cQ$ and $\sigma \in \Sigma$, the free variables of $\delta(q, \sigma)$ are members of the set $\{ v_0, \ldots, v_{\ar(q)}\}$.
\end{definition}

The \emph{arity} of a $\TT$-PA is the maximum arity of the predicates in the relational vocabulary $\cQ$. If the arity is $1$, $\TT$-PA is called \emph{monadic}.

Recall that $\sF^{\TT}_{\atom}(\cQ)$ is the states of the $\TT$-PA. Let $\sF^{\TT}(\cQ)$ be all negation-free Boolean combinations of the states. Then, every syntactic transition function $\delta : \cQ \times \Sigma \to \sF_+(\cV, \cQ)$ induces a \emph{semantic transition function} $\hat{\delta} : \sF_\atom^\TT(\cQ) \times (\Sigma \times \TT) \to \sF^\TT(\cQ)$ by
\begin{equation} \label{eq:delta-formula}
  \hat{\delta}(q(a_1, \ldots, a_{\ar(q)}), \tp{\sigma: a_0}) \equiv \delta(q, \sigma)[v_i \mapsto a_i : i = 0, \ldots, \ar(q)]. \tag{TF}
\end{equation}

Definition \ref{def:A-PA} emphasizes \emph{quantifier-free} formulas, because their use makes the semantic transition function computable without any additional assumptions, and is sufficient for our purpose of representing error runs and proof space languages: both are symmetric w.r.t. $\simeq$, and $\simeq$-equivalence classes of $\TT^d$ are representable by sets of quantifier-free formulas with $d$ variables (Proposition \ref{prop:local-symm-qf}). 

\paragraph{\bfseries Languages Recognized by $\TT$-PA}

We are ready to define the language of a $\TT$-PA and discuss the properties of the class of languages recognized by $\TT$-PA.

\begin{definition}[Configuration]
  Let $A = \tp{\cQ, \Sigma \times \TT, \delta, \varphi_\start, F}$ be a $\TT$-PA. A \emph{configuration} $\cC$ of $A$ is a finite subset of $\sF_\atom^\TT(\cQ)$, identified with the conjunction of the formulas in $\cC$. We say $\cC$ is \emph{initial} if it is a cube in the disjunctive normal form (DNF) of $\varphi_\start$; \emph{accepting} if $q \in F$ for all $q(a_1, \ldots, a_{\ar(q)}) \in \cC$.
\end{definition}

The predicate automaton $A = \tp{\cQ, \Sigma \times \TT, \delta, \varphi_\start, F}$ induces a transition relation on configurations: $\cC \xrightarrow{\sigma : a} \cC'$ iff $\cC'$ is a cube in the DNF of $\bigwedge_{t \in \cC} \hat{\delta}(t, \tp{\sigma : a})$. We write $\cC \to \cC'$ if there exists some $\tp{\sigma : a}$ such that $\cC \xrightarrow{\sigma : a} \cC'$ and call $\cC'$ a \emph{successor} of $\cC$. A word $\tau = \tp{\sigma_1 : a_1} \cdots \tp{\sigma_n : a_n}$ is \emph{accepted} by $A$ if there is a sequence of configurations $\cC_0, \ldots, \cC_n$ such that $\cC_n$ is initial, $\cC_0$ is final, and $\cC_{r-1} \xleftarrow{\sigma_r : a_r} \cC_r$ for $r = n, \ldots, 1$; note that predicate automata read from right to left. The language of $A$, $\Lang(A)$, is the set of words accepted by $A$, and a \emph{$\TT$-predicate automata language} (PAL) is the language of some $\TT$-PA.
$\TT$-PALs have the same closure properties as predicate automata languages \cite{FarzanKP2015}.

\begin{theorem}[Closure Properties] \label{thm:PAL-closedness}
  $\TT$-PALs are closed under intersection and complement, and so are monadic $\TT$-PALs. Any $\TT$-PAL is locally symmetric, i.e., closed under the equivalence relation $\simeq$.
\end{theorem}

\subsection{Error Runs and Proof Space Languages as \texorpdfstring{$\TT$}{T}-Predicate Automata Languages}

By Theorem~\ref{thm:limit-safety}, if the parameterized program $\cP$ admits a limit program $\lmt{\prog}$ over a limit structure $\lmt{\TT}$, then instead of checking the inclusion $\bigcup_{i\in\cI} \Lang_\err(P_i) \subset \Lang(\cH)$ in Theorem~\ref{thm:safety-proof}, we may take $\lmt{\TT}$ as the universal structure and check $\Lang_\err(\lmt{P}) \subset \Lang(\cH)$, thereby sidestepping the construction of an infinite union of languages. In this section, we demonstrate that under certain finiteness conditions, both $\Lang_\err(\lmt{P})$ and $\Lang(\cH)$ are recognized by $\lmt{\TT}$-PA. Our constructions generalize \cite[Propositions 6.3 and 6.4]{FarzanKP2015} and are independent of whether $\lmt{\TT}$ is obtained as a limit structure.

For the language of error runs, our construction appears in the proof of the following theorem:

\begin{theorem}[Error runs as $\TT$-PAL] \label{thm:prog-PAL}
  Let $\TT$ be a topology over a vocabulary $\VNode$. Let $\tp{P, \sem{-}}$ be a locally symmetric concurrent program over $\TT$. If $\TT/{\simeq}$ is finite, then $\Lang_\err(P)$ is a monadic $\TT$-PAL.
\end{theorem}

The proof relies on the fact that if $\TT/{\simeq}$ is finite, then the pre-image of every command under $P$ is definable by a quantifier-free formula, which follows from Proposition~\ref{prop:local-symm-fin}.

To construct a parametric proof space as a $\TT$-PAL, we require the proof space to be generated by a finite basis in normal form:

\begin{definition} \label{def:basis-normal}
  A Hoare triple is in \emph{normal form} if it is of the form $\htri{\phi[\overrightarrow{\sval(a)}]}{\tp{\sigma : b}}{\psi[\overrightarrow{\sval(c)}]}$, where $\{ a_i \}_i \subset \nb(b, \vec{c})$ and $\psi[\vec{x}]$ is syntactically not a conjunction. We say a basis (of a parametric proof space) is in normal form if every Hoare triple in the basis is.
\end{definition}

It is always possible to normalize a Hoare triple with a single command, by breaking down a triple $\htri{\phi}{\tp{\sigma : b}}{\psi_1 \land \psi_2}$ into $\htri{\phi}{\tp{\sigma : b}}{\psi_1}$ and $\htri{\phi}{\tp{\sigma : b}}{\psi_2}$, and adding an existential quantification over data for every redundant node $a_i$ in the precondition. It is reasonable to assume that in an algorithmic refinement scheme (in the style of \cite{FarzanKP2015}), the newly discovered Hoare triples are always on single commands.

\begin{theorem}[Parametric proof space as $\TT$-PAL] \label{thm:proof-PAL}
  Let $\TT$ be a topology over a vocabulary $\VNode$. Let $\cH$ be a $\TT$-proof space generated by a finite basis in normal form. If $\TT^d/{\simeq}$ is finite for every $d \in \NN_+$, then $\Lang(\cH)$ is a $\TT$-PAL.
\end{theorem}
In the special case where $\Lang(\cH)$ is recognized by a \emph{monadic} $\TT$-PA, we also say $\cH$ is monadic.

\subsection{Proof Checking}\label{sec:proof-checking}
As a consequence of Theorem \ref{thm:PAL-closedness}, the inclusion check between the behaviours of a limit program and a parametric proof space can be reduced to checking the emptiness of the language of a $\TT$-PA. Hence, the focus of this section is on this problem. 
General $\NN$-PA emptiness is undecidable \cite[Proposition 6.10]{FarzanKP2015} and this undecidability is inherited by $\TT$-PA. In \cite{FarzanKP2015}, a semi-algorithm for $\NN$-PA emptiness is proposed, which explores reachable configurations starting from an initial configuration; it is sound and complete for non-emptiness. This algorithm is in the style of {\em backward coverability} checks in {\em well-structured transition systems} \cite{AbdullaCJT2000,FinkelS2001}: the search space is equipped with a \emph{covering} pre-order that serves as a pruning strategy. The essence of this algorithm can be generalized to $\TT$-PAs. Note, however, that there are many more sources of undecidability in the richer space of $\TT$-PAs.

\begin{lemma} \label{lem:wsts}
  Let $\TT$ be a $\cV$-structure and $\cQ$ be a finite relational vocabulary disjoint from $\cV[\TT]$. Suppose there is a pre-order $\preceq$ on the set of configurations $2^{\sF^\TT_\atom(\cQ)}$ such that for any $\TT$-predicate automaton $\tp{\cQ, \Sigma \times \TT, \delta, \varphi_\start, F}$, we have
  \begin{enumerate}
    \item $\preceq$ is decidable.
    \item (Essentially Finite Branching) Given any configuration $\cC$, one can effectively compute a finite subset $\sC$ of the successors of $\cC$ such that for any successor $\cC'$ of $\cC$, there is some $\bar{\cC}' \in \sC$ such that $\bar{\cC}' \preceq \cC'$.
    \item (Downward Compatibility) For any configurations $\cC$ and $\bar{\cC}$ such that $\bar{\cC} \preceq \cC$, if $\cC$ is accepting, then $\bar{\cC}$ is accepting, and for any configuration $\cC'$ such that $\cC \to \cC'$, there is some configuration $\bar{\cC}'$ such that $\bar{\cC} \to \bar{\cC}'$.
  \end{enumerate}
  Then there is a semi-algorithm for $\TT$-PA non-emptiness, which becomes a decision procedure for the class of $\TT$-PA whose reachable configurations are well-quasi-ordered by $\preceq$.
\end{lemma}

It remains to find a pre-order that satisfies the conditions in Lemma \ref{lem:wsts}. One can generalize the covering pre-order in \cite{FarzanKP2015}.

\begin{definition}[Covering] \label{def:covering}
  Let $\TT$ be a $\cV$-structure and $\cQ$ be a finite relational vocabulary disjoint from $\cV[\TT]$. We define the \emph{covering pre-order} $\preceq$ 
  as configuration $\cC$ covers $\cC'$, denoted $\cC \preceq \cC'$, if there is an automorphism $\pi$ of $\TT$ such that for all $q(a_1, \ldots, a_{\ar(q)}) \in \cC$, we have $q(\pi(a_1), \ldots, \pi(a_{\ar(q)})) \in \cC'$.
\end{definition}

It is easy to verify that the covering pre-order satisfies downward compatibility; in particular, this would not be true if the automorphism $\pi$ were replaced with a local isomorphism in the definition. However, when we compute with a generic $\TT$ instead of the structure $\NN$, basic operations that one may take for granted become nontrivial: (i) deciding the covering pre-order, and (ii) enumerating the representative successors of a configuration (i.e., the essentially finite branching property).

The following theorem gives a sufficient condition for implementing these operations algorithmically. Condition (1) reduces deciding the covering pre-order to deciding $\simeq$. Conditions (2) and (3) add computability consideration to the conditions in Theorem~\ref{thm:proof-PAL} about $\TT$; in particular, Condition (2) ensures $\simeq$ is decidable.

\begin{theorem}[$\TT$-PA emptiness] \label{thm:A-PA-emp}
  Suppose $\TT$ satisfies the following:
  \begin{enumerate}
    \item $\TT$ is homogeneous.
    \item $\TT^d/{\simeq}$ is finite for every $d \in \NN_+$, and there is an algorithm that takes $d$ and outputs a list of quantifier-free formulas that define the equivalence classes (whose existence is guaranteed by Proposition~\ref{prop:local-symm-fin}).
    \item The satisfiability of any quantifier-free $\cV[\TT]$-formula in $\TT$ is decidable.
  \end{enumerate}
  Then the covering relation satisfies conditions (1)--(3) in Lemma~\ref{lem:wsts}.
\end{theorem}

For example, monadic $\NN$-PAs, where the vocabulary of $\NN$ is empty, satisfy the three conditions in Theorem \ref{thm:A-PA-emp} as well as the well-quasi-order condition in Lemma \ref{lem:wsts}, which makes the decidability result of \cite{FarzanKP2015} a special case of the results here. As another example, with a suitable choice of $\VNode$ (see \Cref{app:example}), the topology in \Cref{sec:example} admits a unique countable homogeneous limit $\TT_\text{forest}$. Monadic $\TT_\text{forest}$-PAs also satisfy all four conditions.

More generally, if $\TT$ is a limit structure that is obtained as the countable Fra\"{i}ss\'{e} limit of a Fra\"{i}ss\'{e} class satisfying the condition in Proposition \ref{prop:fraisse-rep}, then $\TT$ satisfies all conditions in Theorem \ref{thm:A-PA-emp}:

\begin{proposition}[Theorem 7.20 in \cite{Bojanczyk2019}] \label{prop:fraisse-rep}
  Let $\cT$ be a countable Fra\"{i}ss\'{e} class over a finite vocabulary with limit $\TT$. If for every $k \in \NN$ there are (up to isomorphism) finitely many structures in $\cT$ with at most $k$ generators, and a list of these structures can be computed given $k$, then elements in $\TT$ are finitely representable, and $\TT$ has a decidable first-order theory with parameters from the underlying set of $\TT$.
\end{proposition}

As a corollary of Lemma \ref{lem:wsts}, Theorem \ref{thm:A-PA-emp}, and Theorem \ref{thm:bool-c}, we outline conditions under which the safety of parameterized Boolean programs is decidable, thereby completing the discussion around the completeness of parametric proof spaces for parameterized Boolean programs in Section \ref{sec:bool}.
\begin{corollary}[Decidability for Parameterized Boolean Programs] \label{cor:bool-d}
    Fix a structure $\TT$ that satisfies the conditions in Theorem \ref{thm:A-PA-emp}. Additionally, assume for any finite relational vocabulary $\cQ$ disjoint from $\cV[\TT]$ where symbols have arity at most $1$, the covering pre-order is a well-quasi-order on the set of configurations $2^{\sF^\TT_\atom(\cQ)}$. Then the safety of any parameterized Boolean program for which $\TT$ is a limit structure is decidable.
\end{corollary}

Note that the conditions in Corollary \ref{cor:bool-d} are satisfied by the limit structure in Example \ref{ex:star-limit}, which implies the decidability of parameterized Boolean programs over the star topology.

\section{Related Work} \label{sec:related-work}
Safety verification of parameterized programs has been studied extensively. Since the problem is generally undecidable~\cite{AptK1986}, a large body of work exists \cite{AbdullaD2016} on decidable fragments by imposing restrictions on the topology, communication primitives, or the expressivity of safety verification \cite{LaTorreMW2015,EsparzaGM2016,BallCR2001,GermanS1992,AminofKRS+2018,Balasubramanian2021,AbdullaHH2013}. Our focus is not on a new decidable fragment or complexity results. 
As such, we only compare with work that is closely related to our contributions.

\paragraph{\bf Symmetry Reasoning in Program Verification}

Our formulation of symmetries has been inspired by the study of \emph{local symmetries} in networks \cite{GolubitskyS2006,NamjoshiT2012} and in compositional verification of concurrent programs \cite{NamjoshiT2016,NamjoshiT2018}. The insight of reasoning about {\em small neighbourhoods of processes} in place of global reasoning exists in some form in all of these. In \cite{NamjoshiT2016}, it is specifically used as the means of compositional verification for parameterized programs,  where \emph{compositional cutoffs} are derived from local symmetries formalized as a \emph{groupoid}. Our local symmetry induces a groupoid with at most one arrow between every pair of objects (a consequence of \Cref{prop:li-unique}).

In model checking, symmetry reductions \cite{EmersonS1996} target the state explosion problem for concurrent systems with many isomorphic components by constructing and checking a quotient structure in place of the original structure. In \cite{EmersonS1996}, the focus is on token passing rings and correctness properties that are invariant under rotation, for which cut-off results are given. Weaker notions of symmetry, such as \emph{virtual symmetry} \cite{EmersonHT2000}, have also been studied.
In \cite{LinNRS2016}, the inverse problem of identifying symmetries in a finite-state parameterized system (called a \emph{symmetry pattern}) is investigated; symmetry patterns are simulation preorders on states, encoded by transducers. Our work uses a different notion of symmetry that is based on isomorphic neighbourhoods.

\paragraph{\bf Parameterized Program Verification}
When it comes to the verification of parameterized infinite state programs, for a very good reason, the majority of work is in the general area of deductive verification \cite{PadonLSS2017,AshmoreGT2019,v.GleissenthallKBS+2019,ElmasQT2009,KraglQ2018,KraglEHM+2020,FlanaganF2020}; that is, the user is involved, one way or another, in providing the required invariants or verification conditions, and the underlying paradigm is completing/checking this proof. 
In contrast, parametric proof spaces are proposed as a way of discovering such invariants automatically, or at least a subclass of them within the boundaries of decidability of the algorithm in Section \ref{sec:predicate-automata}. 
From this group, \cite{AshmoreGT2019} stands out since it targets a rich class of topologies and models every instance of the topology as a logical structure without resorting to an axiomatization. There is a relative completeness result, where the EPR encoding of the safety problem is shown to be complete relative to a strict subset of our generalized Ashcroft invariants, under a {\em semantic} condition of the invariant.  
In comparison, our conditions for the relative completeness of parametric proof spaces are related to the {\em syntax} of the invariant (i.e. the width) and the presence of (local) symmetry in the topology family.

For algorithmic verification of infinite state programs, there is work aimed at automatically yet {\em incompletely} generating universally quantified invariants \cite{AronsPRX+2001,PnueliRZ2001,EmmiMM2010}. {\em Thread-modular proofs at many levels} \cite{HoenickeMP2017,FarzanKP2024} proposed a {\em complete} algorithm relative to the existence of an Ashcroft invariant with $k$ universal quantifiers, for an input parameterized program $\cP$ over the star topology.  

For parameterized finite-state programs over a rich set of topologies, several decidability and complexity results are presented in \cite{AminofKRS+2018}, including those involving programs that communicate through pairwise rendezvous. It is generally understood that these are expressive enough to model shared memory communication. However, their decidability results \cite[Theorems 10,11]{AminofKRS+2018} are restricted to what they call {\em homogenous} topologies, which differ from our {\em homogenous} topologies; in particular, theirs does not include the tree topologies of our example from Section \ref{sec:example}, which are proven to be decidable in this paper. The first paper to consider a range of topologies was \cite{AbdullaHH2013}, a seminal paper on parameterized verification that presents individual decidability results for a few topologies under broadcast communication.

\paragraph{\bf Automata over Infinite Alphabets}
Our work extends predicate automata introduced in \cite{FarzanKP2015}, which are in turn closely related to \emph{(alternating) register automata (ARA)} \cite{KaminskiF1994,NevenSV2004,DemriL2006,Figueira2012,Bojanczyk2019}. The connection is formalized in \cite{FarzanK2025}, which relates the arity of the automata to the number of available registers. In the most common form, register automata recognize languages over an alphabet $\Sigma \times \NN$, where $\Sigma$ is finite and $\NN$ is the set of equality atoms.
However, to account for a variety of symmetries, our extension, $\TT$-predicate automaton, is parametric on the logical structure $\TT$ and recognizes languages over $\Sigma \times \TT$.
More generally, in \cite{BojanczykBKL2012}, \emph{orbit-finite alternating automata} are proposed; such automata read words over a \emph{orbit-finite nominal set}, which can encode an alphabet $\Sigma \times \TT$ for an oligomorphic $\TT$ \cite[Example 5.8]{Bojanczyk2019}. Our use of the logical structure $\TT$ is not limited to the oligomorphic ones, yet the conditions in both Proposition \ref{prop:fraisse-rep} and Theorem \ref{thm:A-PA-emp} imply oligomorphism.

\bibliographystyle{ACM-Reference-Format}
\bibliography{biblio.bib}


\begin{thebibliography}{44}


\ifx \showCODEN    \undefined \def \showCODEN     #1{\unskip}     \fi
\ifx \showISBNx    \undefined \def \showISBNx     #1{\unskip}     \fi
\ifx \showISBNxiii \undefined \def \showISBNxiii  #1{\unskip}     \fi
\ifx \showISSN     \undefined \def \showISSN      #1{\unskip}     \fi
\ifx \showLCCN     \undefined \def \showLCCN      #1{\unskip}     \fi
\ifx \shownote     \undefined \def \shownote      #1{#1}          \fi
\ifx \showarticletitle \undefined \def \showarticletitle #1{#1}   \fi
\ifx \showURL      \undefined \def \showURL       {\relax}        \fi
\providecommand\bibfield[2]{#2}
\providecommand\bibinfo[2]{#2}
\providecommand\natexlab[1]{#1}
\providecommand\showeprint[2][]{arXiv:#2}

\bibitem[Abdulla et~al\mbox{.}(2000)]%
        {AbdullaCJT2000}
\bibfield{author}{\bibinfo{person}{Parosh~Aziz Abdulla}, \bibinfo{person}{K{\=a}rlis {\v C}er{\=a}ns}, \bibinfo{person}{Bengt Jonsson}, {and} \bibinfo{person}{Yih-Kuen Tsay}.} \bibinfo{year}{2000}\natexlab{}.
\newblock \showarticletitle{Algorithmic {{Analysis}} of {{Programs}} with {{Well Quasi-ordered Domains}}}.
\newblock \bibinfo{journal}{\emph{Information and Computation}} \bibinfo{volume}{160}, \bibinfo{number}{1} (\bibinfo{date}{July} \bibinfo{year}{2000}), \bibinfo{pages}{109--127}.
\newblock
\showISSN{0890-5401}
\href{https://doi.org/10.1006/inco.1999.2843}{doi:\nolinkurl{10.1006/inco.1999.2843}}


\bibitem[Abdulla and Delzanno(2016)]%
        {AbdullaD2016}
\bibfield{author}{\bibinfo{person}{Parosh~A. Abdulla} {and} \bibinfo{person}{Giorgio Delzanno}.} \bibinfo{year}{2016}\natexlab{}.
\newblock \showarticletitle{Parameterized Verification}.
\newblock \bibinfo{journal}{\emph{International Journal on Software Tools for Technology Transfer}} \bibinfo{volume}{18}, \bibinfo{number}{5} (\bibinfo{date}{Oct.} \bibinfo{year}{2016}), \bibinfo{pages}{469--473}.
\newblock
\showISSN{1433-2787}
\href{https://doi.org/10.1007/s10009-016-0424-3}{doi:\nolinkurl{10.1007/s10009-016-0424-3}}


\bibitem[Abdulla et~al\mbox{.}(2013)]%
        {AbdullaHH2013}
\bibfield{author}{\bibinfo{person}{Parosh~Aziz Abdulla}, \bibinfo{person}{Fr{\'e}d{\'e}ric Haziza}, {and} \bibinfo{person}{Luk{\'a}{\v s} Hol{\'i}k}.} \bibinfo{year}{2013}\natexlab{}.
\newblock \showarticletitle{All for the {{Price}} of {{Few}}}. In \bibinfo{booktitle}{\emph{Verification, {{Model Checking}}, and {{Abstract Interpretation}}}}, \bibfield{editor}{\bibinfo{person}{Roberto Giacobazzi}, \bibinfo{person}{Josh Berdine}, {and} \bibinfo{person}{Isabella Mastroeni}} (Eds.). \bibinfo{publisher}{Springer}, \bibinfo{address}{Berlin, Heidelberg}, \bibinfo{pages}{476--495}.
\newblock
\showISBNx{978-3-642-35873-9}
\href{https://doi.org/10.1007/978-3-642-35873-9_28}{doi:\nolinkurl{10.1007/978-3-642-35873-9_28}}


\bibitem[Aminof et~al\mbox{.}(2018)]%
        {AminofKRS+2018}
\bibfield{author}{\bibinfo{person}{Benjamin Aminof}, \bibinfo{person}{Tomer Kotek}, \bibinfo{person}{Sasha Rubin}, \bibinfo{person}{Francesco Spegni}, {and} \bibinfo{person}{Helmut Veith}.} \bibinfo{year}{2018}\natexlab{}.
\newblock \showarticletitle{Parameterized Model Checking of Rendezvous Systems}.
\newblock \bibinfo{journal}{\emph{Distributed Computing}} \bibinfo{volume}{31}, \bibinfo{number}{3} (\bibinfo{date}{June} \bibinfo{year}{2018}), \bibinfo{pages}{187--222}.
\newblock
\showISSN{1432-0452}
\href{https://doi.org/10.1007/s00446-017-0302-6}{doi:\nolinkurl{10.1007/s00446-017-0302-6}}


\bibitem[Apt and Kozen(1986)]%
        {AptK1986}
\bibfield{author}{\bibinfo{person}{Krzysztof~R. Apt} {and} \bibinfo{person}{Dexter~C. Kozen}.} \bibinfo{year}{1986}\natexlab{}.
\newblock \showarticletitle{Limits for Automatic Verification of Finite-State Concurrent Systems}.
\newblock \bibinfo{journal}{\emph{Inform. Process. Lett.}} \bibinfo{volume}{22}, \bibinfo{number}{6} (\bibinfo{date}{May} \bibinfo{year}{1986}), \bibinfo{pages}{307--309}.
\newblock
\showISSN{0020-0190}
\href{https://doi.org/10.1016/0020-0190(86)90071-2}{doi:\nolinkurl{10.1016/0020-0190(86)90071-2}}


\bibitem[Arons et~al\mbox{.}(2001)]%
        {AronsPRX+2001}
\bibfield{author}{\bibinfo{person}{Tamarah Arons}, \bibinfo{person}{Amir Pnueli}, \bibinfo{person}{Sitvanit Ruah}, \bibinfo{person}{Ying Xu}, {and} \bibinfo{person}{Lenore Zuck}.} \bibinfo{year}{2001}\natexlab{}.
\newblock \showarticletitle{Parameterized {{Verification}} with {{Automatically Computed Inductive Assertions}}?}. In \bibinfo{booktitle}{\emph{Computer {{Aided Verification}}}}, \bibfield{editor}{\bibinfo{person}{G{\'e}rard Berry}, \bibinfo{person}{Hubert Comon}, {and} \bibinfo{person}{Alain Finkel}} (Eds.). \bibinfo{publisher}{Springer}, \bibinfo{address}{Berlin, Heidelberg}, \bibinfo{pages}{221--234}.
\newblock
\showISBNx{978-3-540-44585-2}
\href{https://doi.org/10.1007/3-540-44585-4_19}{doi:\nolinkurl{10.1007/3-540-44585-4_19}}


\bibitem[Ashcroft(1975)]%
        {Ashcroft1975}
\bibfield{author}{\bibinfo{person}{E.~A. Ashcroft}.} \bibinfo{year}{1975}\natexlab{}.
\newblock \showarticletitle{Proving Assertions about Parallel Programs}.
\newblock \bibinfo{journal}{\emph{J. Comput. System Sci.}} \bibinfo{volume}{10}, \bibinfo{number}{1} (\bibinfo{date}{Feb.} \bibinfo{year}{1975}), \bibinfo{pages}{110--135}.
\newblock
\showISSN{0022-0000}
\href{https://doi.org/10.1016/S0022-0000(75)80018-3}{doi:\nolinkurl{10.1016/S0022-0000(75)80018-3}}


\bibitem[Ashmore et~al\mbox{.}(2019)]%
        {AshmoreGT2019}
\bibfield{author}{\bibinfo{person}{Rylo Ashmore}, \bibinfo{person}{Arie Gurfinkel}, {and} \bibinfo{person}{Richard Trefler}.} \bibinfo{year}{2019}\natexlab{}.
\newblock \showarticletitle{Local {{Reasoning}} for {{Parameterized First Order Protocols}}}. In \bibinfo{booktitle}{\emph{{{NASA Formal Methods}}}}, \bibfield{editor}{\bibinfo{person}{Julia~M. Badger} {and} \bibinfo{person}{Kristin~Yvonne Rozier}} (Eds.). \bibinfo{publisher}{Springer International Publishing}, \bibinfo{address}{Cham}, \bibinfo{pages}{36--53}.
\newblock
\showISBNx{978-3-030-20652-9}
\href{https://doi.org/10.1007/978-3-030-20652-9_3}{doi:\nolinkurl{10.1007/978-3-030-20652-9_3}}


\bibitem[Balasubramanian(2021)]%
        {Balasubramanian2021}
\bibfield{author}{\bibinfo{person}{A.~R. Balasubramanian}.} \bibinfo{year}{2021}\natexlab{}.
\newblock \showarticletitle{Parameterized Verification of Coverability in Infinite State Broadcast Networks}.
\newblock \bibinfo{journal}{\emph{Information and Computation}}  \bibinfo{volume}{278} (\bibinfo{date}{June} \bibinfo{year}{2021}), \bibinfo{pages}{104592}.
\newblock
\showISSN{0890-5401}
\href{https://doi.org/10.1016/j.ic.2020.104592}{doi:\nolinkurl{10.1016/j.ic.2020.104592}}


\bibitem[Ball et~al\mbox{.}(2001)]%
        {BallCR2001}
\bibfield{author}{\bibinfo{person}{Thomas Ball}, \bibinfo{person}{Sagar Chaki}, {and} \bibinfo{person}{Sriram~K. Rajamani}.} \bibinfo{year}{2001}\natexlab{}.
\newblock \showarticletitle{Parameterized {{Verification}} of {{Multithreaded Software Libraries}}}.
\newblock In \bibinfo{booktitle}{\emph{Tools and {{Algorithms}} for the {{Construction}} and {{Analysis}} of {{Systems}}}}, \bibfield{editor}{\bibinfo{person}{Gerhard Goos}, \bibinfo{person}{Juris Hartmanis}, \bibinfo{person}{Jan Van~Leeuwen}, \bibinfo{person}{Tiziana Margaria}, {and} \bibinfo{person}{Wang Yi}} (Eds.). Vol.~\bibinfo{volume}{2031}. \bibinfo{publisher}{Springer Berlin Heidelberg}, \bibinfo{address}{Berlin, Heidelberg}, \bibinfo{pages}{158--173}.
\newblock
\showISBNx{978-3-540-41865-8 978-3-540-45319-2}
\href{https://doi.org/10.1007/3-540-45319-9_12}{doi:\nolinkurl{10.1007/3-540-45319-9_12}}


\bibitem[Boja{\'n}czyk(2019)]%
        {Bojanczyk2019}
\bibfield{author}{\bibinfo{person}{Miko{\l}aj Boja{\'n}czyk}.} \bibinfo{year}{2019}\natexlab{}.
\newblock \bibinfo{title}{Slightly {{Infinite Sets}}}.  (\bibinfo{year}{2019}).
\newblock


\bibitem[Bojanczyk et~al\mbox{.}(2012)]%
        {BojanczykBKL2012}
\bibfield{author}{\bibinfo{person}{Mikolaj Bojanczyk}, \bibinfo{person}{Laurent Braud}, \bibinfo{person}{Bartek Klin}, {and} \bibinfo{person}{Slawomir Lasota}.} \bibinfo{year}{2012}\natexlab{}.
\newblock \showarticletitle{Towards Nominal Computation}. In \bibinfo{booktitle}{\emph{Proceedings of the 39th Annual {{ACM SIGPLAN-SIGACT}} Symposium on {{Principles}} of Programming Languages}} \emph{(\bibinfo{series}{{{POPL}} '12})}. \bibinfo{publisher}{Association for Computing Machinery}, \bibinfo{address}{New York, NY, USA}, \bibinfo{pages}{401--412}.
\newblock
\showISBNx{978-1-4503-1083-3}
\href{https://doi.org/10.1145/2103656.2103704}{doi:\nolinkurl{10.1145/2103656.2103704}}


\bibitem[Demri and Lazic(2006)]%
        {DemriL2006}
\bibfield{author}{\bibinfo{person}{S. Demri} {and} \bibinfo{person}{R. Lazic}.} \bibinfo{year}{2006}\natexlab{}.
\newblock \showarticletitle{{{LTL}} with the {{Freeze Quantifier}} and {{Register Automata}}}. In \bibinfo{booktitle}{\emph{21st {{Annual IEEE Symposium}} on {{Logic}} in {{Computer Science}} ({{LICS}}'06)}}. \bibinfo{pages}{17--26}.
\newblock
\showISSN{1043-6871}
\href{https://doi.org/10.1109/LICS.2006.31}{doi:\nolinkurl{10.1109/LICS.2006.31}}


\bibitem[Elmas et~al\mbox{.}(2009)]%
        {ElmasQT2009}
\bibfield{author}{\bibinfo{person}{Tayfun Elmas}, \bibinfo{person}{Shaz Qadeer}, {and} \bibinfo{person}{Serdar Tasiran}.} \bibinfo{year}{2009}\natexlab{}.
\newblock \showarticletitle{A Calculus of Atomic Actions}. In \bibinfo{booktitle}{\emph{Proceedings of the 36th Annual {{ACM SIGPLAN-SIGACT}} Symposium on {{Principles}} of Programming Languages}} \emph{(\bibinfo{series}{{{POPL}} '09})}. \bibinfo{publisher}{Association for Computing Machinery}, \bibinfo{address}{New York, NY, USA}, \bibinfo{pages}{2--15}.
\newblock
\showISBNx{978-1-60558-379-2}
\href{https://doi.org/10.1145/1480881.1480885}{doi:\nolinkurl{10.1145/1480881.1480885}}


\bibitem[Emerson et~al\mbox{.}(2000)]%
        {EmersonHT2000}
\bibfield{author}{\bibinfo{person}{E.A. Emerson}, \bibinfo{person}{J.W. Havlicek}, {and} \bibinfo{person}{R.J. Trefler}.} \bibinfo{year}{2000}\natexlab{}.
\newblock \showarticletitle{Virtual Symmetry Reduction}. In \bibinfo{booktitle}{\emph{Proceedings {{Fifteenth Annual IEEE Symposium}} on {{Logic}} in {{Computer Science}} ({{Cat}}. {{No}}.{{99CB36332}})}}. \bibinfo{pages}{121--131}.
\newblock
\showISSN{1043-6871}
\href{https://doi.org/10.1109/LICS.2000.855761}{doi:\nolinkurl{10.1109/LICS.2000.855761}}


\bibitem[Emerson and Namjoshi(1995)]%
        {EmersonN1995}
\bibfield{author}{\bibinfo{person}{E.~Allen Emerson} {and} \bibinfo{person}{Kedar~S. Namjoshi}.} \bibinfo{year}{1995}\natexlab{}.
\newblock \showarticletitle{Reasoning about Rings}. In \bibinfo{booktitle}{\emph{Proceedings of the 22nd {{ACM SIGPLAN-SIGACT}} Symposium on {{Principles}} of Programming Languages}} \emph{(\bibinfo{series}{{{POPL}} '95})}. \bibinfo{publisher}{Association for Computing Machinery}, \bibinfo{address}{New York, NY, USA}, \bibinfo{pages}{85--94}.
\newblock
\showISBNx{978-0-89791-692-9}
\href{https://doi.org/10.1145/199448.199468}{doi:\nolinkurl{10.1145/199448.199468}}


\bibitem[Emerson and Sistla(1996)]%
        {EmersonS1996}
\bibfield{author}{\bibinfo{person}{E.~Allen Emerson} {and} \bibinfo{person}{A.~Prasad Sistla}.} \bibinfo{year}{1996}\natexlab{}.
\newblock \showarticletitle{Symmetry and Model Checking}.
\newblock \bibinfo{journal}{\emph{Formal Methods in System Design}} \bibinfo{volume}{9}, \bibinfo{number}{1} (\bibinfo{date}{Aug.} \bibinfo{year}{1996}), \bibinfo{pages}{105--131}.
\newblock
\showISSN{1572-8102}
\href{https://doi.org/10.1007/BF00625970}{doi:\nolinkurl{10.1007/BF00625970}}


\bibitem[Emmi et~al\mbox{.}(2010)]%
        {EmmiMM2010}
\bibfield{author}{\bibinfo{person}{Michael Emmi}, \bibinfo{person}{Rupak Majumdar}, {and} \bibinfo{person}{Roman Manevich}.} \bibinfo{year}{2010}\natexlab{}.
\newblock \showarticletitle{Parameterized Verification of Transactional Memories}. In \bibinfo{booktitle}{\emph{Proceedings of the 31st {{ACM SIGPLAN Conference}} on {{Programming Language Design}} and {{Implementation}}}} \emph{(\bibinfo{series}{{{PLDI}} '10})}. \bibinfo{publisher}{Association for Computing Machinery}, \bibinfo{address}{New York, NY, USA}, \bibinfo{pages}{134--145}.
\newblock
\showISBNx{978-1-4503-0019-3}
\href{https://doi.org/10.1145/1806596.1806613}{doi:\nolinkurl{10.1145/1806596.1806613}}


\bibitem[Esparza et~al\mbox{.}(2016)]%
        {EsparzaGM2016}
\bibfield{author}{\bibinfo{person}{Javier Esparza}, \bibinfo{person}{Pierre Ganty}, {and} \bibinfo{person}{Rupak Majumdar}.} \bibinfo{year}{2016}\natexlab{}.
\newblock \showarticletitle{Parameterized {{Verification}} of {{Asynchronous Shared-Memory Systems}}}.
\newblock \bibinfo{journal}{\emph{J. ACM}} \bibinfo{volume}{63}, \bibinfo{number}{1} (\bibinfo{date}{Feb.} \bibinfo{year}{2016}), \bibinfo{pages}{10:1--10:48}.
\newblock
\showISSN{0004-5411}
\href{https://doi.org/10.1145/2842603}{doi:\nolinkurl{10.1145/2842603}}


\bibitem[Farzan and Kincaid(2025)]%
        {FarzanK2025}
\bibfield{author}{\bibinfo{person}{Azadeh Farzan} {and} \bibinfo{person}{Zachary Kincaid}.} \bibinfo{year}{2025}\natexlab{}.
\newblock \showarticletitle{The {{Beauty}} of {{Predicate Automata}}}.
\newblock In \bibinfo{booktitle}{\emph{On the Pursuit of Insight and Elegance: Essays on the Occasion of {{Andreas Podelski}}'s 65th Birthday}}. Vol.~\bibinfo{volume}{14765}. \bibinfo{publisher}{LNCS}.
\newblock


\bibitem[Farzan et~al\mbox{.}(2015)]%
        {FarzanKP2015}
\bibfield{author}{\bibinfo{person}{Azadeh Farzan}, \bibinfo{person}{Zachary Kincaid}, {and} \bibinfo{person}{Andreas Podelski}.} \bibinfo{year}{2015}\natexlab{}.
\newblock \showarticletitle{Proof {{Spaces}} for {{Unbounded Parallelism}}}. In \bibinfo{booktitle}{\emph{Proceedings of the 42nd {{Annual ACM SIGPLAN-SIGACT Symposium}} on {{Principles}} of {{Programming Languages}}}}. \bibinfo{publisher}{ACM}, \bibinfo{address}{Mumbai India}, \bibinfo{pages}{407--420}.
\newblock
\showISBNx{978-1-4503-3300-9}
\href{https://doi.org/10.1145/2676726.2677012}{doi:\nolinkurl{10.1145/2676726.2677012}}


\bibitem[Farzan et~al\mbox{.}(2024)]%
        {FarzanKP2024}
\bibfield{author}{\bibinfo{person}{Azadeh Farzan}, \bibinfo{person}{Dominik Klumpp}, {and} \bibinfo{person}{Andreas Podelski}.} \bibinfo{year}{2024}\natexlab{}.
\newblock \showarticletitle{Commutativity {{Simplifies Proofs}} of {{Parameterized Programs}}}.
\newblock \bibinfo{journal}{\emph{Proceedings of the ACM on Programming Languages}} \bibinfo{volume}{8}, \bibinfo{number}{POPL} (\bibinfo{date}{Jan.} \bibinfo{year}{2024}), \bibinfo{pages}{2485--2513}.
\newblock
\showISSN{2475-1421}
\href{https://doi.org/10.1145/3632925}{doi:\nolinkurl{10.1145/3632925}}


\bibitem[Figueira(2012)]%
        {Figueira2012}
\bibfield{author}{\bibinfo{person}{Diego Figueira}.} \bibinfo{year}{2012}\natexlab{}.
\newblock \showarticletitle{Alternating Register Automata on Finite Words and Trees}.
\newblock \bibinfo{journal}{\emph{Logical Methods in Computer Science}}  \bibinfo{volume}{Volume 8, Issue 1} (\bibinfo{date}{March} \bibinfo{year}{2012}), \bibinfo{pages}{907}.
\newblock
\showISSN{1860-5974}
\href{https://doi.org/10.2168/LMCS-8(1:22)2012}{doi:\nolinkurl{10.2168/LMCS-8(1:22)2012}}


\bibitem[Finkel and Schnoebelen(2001)]%
        {FinkelS2001}
\bibfield{author}{\bibinfo{person}{A. Finkel} {and} \bibinfo{person}{{\relax Ph}. Schnoebelen}.} \bibinfo{year}{2001}\natexlab{}.
\newblock \showarticletitle{Well-Structured Transition Systems Everywhere!}
\newblock \bibinfo{journal}{\emph{Theoretical Computer Science}} \bibinfo{volume}{256}, \bibinfo{number}{1} (\bibinfo{date}{April} \bibinfo{year}{2001}), \bibinfo{pages}{63--92}.
\newblock
\showISSN{0304-3975}
\href{https://doi.org/10.1016/S0304-3975(00)00102-X}{doi:\nolinkurl{10.1016/S0304-3975(00)00102-X}}


\bibitem[Flanagan and Freund(2020)]%
        {FlanaganF2020}
\bibfield{author}{\bibinfo{person}{Cormac Flanagan} {and} \bibinfo{person}{Stephen~N. Freund}.} \bibinfo{year}{2020}\natexlab{}.
\newblock \showarticletitle{The Anchor Verifier for Blocking and Non-Blocking Concurrent Software}.
\newblock \bibinfo{journal}{\emph{Software Artifact for "The Anchor Verifier for Blocking and Non-Blocking Concurrent Software"}} \bibinfo{volume}{4}, \bibinfo{number}{OOPSLA} (\bibinfo{date}{Nov.} \bibinfo{year}{2020}), \bibinfo{pages}{156:1--156:29}.
\newblock
\href{https://doi.org/10.1145/3428224}{doi:\nolinkurl{10.1145/3428224}}


\bibitem[German and Sistla(1992)]%
        {GermanS1992}
\bibfield{author}{\bibinfo{person}{Steven~M. German} {and} \bibinfo{person}{A.~Prasad Sistla}.} \bibinfo{year}{1992}\natexlab{}.
\newblock \showarticletitle{Reasoning about Systems with Many Processes}.
\newblock \bibinfo{journal}{\emph{J. ACM}} \bibinfo{volume}{39}, \bibinfo{number}{3} (\bibinfo{date}{July} \bibinfo{year}{1992}), \bibinfo{pages}{675--735}.
\newblock
\showISSN{0004-5411}
\href{https://doi.org/10.1145/146637.146681}{doi:\nolinkurl{10.1145/146637.146681}}


\bibitem[Golubitsky and Stewart(2006)]%
        {GolubitskyS2006}
\bibfield{author}{\bibinfo{person}{Martin Golubitsky} {and} \bibinfo{person}{Ian Stewart}.} \bibinfo{year}{2006}\natexlab{}.
\newblock \showarticletitle{Nonlinear Dynamics of Networks: The Groupoid Formalism}.
\newblock \bibinfo{journal}{\emph{Bull. Amer. Math. Soc.}} \bibinfo{volume}{43}, \bibinfo{number}{03} (\bibinfo{date}{May} \bibinfo{year}{2006}), \bibinfo{pages}{305--365}.
\newblock
\showISSN{0273-0979}
\href{https://doi.org/10.1090/S0273-0979-06-01108-6}{doi:\nolinkurl{10.1090/S0273-0979-06-01108-6}}


\bibitem[Hodges(1997)]%
        {Hodges1997}
\bibfield{author}{\bibinfo{person}{Wilfrid Hodges}.} \bibinfo{year}{1997}\natexlab{}.
\newblock \bibinfo{booktitle}{\emph{A Shorter Model Theory}}.
\newblock \bibinfo{publisher}{Cambridge University Press}, \bibinfo{address}{Cambridge ; New York}.
\newblock
\showISBNx{978-0-521-58713-6}
\showLCCN{QA9.7 .H65 1997}


\bibitem[Hoenicke et~al\mbox{.}(2017)]%
        {HoenickeMP2017}
\bibfield{author}{\bibinfo{person}{Jochen Hoenicke}, \bibinfo{person}{Rupak Majumdar}, {and} \bibinfo{person}{Andreas Podelski}.} \bibinfo{year}{2017}\natexlab{}.
\newblock \showarticletitle{Thread Modularity at Many Levels: A Pearl in Compositional Verification}. In \bibinfo{booktitle}{\emph{Proceedings of the 44th {{ACM SIGPLAN Symposium}} on {{Principles}} of {{Programming Languages}}}}. \bibinfo{publisher}{ACM}, \bibinfo{address}{Paris France}, \bibinfo{pages}{473--485}.
\newblock
\showISBNx{978-1-4503-4660-3}
\href{https://doi.org/10.1145/3009837.3009893}{doi:\nolinkurl{10.1145/3009837.3009893}}


\bibitem[Kaminski and Francez(1994)]%
        {KaminskiF1994}
\bibfield{author}{\bibinfo{person}{Michael Kaminski} {and} \bibinfo{person}{Nissim Francez}.} \bibinfo{year}{1994}\natexlab{}.
\newblock \showarticletitle{Finite-Memory Automata}.
\newblock \bibinfo{journal}{\emph{Theoretical Computer Science}} \bibinfo{volume}{134}, \bibinfo{number}{2} (\bibinfo{date}{Nov.} \bibinfo{year}{1994}), \bibinfo{pages}{329--363}.
\newblock
\showISSN{0304-3975}
\href{https://doi.org/10.1016/0304-3975(94)90242-9}{doi:\nolinkurl{10.1016/0304-3975(94)90242-9}}


\bibitem[Karp and Miller(1969)]%
        {KarpM1969}
\bibfield{author}{\bibinfo{person}{Richard~M. Karp} {and} \bibinfo{person}{Raymond~E. Miller}.} \bibinfo{year}{1969}\natexlab{}.
\newblock \showarticletitle{Parallel Program Schemata}.
\newblock \bibinfo{journal}{\emph{J. Comput. System Sci.}} \bibinfo{volume}{3}, \bibinfo{number}{2} (\bibinfo{date}{May} \bibinfo{year}{1969}), \bibinfo{pages}{147--195}.
\newblock
\showISSN{0022-0000}
\href{https://doi.org/10.1016/S0022-0000(69)80011-5}{doi:\nolinkurl{10.1016/S0022-0000(69)80011-5}}


\bibitem[Kincaid(2016)]%
        {Kincaid2016}
\bibfield{author}{\bibinfo{person}{Zachary Kincaid}.} \bibinfo{year}{2016}\natexlab{}.
\newblock \emph{\bibinfo{title}{Parallel {{Proofs}} for {{Parallel Programs}}}}.
\newblock \bibinfo{thesistype}{Ph.\,D. Dissertation}.
\newblock


\bibitem[Kragl et~al\mbox{.}(2020)]%
        {KraglEHM+2020}
\bibfield{author}{\bibinfo{person}{Bernhard Kragl}, \bibinfo{person}{Constantin Enea}, \bibinfo{person}{Thomas~A. Henzinger}, \bibinfo{person}{Suha~Orhun Mutluergil}, {and} \bibinfo{person}{Shaz Qadeer}.} \bibinfo{year}{2020}\natexlab{}.
\newblock \showarticletitle{Inductive Sequentialization of Asynchronous Programs}. In \bibinfo{booktitle}{\emph{Proceedings of the 41st {{ACM SIGPLAN Conference}} on {{Programming Language Design}} and {{Implementation}}}} \emph{(\bibinfo{series}{{{PLDI}} 2020})}. \bibinfo{publisher}{Association for Computing Machinery}, \bibinfo{address}{New York, NY, USA}, \bibinfo{pages}{227--242}.
\newblock
\showISBNx{978-1-4503-7613-6}
\href{https://doi.org/10.1145/3385412.3385980}{doi:\nolinkurl{10.1145/3385412.3385980}}


\bibitem[Kragl and Qadeer(2018)]%
        {KraglQ2018}
\bibfield{author}{\bibinfo{person}{Bernhard Kragl} {and} \bibinfo{person}{Shaz Qadeer}.} \bibinfo{year}{2018}\natexlab{}.
\newblock \showarticletitle{Layered {{Concurrent Programs}}}. In \bibinfo{booktitle}{\emph{Computer {{Aided Verification}}}}, \bibfield{editor}{\bibinfo{person}{Hana Chockler} {and} \bibinfo{person}{Georg Weissenbacher}} (Eds.). \bibinfo{publisher}{Springer International Publishing}, \bibinfo{address}{Cham}, \bibinfo{pages}{79--102}.
\newblock
\showISBNx{978-3-319-96145-3}
\href{https://doi.org/10.1007/978-3-319-96145-3_5}{doi:\nolinkurl{10.1007/978-3-319-96145-3_5}}


\bibitem[La~Torre et~al\mbox{.}(2015)]%
        {LaTorreMW2015}
\bibfield{author}{\bibinfo{person}{Salvatore La~Torre}, \bibinfo{person}{Anca Muscholl}, {and} \bibinfo{person}{Igor Walukiewicz}.} \bibinfo{year}{2015}\natexlab{}.
\newblock \showarticletitle{Safety of {{Parametrized Asynchronous Shared-Memory Systems}} Is {{Almost Always Decidable}}}. In \bibinfo{booktitle}{\emph{26th {{International Conference}} on {{Concurrency Theory}} ({{CONCUR}} 2015)}} \emph{(\bibinfo{series}{Leibniz {{International Proceedings}} in {{Informatics}} ({{LIPIcs}})}, Vol.~\bibinfo{volume}{42})}, \bibfield{editor}{\bibinfo{person}{Luca Aceto} {and} \bibinfo{person}{David {de Frutos Escrig}}} (Eds.). \bibinfo{publisher}{Schloss Dagstuhl -- Leibniz-Zentrum f\"ur Informatik}, \bibinfo{address}{Dagstuhl, Germany}, \bibinfo{pages}{72--84}.
\newblock
\showISBNx{978-3-939897-91-0}
\showISSN{1868-8969}
\href{https://doi.org/10.4230/LIPIcs.CONCUR.2015.72}{doi:\nolinkurl{10.4230/LIPIcs.CONCUR.2015.72}}


\bibitem[Libkin(2004)]%
        {Libkin2004}
\bibfield{author}{\bibinfo{person}{Leonid Libkin}.} \bibinfo{year}{2004}\natexlab{}.
\newblock \bibinfo{booktitle}{\emph{Elements of {{Finite Model Theory}}}}.
\newblock \bibinfo{publisher}{Springer Berlin Heidelberg}, \bibinfo{address}{Berlin, Heidelberg}.
\newblock
\showISBNx{978-3-642-05948-3 978-3-662-07003-1}
\href{https://doi.org/10.1007/978-3-662-07003-1}{doi:\nolinkurl{10.1007/978-3-662-07003-1}}


\bibitem[Lin et~al\mbox{.}(2016)]%
        {LinNRS2016}
\bibfield{author}{\bibinfo{person}{Anthony~W. Lin}, \bibinfo{person}{Truong~Khanh Nguyen}, \bibinfo{person}{Philipp R{\"u}mmer}, {and} \bibinfo{person}{Jun Sun}.} \bibinfo{year}{2016}\natexlab{}.
\newblock \showarticletitle{Regular {{Symmetry Patterns}}}. In \bibinfo{booktitle}{\emph{Verification, {{Model Checking}}, and {{Abstract Interpretation}}}}, \bibfield{editor}{\bibinfo{person}{Barbara Jobstmann} {and} \bibinfo{person}{K.~Rustan~M. Leino}} (Eds.). \bibinfo{publisher}{Springer}, \bibinfo{address}{Berlin, Heidelberg}, \bibinfo{pages}{455--475}.
\newblock
\showISBNx{978-3-662-49122-5}
\href{https://doi.org/10.1007/978-3-662-49122-5_22}{doi:\nolinkurl{10.1007/978-3-662-49122-5_22}}


\bibitem[Namjoshi and Trefler(2012)]%
        {NamjoshiT2012}
\bibfield{author}{\bibinfo{person}{Kedar~S. Namjoshi} {and} \bibinfo{person}{Richard~J. Trefler}.} \bibinfo{year}{2012}\natexlab{}.
\newblock \showarticletitle{Local {{Symmetry}} and {{Compositional Verification}}}.
\newblock In \bibinfo{booktitle}{\emph{Verification, {{Model Checking}}, and {{Abstract Interpretation}}}}, \bibfield{editor}{\bibinfo{person}{Viktor Kuncak} {and} \bibinfo{person}{Andrey Rybalchenko}} (Eds.). Vol.~\bibinfo{volume}{7148}. \bibinfo{publisher}{Springer Berlin Heidelberg}, \bibinfo{address}{Berlin, Heidelberg}, \bibinfo{pages}{348--362}.
\newblock
\showISBNx{978-3-642-27939-3 978-3-642-27940-9}
\href{https://doi.org/10.1007/978-3-642-27940-9_23}{doi:\nolinkurl{10.1007/978-3-642-27940-9_23}}


\bibitem[Namjoshi and Trefler(2016)]%
        {NamjoshiT2016}
\bibfield{author}{\bibinfo{person}{Kedar~S. Namjoshi} {and} \bibinfo{person}{Richard~J. Trefler}.} \bibinfo{year}{2016}\natexlab{}.
\newblock \showarticletitle{Parameterized {{Compositional Model Checking}}}. In \bibinfo{booktitle}{\emph{Tools and {{Algorithms}} for the {{Construction}} and {{Analysis}} of {{Systems}}}}, \bibfield{editor}{\bibinfo{person}{Marsha Chechik} {and} \bibinfo{person}{Jean-Fran{\c c}ois Raskin}} (Eds.). \bibinfo{publisher}{Springer}, \bibinfo{address}{Berlin, Heidelberg}, \bibinfo{pages}{589--606}.
\newblock
\showISBNx{978-3-662-49674-9}
\href{https://doi.org/10.1007/978-3-662-49674-9_39}{doi:\nolinkurl{10.1007/978-3-662-49674-9_39}}


\bibitem[Namjoshi and Trefler(2018)]%
        {NamjoshiT2018}
\bibfield{author}{\bibinfo{person}{Kedar~S. Namjoshi} {and} \bibinfo{person}{Richard~J. Trefler}.} \bibinfo{year}{2018}\natexlab{}.
\newblock \showarticletitle{Symmetry {{Reduction}} for the {{Local Mu-Calculus}}}. In \bibinfo{booktitle}{\emph{Tools and {{Algorithms}} for the {{Construction}} and {{Analysis}} of {{Systems}}}}, \bibfield{editor}{\bibinfo{person}{Dirk Beyer} {and} \bibinfo{person}{Marieke Huisman}} (Eds.). \bibinfo{publisher}{Springer International Publishing}, \bibinfo{address}{Cham}, \bibinfo{pages}{379--395}.
\newblock
\showISBNx{978-3-319-89963-3}
\href{https://doi.org/10.1007/978-3-319-89963-3_22}{doi:\nolinkurl{10.1007/978-3-319-89963-3_22}}


\bibitem[Neven et~al\mbox{.}(2004)]%
        {NevenSV2004}
\bibfield{author}{\bibinfo{person}{Frank Neven}, \bibinfo{person}{Thomas Schwentick}, {and} \bibinfo{person}{Victor Vianu}.} \bibinfo{year}{2004}\natexlab{}.
\newblock \showarticletitle{Finite State Machines for Strings over Infinite Alphabets}.
\newblock \bibinfo{journal}{\emph{ACM Trans. Comput. Logic}} \bibinfo{volume}{5}, \bibinfo{number}{3} (\bibinfo{date}{July} \bibinfo{year}{2004}), \bibinfo{pages}{403--435}.
\newblock
\showISSN{1529-3785}
\href{https://doi.org/10.1145/1013560.1013562}{doi:\nolinkurl{10.1145/1013560.1013562}}


\bibitem[Padon et~al\mbox{.}(2017)]%
        {PadonLSS2017}
\bibfield{author}{\bibinfo{person}{Oded Padon}, \bibinfo{person}{Giuliano Losa}, \bibinfo{person}{Mooly Sagiv}, {and} \bibinfo{person}{Sharon Shoham}.} \bibinfo{year}{2017}\natexlab{}.
\newblock \showarticletitle{Paxos Made {{EPR}}: Decidable Reasoning about Distributed Protocols}.
\newblock \bibinfo{journal}{\emph{Proceedings of the ACM on Programming Languages}} \bibinfo{volume}{1}, \bibinfo{number}{OOPSLA} (\bibinfo{date}{Oct.} \bibinfo{year}{2017}), \bibinfo{pages}{1--31}.
\newblock
\showISSN{2475-1421}
\href{https://doi.org/10.1145/3140568}{doi:\nolinkurl{10.1145/3140568}}


\bibitem[Pnueli et~al\mbox{.}(2001)]%
        {PnueliRZ2001}
\bibfield{author}{\bibinfo{person}{Amir Pnueli}, \bibinfo{person}{Sitvanit Ruah}, {and} \bibinfo{person}{Lenore Zuck}.} \bibinfo{year}{2001}\natexlab{}.
\newblock \showarticletitle{Automatic {{Deductive Verification}} with {{Invisible Invariants}}}. In \bibinfo{booktitle}{\emph{Tools and {{Algorithms}} for the {{Construction}} and {{Analysis}} of {{Systems}}}}, \bibfield{editor}{\bibinfo{person}{Tiziana Margaria} {and} \bibinfo{person}{Wang Yi}} (Eds.). \bibinfo{publisher}{Springer}, \bibinfo{address}{Berlin, Heidelberg}, \bibinfo{pages}{82--97}.
\newblock
\showISBNx{978-3-540-45319-2}
\href{https://doi.org/10.1007/3-540-45319-9_7}{doi:\nolinkurl{10.1007/3-540-45319-9_7}}


\bibitem[{v. Gleissenthall} et~al\mbox{.}(2019)]%
        {v.GleissenthallKBS+2019}
\bibfield{author}{\bibinfo{person}{Klaus {v. Gleissenthall}}, \bibinfo{person}{Rami~G{\"o}khan K{\i}c{\i}}, \bibinfo{person}{Alexander Bakst}, \bibinfo{person}{Deian Stefan}, {and} \bibinfo{person}{Ranjit Jhala}.} \bibinfo{year}{2019}\natexlab{}.
\newblock \showarticletitle{Pretend Synchrony: Synchronous Verification of Asynchronous Distributed Programs}.
\newblock \bibinfo{journal}{\emph{Replication Package for Article: "Pretend Synchrony: Synchronous Verification of Asynchronous Distributed Programs"}} \bibinfo{volume}{3}, \bibinfo{number}{POPL} (\bibinfo{date}{Jan.} \bibinfo{year}{2019}), \bibinfo{pages}{59:1--59:30}.
\newblock
\href{https://doi.org/10.1145/3290372}{doi:\nolinkurl{10.1145/3290372}}


\end{thebibliography}

\appendix
{
\crefalias{section}{appendix}
\crefalias{subsection}{appendix}
\section{Details on Section \ref{sec:example}} \label{app:example}

\paragraph{\bfseries Vocabulary and Limit}

Fix a height $h \in \NN_+$. Let $\VNode := \{ u, d \} \cup \{ p_i \mid i \in \{ -1, \ldots, h - 1 \} \}$, where
\begin{itemize}
    \item $u$ and $d$ are unary function symbols that stand for ``up'' and ``down'' respectively; for a rectangle node $e$, we set $u(e) = d(e) = e$.
    \item For every $i \in \{ -1, \ldots, h - 1 \}$, $p_i$ is a binary predicate symbol that stands for the test of whether the lowest common ancestor (LCA) of two nodes has depth $i$; for two nodes belonging to two different forests, we consider the depth of their LCA to be $-1$.
\end{itemize}

For any $k, l \in \NN_+$, let the $\VNode$-structure $\TT_{k,l}$ be the forest consisting of $l$ trees of height $h$ with a branching factor of $k$ for the rectangle nodes, in the style of \Cref{fig:example} (except that the height of the tree is $h$ instead of $5$). Let $\cT = \{ \TT_{k,l} \}_{(k,l) \in \cI}$, where $\cI := \NN_+ \times \NN_+$.

One can verify that the condition of \Cref{prop:fraisse} is satisfied, so $\cT$ admits a limit $\TT_{\mathrm{forest}}$ that is constructible as a Fra\"{i}ss\'{e} limit. Note that the choice of the vocabulary $\VNode$ matters for the existence of a Fra\"{i}ss\'{e} limit, although the function symbols $u$ and $d$ are sufficient to describe a forest, and a limit in the sense of \Cref{def:lim-struct} would still exist with $\VNode = \{ u, d \}$. Without the binary predicate symbols (even if we replace them with unary predicates that indicate the depth of each node), $\cls{\cT}$ would not be a Fra\"{i}ss\'{e} class, as the \emph{amalgamation property} would be violated.

\paragraph{\bfseries Well-quasi-order} We first show two lemmas. The first one is inspired by the proof of \cite[Lemma 1.15]{Bojanczyk2019}.

\begin{lemma} \label{lem:covering-refl}
  Let $\cQ$ be a finite relational vocabulary and $\sC \subset 2^{\sF^\TT_\atom(\cQ)}$ be a set of $\TT$-PA configurations. The covering relation $\preceq$ is a wqo on $\sC$ iff there is a wqo set $(X, \preccurlyeq)$ and a map $f : \sC \to X$ that reflects the quasi-order:
  \[ \forall \cC, \cC' \in \sC.\, f(\cC) \preccurlyeq f(\cC') \implies \cC \preceq \cC'. \]
\end{lemma}
\begin{proof}
  The $\Rightarrow$ direction is trivial. For the $\Leftarrow$ direction, suppose there is a wqo set $(X, \preccurlyeq)$ and a map $f : \sC \to X$ that reflects the quasi-order. Let $\cC_1, \cC_2, \ldots$ be an infinite sequence over $\sC$. If this is an antichain, then so is $f(\cC_1), f(\cC_2), \ldots$, contradicting $\preccurlyeq$ being a wqo. It remains to show it cannot be a strictly descending chain. Suppose $\cC \prec \cC'$, i.e., $\cC \preceq \cC'$ but $\cC' \not\preceq \cC$. Take automorphism $\pi$ such that for any $q(\ti_1, \ldots, \ti_n) \in \cC$, we have $q(\pi(\ti_1), \ldots, \pi(\ti_n)) \in \cC$. Then $\pi$ induces an injection from $\cC$ into $\cC'$. If it is a bijection, then the automorphism $\pi^{-1}$ witnesses $\cC' \preceq \cC$, a contradiction. Thus, we have $\cC \prec \cC' \Rightarrow |\cC| < |\cC'|$, which shows $\cC_1, \cC_2, \ldots$ cannot be a strictly descending chain.
\end{proof}

\begin{lemma} \label{lem:wqo-nullary}
  Let $\cQ$ be a finite relational vocabulary and $\sC \subset 2^{\sF^\TT_\atom(\cQ)}$ be a set of $\TT$-PA configurations. Let $\sC'$ be $\sC$ with all nullary symbols removed from each configuration. If the covering relation wqos $\sC'$, then it also wqos $\sC$.
\end{lemma}
\begin{proof}
  Suppose the covering relation wqos $\sC'$. Let $\cQ_0$ be all nullary symbols in $\cQ$. Then $2^{\cQ_0}$ is wqo by the subset relation. The product $\sC' \times 2^{\cQ_0}$ is wqo by the product order.

  Define map $f : \sC \to \sC' \times 2^{\cQ_0}$ by separating nullary symbols from other atomic formulas. Then $f$ reflects the quasi-order. By Lemma \ref{lem:covering-refl}, $\sC$ is wqo by the covering relation.
\end{proof}

We are ready to show that the covering relation (\Cref{def:covering}) wqos monadic configurations of a $\TT_{\mathrm{forest}}$-PA:
\begin{theorem}
  Let $\TT := \TT_{\mathrm{forest}}$ and $\cQ$ be a finite relational vocabulary (disjoint from $\VNode[\TT]$) where symbols have arity at most~$1$. Let $\sC \subset 2^{\sF^\TT_\atom(\cQ)}$ be a set of $\TT$-PA configurations. Then the covering relation is a well-quasi-order on $\sC$.
\end{theorem}
\begin{proof}
  By \Cref{lem:wqo-nullary}, we may assume that symbols in $\cQ$ are all unary. By \Cref{lem:covering-refl}, it suffices to find a wqo set $(X, \preccurlyeq)$ and a map $f : \sC \to X$ that reflects the quasi-order. Let $L := 2^{\cQ} \times \{-1, \ldots, h- 1\}$, wqo'd by the identity relation. Take $X$ to be the set of $L$-labelled trees and $\preccurlyeq$ be the $L$-labelled topological minor. By Kruskal's tree theorem, $(X, \preccurlyeq)$ is a wqo set.

  For any configuration $\cC \in \sC$, define $f(\cC)$ as the $L$-labelled tree obtained by
  \begin{itemize}
    \item taking the minimal sub-forest of $\TT_{\mathrm{forest}}$ containing all nodes in $\cC$ and their roots,
    \item labeling every node $a$ in the sub-forest with $(\{ q \in \cQ \mid q(a) \in \cC \}, d)$, where $d$ is the depth of $a$, and
    \item adding an extra node labelled $(\varnothing, -1)$ and connecting it to every root in the sub-forest.
  \end{itemize}

  One can verify that $f$ does reflect the quasi-order.
\end{proof}
\section{Proofs}

Throughout the paper, we distinguish the use of parentheses and (square) brackets: the former for (concrete or symbolic) function application and the latter for formula substitution. For instance, if $t[u_1, \ldots, u_n]$ is a $\VNode$-term and $(a_1, \ldots, a_n)$ is a tuple of nodes, then $t(\vec{a})$ is a node and $t[\vec{a}]$ is a formula. $\sval(a_1)$ is a formula as well, since $\sval$ is a concrete function symbol.

We use $\concat$ to denote the concatenation of two tuples. For a topology or a universal structure $\TT$ and any word $\rho$ over $\Sigma \times \TT$, we use $\nodes(\rho)$ to denote the set of nodes that appear in $\rho$.

\subsection{Proofs from Section \ref{sec:programs}}

\begin{proposition}[\Cref{prop:li-unique}]
    For any $\vec{a} \in \TT^d$ and $\vec{b} \in \TT'^{d}$, there is at most one isomorphism from $\nb(\vec{a})$ to $\nb(\vec{b})$ that maps $\vec{a}$ to $\vec{b}$.
\end{proposition}
\begin{proof}
    Let $\cV$ be the common vocabulary of $\TT$ and $\TT'$. Let $f$ and $g$ be isomorphisms from $\nb(\vec{a})$ to $\nb(\vec{b})$ such that $f(\vec{a}) = \vec{b}$ and $g(\vec{a}) = \vec{b}$. Let $c \in \nb(\vec{a})$. Then there is some $\cV$-term $t[\vec{u}]$ such that $c = t(\vec{a})$. It follows that $f(c) = f(t(\vec{a})) = t(f(\vec{a})) = t(\vec{b})$. Similarly, we have $g(c) = t(\vec{b})$. Therefore, $f$ and $g$ are identical.
\end{proof}

\begin{proposition}[\Cref{prop:local-symm-qf}]
    Let $\TT$ and $\TT'$ be structures over the vocabulary $\cV$. For any $\vec{a} \in \TT^d$ and $\vec{b} \in {\TT'}^d$, we have $\vec{a} \simeq \vec{b}$ if and only if for every quantifier-free $\cV$-formula $\varphi[x_1, \ldots, x_d]$, $\TT \models \varphi[\vec{a}] \iff \TT' \models \varphi[\vec{b}]$.
\end{proposition}
\begin{proof}
    $\Rightarrow$: Suppose $\vec{a} \simeq \vec{b}$. Then there is an isomorphism $\varphi : \nb(\vec{a}) \to \nb(\vec{b})$, which induces an embedding $\nb(\vec{a}) \hookrightarrow \TT'$. \\ Quantifier-free formulas are absolute under embeddings.

    $\Leftarrow$: Suppose for every quantifier-free $\cV$-formula $\varphi[x_1, \ldots, x_d]$, $\TT \models \varphi[\vec{a}] \iff \TT' \models \varphi[\vec{b}]$. Define a relation $f \in \nb(\vec{a}) \times \nb(\vec{b}) := \{ (t(\vec{a}), t(\vec{b})) \mid t \text{ is a } \cV \text{-term} \}$. It is easy to show that $f$ is (the graph of) a local isomorphism from $\vec{a}$ to $\vec{b}$. For instance, if $t_1, t_2$ are $\cV$-terms and $t_1(\vec{a}) = t_2(\vec{a})$, setting $\varphi := t_1[\vec{u}] = t_2[\vec{u}]$, we have $t_1(\vec{b}) = t_2(\vec{b})$, which shows well-definedness.
\end{proof}

\begin{proposition}[\Cref{prop:local-symm-fin}]
    For any $d \in \NN_+$, if the quotient $\TT^d/{\simeq}$ is finite, then every equivalence class of $\simeq$ is definable by a quantifier-free $\cV$-formula.
\end{proposition}
\begin{proof}
    Write $\TT^d / {\simeq}$ as $\{ E_1, \ldots, E_n \}$. For $i = 1, \ldots, n$, let $\cF_i$ be the set of all quantifier-free formulas with $d$ free variables satisfied by the tuples in $E_i$. Let $i, j \in \{1, \ldots, n\}$ be distinct. By \Cref{prop:local-symm-qf}, there is a quantifier-free formula $\varphi$ such that $\varphi \in \cF_i$ and $\varphi \notin \cF_j$ or $\varphi \in \cF_j$ and $\varphi \notin \cF_i$. In the former case, set $\psi_{i,j} := \varphi$. In the latter case, set $\psi_{i,j} := \neg \varphi$. Then, $E_i$ is defined by $\bigwedge_{j \neq i} \psi_{i,j}$.
\end{proof}

When it comes to a relational vocabulary, the sets of formulas $\cF_1, \ldots, \cF_n$ are also known as \emph{rank-$0$ $d$-types} \cite[Definition 3.14]{Libkin2004}.

\subsection{Proofs from Section \ref{sec:proofs}}

\begin{lemma} \label{lem:pullback}
    Let $\TT$ be a universal structure. Let $\tp{\sigma_1 : v_1} \cdots \tp{\sigma_n : v_n} \in (\TT \times \Sigma)^*$, $U$ be any neighbourhood containing $\vec{v}$, and $\beta$ be a local isomorphism from $\vec{v}$ to another tuple over $\TT$.
    Then
    \begin{align*}
        & \beta^* \Global_{\beta(U)}\sem{\tp{\sigma_1 : \beta(v_1)} \cdots \tp{\sigma_n : \beta(v_n)}} \\
        ={}& \Global_U \sem{\tp{\sigma_1 : v_1} \cdots \tp{\sigma_n : v_n}}.
    \end{align*}
\end{lemma}
\begin{proof}
    It is easy to verify that
    \begin{align*}
        & \beta^* \Global_{\beta(U)}\sem{\tp{\sigma_1 : \beta(v_1)} \cdots \tp{\sigma_n : \beta(v_n)}} \\
        ={}& \beta^* \Global_{\beta(U)} \sem{\tp{\sigma_1 : \beta(v_1)}} \circ \cdots \circ \beta^* \Global_{\beta(U)} \tp{\sigma_n : \beta(v_n)}.
    \end{align*}
    On the other hand,
    \begin{align*}
        & \Global_U \sem{\tp{\sigma_1 : v_1} \cdots \tp{\sigma_n : v_n}} \\
        ={}& \Global_U \sem{\tp{\sigma_1 : v_1}} \circ \cdots \circ \Global_U \sem{\tp{\sigma_n : v_n}}.
    \end{align*}
    So, it suffices to show that
    \[ \beta^* \Global_{\beta(U)} \sem{\tp{\sigma_1 : \beta(v_1)}} \subset \Global_U \sem{\tp{\sigma_1 : v_1}}. \]
    (For the other direction of the inclusion, note that $\beta^{-1}$ is a local isomorphism from $\beta(\vec{v})$ to $\vec{v}$.)

    Let $(\val_1, \val'_1) \in \beta^* \Global_{\beta(U)} \sem{\tp{\sigma_1 : \beta(v_1)}}$. Set $\val := (\beta^{-1})^* \val_1$ and $\val' := (\beta^{-1})^* \val'_1$. Then $(\val, \val') \in \Global_{\beta(U)} \sem{\tp{\sigma_1 : \beta(v_1)}}$. By definition, we have
    \begin{gather}
        (\val|_{\nb(\beta(v_1))}, \val'|_{\nb(\beta(v_1))}) \in \sem{\tp{\sigma_1 : \beta(v_1)}}, \label{eq:pullback:nb} \\
        \val|_{\setcomp{\nb(\beta(v_1))}} = \val'|_{\setcomp{\nb(\beta(v_1))}}. \label{eq:pullback:outside}
    \end{gather}
    Let $\gamma$ be the restriction of $\beta$ to $\nb(v_1) \to \nb(\beta(v_1))$, which is a local isomorphism from $v_1$ to $\beta(v_1)$. By \eqref{eq:pullback:nb}, we have \[ (\gamma^* (\val|_{\nb(\beta(v_1))}), \gamma^* (\val'|_{\nb(\beta(v_1))})) \in \sem{\tp{\sigma_1 : v_1}}, \]
    where the LHS is exactly $(\val_1|_{\nb(v_1)}, \val'_1|_{\nb(v_1)})$. On the other hand, it follows from \eqref{eq:pullback:outside} that
    \[ \val_1|_{\setcomp{\nb(v_1)}} = \val'_1|_{\setcomp{\nb(v_1)}}. \]
    Hence, $(\val_1, \val'_1) \in \Global_U \sem{\tp{\sigma_1 : v_1}}$.
\end{proof}

\begin{theorem}[\Cref{thm:s-param-sound}]
    The rule \textsc{S-param} is sound for any tuples $\vec{u}$, $\vec{v}$, $\vec{w}$, $\vec{u}'$, $\vec{v}'$, and $\vec{w}'$ from $\TT$.
\end{theorem}
\begin{proof}
    Suppose $\htri{\phi[\overrightarrow{\iv{u}}]}{\tp{\sigma_1 : v_1} \cdots \tp{\sigma_n : v_n}}{\psi[\overrightarrow{\iv{w}}]}$ is valid w.r.t. $\sem{-}$ (the semantic function of indexed commands over the universal structure). Let $U = \nb(\vec{u} \concat \vec{v} \concat \vec{w})$ and $\beta$ be a local isomorphism from $U$. We want to show that $\htri{\phi[\overrightarrow{\iv{\beta(u)}}]}{\tp{\sigma_1 : \beta(v_1)} \cdots \tp{\sigma_n : \beta(v_n)}}{\psi[\overrightarrow{\iv{\beta(w)}}]}$ is also valid.

    Note that $\nb(\beta(\vec{u}) \concat \beta(\vec{v}) \concat \beta(\vec{w})) = \nb(\beta(\vec{u} \concat \vec{v} \concat \vec{w})) = \beta(U)$.
    Let $(\val, \val') \in \Global_{\beta(U)}\sem{\tp{\sigma_1 : \beta(v_1)} \cdots \tp{\sigma_n : \beta(v_n)}}$. Suppose $\val \models \phi[\overrightarrow{\iv{\beta(u)}}]$. Then $(\beta^* \val, \beta^* \val') \in \Global_U \sem{\tp{\sigma_1 : v_1} \cdots \tp{\sigma_n : v_n}}$ by \Cref{lem:pullback}, and $\beta^* \val \models \phi[\overrightarrow{\iv{u}}]$. Thus, $\beta^* \val' \models \psi[\overrightarrow{\iv{w}}]$, and therefore $\val' \models \psi[\overrightarrow{\iv{\beta(w)}}]$.
\end{proof}

\begin{theorem}[\Cref{thm:psl-sym}]
    A proof space language is closed under the equivalence relation $\simeq$.
\end{theorem}
\begin{proof}
    Let $\cH$ be a $\TT$-proof space. Suppose $\tp{\sigma_1 : v_1} \cdots \tp{\sigma_n : v_n} \in \Lang(\cH)$. Then $\htri{\top}{\tp{\sigma_1 : v_1} \cdots \tp{\sigma_n : v_n}}{\bot} \in \cH$. Suppose $\rho \in (\Sigma \times \TT)^*$ and $\rho \simeq \tp{\sigma_1 : v_1} \cdots \tp{\sigma_n : v_n}$. Then there is a tuple $\vec{v}' \in \TT^n$ such that $\vec{v}' \simeq \vec{v}$ and $\rho = \tp{\sigma_1 : v'_1} \cdots \tp{\sigma_n : v'_n}$. By \textsc{S-Param}, we have $\htri{\top}{\rho}{\bot} \in \cH$. Thus, $\rho \in \Lang(\cH)$.
\end{proof}

\begin{lemma} \label{lem:same-validity}
    Let $\{\tp{P_i, \sem{-}_i}\}_{i\in\cI}$ be a parameterized program over topologies $\{\TT_i\}_{i\in\cI}$ with a universal structure $\TT$. For any $i \in \cI$, the validity of any Hoare triple over $(\Sigma \times \TT_i)^*$ is the same w.r.t. $\sem{-}_i$ and $\sem{-}$.
\end{lemma}
\begin{proof}
    Let $i \in \cI$. For any $\tp{\sigma : v} \in \Sigma \times \TT_i$, we have $\sem{\sigma : v} = \sem{\sigma : v}_i$. Thus, for any $i \in \cI$ and $\rho \in (\Sigma \times \TT_i)^*$, for $V = \nb(\nodes(\rho))$, we have $\Global_V\sem{\rho} = \Global_V\sem{\rho}_i$.
\end{proof}

\begin{theorem}[\Cref{thm:safety-proof}]
A parameterized program $\cP = \{ \tp{P_i, \sem{-}_i} \}_{i\in \cI}$ is safe if there is a universal structure $\TT$ and a valid $\TT$-proof space $\cH$ such that $\bigcup_{i\in \cI} \Lang_\err(P_i) \subset \Lang(\cH)$.
\end{theorem}
\begin{proof}
    Let $\cP = \{ \tp{P_i, \sem{-}_i} \}_{i\in \cI}$ be a parameterized program. Suppose there is a universal structure $\TT$ and a valid $\TT$-proof space $\cH$ such that $\bigcup_{i\in \cI} \Lang_\err(P_i) \subset \Lang(\cH)$. Let $i \in \cI$ and $\rho \in \Lang_\err(P_i)$. Then $\rho \in \Lang(\cH)$. By the soundness of parametric proof spaces and \Cref{lem:same-validity}, the Hoare triple $\htri{\top}{\rho}{\bot}$ is valid w.r.t. $\sem{-}_i$, so $\rho$ is infeasible. Thus, every member of $\cP$ is safe, and therefore $\cP$ is safe.
\end{proof}
\subsection{Proofs from Section \ref{sec:completeness}} \label{app:completeness}

\begin{proposition}[\Cref{prop:ext-ass}]
    Given a $\VNode$-structure $\TT$, for any \QFAshcroft{} formula $\varphi[u_1, \ldots, u_n]$ and $\vec{a} \in \TT^n$, there is a unique (up to equivalence) extended assertion $\phi[\overrightarrow{\sval(b)}, \overrightarrow{\sloc(b)}]$, where $\vec{b}$ is a tuple from $\nb(\vec{a})$, such that for any local isomorphism $\beta$ from $\vec{a}$, the extended assertion $\phi[\overrightarrow{\sval(\beta(b))}, \overrightarrow{\sloc(\beta(b))}]$ is equivalent to $\varphi[\beta(\vec{a})]$.
\end{proposition}

\begin{proof}
    Let $\varphi[u_1, \ldots, u_n]$ be a \QFAshcroft{} formula and $\vec{a} \in \TT^n$. One can find $\VNode$-terms $t_1[\vec{u}], \ldots, t_m[\vec{u}]$ such that $\varphi[\vec{u}]$ is (logically) equivalent to a Boolean combination of formulas of the form
    \begin{enumerate}
        \item $\phi_\Node[\vec{u}]$,
        \item $\phi_\Data[\sval(t_1[\vec{u}]), \ldots, \sval(t_m[\vec{u}])]$, and
        \item $\phi_\Loc[\sloc(t_1[\vec{u}]), \ldots, \sloc(t_m[\vec{u}])]$.
    \end{enumerate}
    where $\phi_\Node$ is a quantifier-free $\VNode$-formula, $\phi_\Data$ is a $\VData$-formula, and $\phi_\Loc$ is a $\VLoc$-formula. Thus, we can write
    \begin{multline*}
        \varphi[\vec{u}] \equiv \bigvee_{r=1}^k (\phi_\Node^{(r)}[\vec{u}] \\
        \land \phi_\Data^{(r)}[\sval(t_1[\vec{u}]), \ldots, \sval(t_m[\vec{u}])] \\
        \land \phi_\Loc^{(r)}[\sloc(t_1[\vec{u}]), \ldots, \sloc(t_m[\vec{u}])]).
    \end{multline*}

    Let $R := \{ r \in \{1, \ldots, k\} \mid \phi^{(r)}_\Node[\vec{a}] \equiv \top \}$. Define
    \begin{align*}
        \phi[x_1, \ldots, x_m, l_1, \ldots, l_m] &:= \bigvee_{r\in R} \phi_\Data^{(r)}[\vec{x}]
        \land \phi_\Loc^{(r)}[\vec{l}],\\
        \vec{b} &:= (t_1(\vec{a}), \ldots, t_m(\vec{a})) \in \TT^m.
    \end{align*}
    Then $\phi[\overrightarrow{\sval(b)}, \overrightarrow{\sloc(b)}]$ is an extended assertion that is equivalent to $\varphi[\vec{a}]$. The uniqueness follows easily.

    Let $\beta$ be a local isomorphism from $\vec{a}$. Since local isomorphisms preserve the truth value of quantifier-free $\VNode$-formulas, we have
    \begin{multline*}
        \varphi[\beta(\vec{a})] \equiv \bigvee_{r\in R} (\phi_\Data^{(r)}[\sval(t_1(\beta(\vec{a}))), \ldots, \sval(t_m(\beta(\vec{a})))] \\
        \land \phi_\Loc^{(r)}[\sloc(t_1(\beta(\vec{a}))), \ldots, \sloc(t_m(\beta(\vec{a})))]).
    \end{multline*}
    Since $t_i(\beta(\vec{a})) = \beta(b_i)$, this is equivalent to $\phi[\overrightarrow{\sval(\beta(b))}, \overrightarrow{\sloc(\beta(b))}]$.
\end{proof}

\begin{lemma}[\Cref{lem:ashcroft-ext}]
    Let $\prog$ be a program over a finite topology $\TT_i \subset \TT$. For any Ashcroft assertion $\Phi$, the extended Hoare triples in $H_{\Phi}^{\TT_i}$ are valid if $\Phi$ is an Ashcroft invariant for $\prog$.
\end{lemma}
\begin{proof}
    The extended Hoare triples in $H_{\Phi}^{\TT_i}$ are valid if the following extended Hoare triples are valid: for all indexed command $\tp{\sigma : a_0} \in \Sigma \times \TT_i$,
    \begin{gather*}
        \textstyle
        \htri{\bigwedge_{a \in \TT_i} \sloc(a) = l_\init}{\epsilon}{\bigwedge_{\vec{a} \in \TT_i^k} \varphi[\vec{a}]} \\
        \textstyle
        \htri{\bigwedge_{\vec{a} \in \TT_i^k} \varphi[\vec{a}]}{\tp{\sigma : a_0}}{\bigwedge_{\vec{a} \in \TT_i^k} \varphi[\vec{a}]} \\
        \textstyle
        \htri{\bigwedge_{\vec{a} \in \TT_i^k} \varphi[\vec{a}]}{\epsilon}{\bigwedge_{a \in \TT_i} \sloc(a) \neq l_\err}
    \end{gather*}
    These are just a restatement of the initialization, continuation, and safety condition of $\Phi$, by replacing every universal quantification with a conjunction.
\end{proof}

\begin{lemma}[\Cref{lem:fin-gen-ext}]
    Let $\cP = \{\prog_i\}_{i\in\cI}$ be a parameterized program over topologies $\{ \TT_i\}_{i\in\cI}$ with universal structure $\TT$. Suppose there is a valid finitely generated extended $\TT$-proof space $\cH$ such that for every $i \in \cI$, $\rho \in \Lang_\err(P_i)$, and $b \in \TT_i$, there is some finite subset $U$ of $\TT_i$ such that $\textstyle \htri{\bigwedge_{a \in U} \sloc(a) = l_\init}{\rho}{\sloc(b) \neq l_\err} \in \cH$. 
    Then $\cP$ can be proved safe by a finitely generated $\TT$-proof space.
\end{lemma}
\begin{proof}[Proof Sketch]
    Let $H$ be a finite basis of $\cH$. Let $H'$ contain, for every extended Hoare triple $\htri{\phi[\overrightarrow{\sval(a)},\overrightarrow{\sloc(a)}]}{\rho}{\psi[\overrightarrow{\sval(c)},\overrightarrow{\sloc(c)}]}$ in $H$, for every pair of location maps $\loc$ and $\loc'$ on $\{a_i\}_i \cup \nodes(\rho) \cup \{c_j\}_j$ such that $\loc \xrightarrow{\rho} \loc'$, the Hoare triple
    \[
        \htri{\phi[\overrightarrow{\sval(a)},\overrightarrow{\loc(a)}]}{\rho}{\psi[\overrightarrow{\sval(c)},\overrightarrow{\loc'(c)}]},
    \]
    i.e., substituting $\loc$ and $\loc'$ for $\sloc$ in the pre- and postcondition respectively and simplify, which can be shown to produce a formula in \Assertions{}. Then $H'$ can be shown to be a finite set of valid Hoare triples. Our goal is to show that $\cP$ can be proved safe by the $\TT$-proof space $\cH'$ generated by $H'$, which relies on the following claim that can be shown by induction over the derivation:
    \begin{claim} \label{clm:ext-htri-to-htri}
        For any extended Hoare triple $\htri{\phi[\overrightarrow{\sval(a)},\overrightarrow{\sloc(a)}]}{\rho}{\psi[\overrightarrow{\sval(c)},\overrightarrow{\sloc(c)}]}$ in $\cH$, for any pair of location maps $\loc$ and $\loc'$ on $\{a_i\}_i \cup \nodes(\rho) \cup \{c_j\}_j$ such that $\loc \xrightarrow{\rho} \loc'$, the Hoare triple \\$\htri{\phi[\overrightarrow{\sval(a)},\overrightarrow{\loc(a)}]}{\rho}{\psi[\overrightarrow{\sval(c)},\overrightarrow{\loc'(c)}]}$ is in $\cH'$.
    \end{claim}

    Let $i \in \cI$ and $\rho \in \Lang_\err(P_i)$. We want to show that $\htri{\top}{\rho}{\bot}$ is in $\cH'$. Let $\loc$ be an initial location map on $\nodes(\rho)$ and let $\loc \xrightarrow{\rho} \loc'$. Since $\rho$ is an error run, there is some $b \in \TT_i$ such that $\loc'(b) = l_{\err}$. Then, by assumption, there is some finite subset $U$ of $\TT_i$ such that $\textstyle \htri{\bigwedge_{a \in U} \sloc(a) = l_\init}{\rho}{\sloc(b) \neq l_\err} \in \cH$. By \Cref{clm:ext-htri-to-htri}, it follows that $\textstyle \htri{\bigwedge_{a \in U} \loc(a) = l_\init}{\rho}{\loc(b) \neq l_\err} \in \cH'$. Note that $\left(\bigwedge_{a \in U} \loc(a) = l_\init\right) \equiv \top$ and $\left(\loc(b) \neq l_\err\right) \equiv \bot$.
\end{proof}

\begin{lemma} \label{lem:symm}
    Let $\{ \TT_i \}_{i\in\cI}$ be a family of topologies with universal structure $\TT$. Suppose $\{ \TT_i \}_{i\in\cI}$ has a basis $\{ \TT_j \}_{j\in\cJ}$ of rank $k+1$ where every $\TT_j$ is finite. Let $\cH$ be the extended $\TT$-proof space generated by $\bigcup_{j\in\cJ} H_{\Phi}^{\TT_j}$. Let $i \in \cI$.
    \begin{itemize}
        \item For any $\vec{a} \in \TT_i^k$, there is a finite subset $U$ of $\TT_i$ such that
        \[ \textstyle \htri{\bigwedge_{a \in U} \sloc(a) = l_\init}{\epsilon}{\varphi[\vec{a}]} \in \cH. \]
        \item For any $\tp{\sigma : a_0} \in \Sigma \times \TT_i$ and $\vec{a} \in \TT_i^k$, there is a finite subset $U$ of $\TT_i^k$ such that
        \[ \textstyle \htri{\bigwedge_{\vec{a} \in U} \varphi[\vec{a}]}{\tp{\sigma : a_0}}{\varphi[\vec{a}]} \in \cH. \]
        \item For any $b \in \TT_i$, there is a finite subset $U$ of $\TT_i^k$ such that
        \[ \textstyle \htri{\bigwedge_{\vec{a} \in U} \varphi[\vec{a}]}{\epsilon}{\sloc(b) \neq l_\err} \in \cH. \]
    \end{itemize}
\end{lemma}
\begin{proof}
    We show the second bullet point. The two others are similar. Let $\tp{\sigma : a_0} \in \Sigma \times \TT_i$ and $\vec{a} \in \TT_i^k$. Then there is some $j \in \cJ$ and embedding $f : \TT_j \hookrightarrow \TT_i$ such that $a_0,\vec{a} \in f(\TT_j)^{k+1}$. Let $\beta : \TT_j \to f(\TT_j)$ be $f$ with codomain restricted to its image. Then $\beta$ is an isomorphism, and
    \begin{align*}
        {}&{} \textstyle \htri{\bigwedge_{\vec{b} \in \TT_j^k} \varphi[\vec{b}]}{\tp{\sigma : \beta^{-1}(a_0)}}{\varphi[\beta^{-1}(\vec{a})]} \\
        \xRightarrow{\textsc{S-Param}} & \textstyle \htri{\bigwedge_{\vec{b} \in \TT_j^k} \varphi[\beta(\vec{b})]}{\tp{\sigma : a_0}}{\varphi[\vec{a}]} \\
        \iff & \textstyle \htri{\bigwedge_{\vec{a} \in f(\TT_j)^k} \varphi[\vec{a}]}{\tp{\sigma : a_0}}{\varphi[\vec{a}]}.
    \end{align*}
    Thus, we may take $U = f(\TT_j)^k$.
\end{proof}

\begin{theorem}[\Cref{thm:relative-completeness}]
   Let $\cP = \{\prog_i\}_{i\in\cI}$ be a parameterized program over a family $\cT = \{\TT_i\}_{i\in\cI}$ of topologies with universal structure $\TT$. If there is an Ashcroft invariant $\Phi$ of width $k$ for $\cP$, and $\cT$ has a finite basis of rank $k+1$ consisting of finite topologies, then there is a finitely generated $\TT$-proof space that proves $\cP$ safe.
\end{theorem}
\begin{proof}
   Suppose $\Phi$ is an Ashcroft invariant of width $k$ for $\cP$, and $\cT$ has a finite basis $\{\TT_j\}_{j\in\cJ}$ of rank $k+1$ where every $\TT_j$ is finite. Let $\cH$ be the extended $\TT$-proof space generated by the finite basis $\bigcup_{j\in\cJ} H_{\Phi}^{\TT_j}$, which is valid by Lemma \ref{lem:ashcroft-ext}. By Lemma \ref{lem:fin-gen-ext}, it suffices to show that for any $i \in \cI$, $\rho \in \Lang_\err(P_i)$, and $b \in \TT_i$, there is some finite subset $U$ of $\TT_i$ such that
   \[ \textstyle \htri{\bigwedge_{a \in U} \sloc(a) = l_\init}{\rho}{\sloc(b) \neq l_\err} \in \cH. \]
   This follows from Lemma \ref{lem:symm} and the conjunction rule, and the sequencing rule.
\end{proof}

\begin{theorem}[\Cref{thm:subprogram-safety}]
    For any parameterized programs $\cP$ and $\cP'$, if $\cP \subset \cP'$ and every program of $\cP'$ is a subprogram of some program in $\cP$, then $\cP$ is safe if and only if $\cP'$ is safe; if furthermore, $\TT$ is a universal structure for $\cP$ and $\cP'$, then any $\TT$-proof space $\cH$ proves the safety of $\cP$ if and only if it proves the safety of $\cP'$.
\end{theorem}
\begin{proof}
    Let $\cP$ and $\cP'$ be parameterized programs such that $\cP \subset \cP'$ and every program of $\cP'$ is a subprogram of some program in $\cP$. Since $\cP \subset \cP'$, if $\cP'$ is safe, then $\cP$ is also safe. Suppose $\cP'$ is unsafe and $\rho$ is a feasible error run of some program $\prog' \in \cP'$. Then there is a program $\prog \in \cP$ that is a superprogram of $\prog'$. Observe that $\rho$ is also a feasible error run of $\prog$, so $\cP$ is unsafe as well.

    Let $\TT$ be a universal structure for $\cP$ and $\cP'$ and $\cH$ be a $\TT$-proof space. Note that the set of error runs of $\cP$ and $\cP'$ are equal. Thus, $\Lang(\cH)$ includes the former if and only if it includes the latter.
\end{proof}

For any first-order formula $\varphi$, we define the term-height of $\varphi$ as the maximum height of the terms appearing in $\varphi$. A variable has term-height $0$.

\begin{lemma} \label{lem:local-symm-fin-card}
    Let $\cV$ be a finite vocabulary and $\TT$ be a $\cV$-structure. If $\vec{a} \in \TT^k$ generates a substructure of size $n < \omega$, then there is a quantifier-free formula $\chi[u_1, \ldots, u_k]$ with term-height up to $n$ such that for any $\cV$-structure $\TT'$ and tuple $\vec{b} \in {\TT'}^k$, we have $\TT' \models \chi(\vec{b}) \iff \vec{a} \simeq \vec{b}$.
\end{lemma}
\begin{proof}
  Let $\SS$ be the substructure generated by $\vec{a}$. For every $b \in \SS$, let $t_b [\vec{u}]$ be a $\cV$-term of minimal height such that $t_b (\vec{a}) = b$. Let $\cC$ consist of the following formulas:
  \begin{itemize}
    \item $t_{a_i} = u_i$ for each $i \in \{ 1, \ldots, k\}$;
    \item $t_{b_1} \neq t_{b_2}$ for all distinct $b_1, b_2 \in \SS$;
    \item for all relation symbol $R \in \cV$ and $\vec{b} \in \SS^r$ where $r = \ar(R)$, if $R(\vec{b})$ is true, then $R (t_{b_1}, \ldots, t_{b_r})$, otherwise $\neg R (t_{b_1}, \ldots, t_{b_r})$; and
    \item $f (t_{b_1}, \ldots, t_{b_r}) = t_c$ for all function symbol $f \in \cV$, $\vec{b} \in \SS^r$ where $r = \ar(f)$, and $c = f(b_1, \ldots, b_r)$.
  \end{itemize}
  Since $\SS$ and $\cV$ are finite, the set $\cC$ is finite. Let $\chi$ be the conjunction of all formulas in $\cC$. Since $\SS$ has size $n$, every $t_b$ has term-height up to $n - 1$, and it follows that $\chi$ has term-height up to $n$. It is straightforward to verify that this choice of $\chi$ satisfies the requirement.
\end{proof}

\begin{proposition}[\Cref{prop:topo-sub}]
    For any $k \in \NN_+$ and family of topologies $\cT$ over a common vocabulary $\cV$, if $\substructures_k(\TT) \subset \cT$ for any $\TT \in \cT$, then $\bigcup_{\TT\in\cT} \substructures_k(\TT)$ is a basis for $\cT$ of rank $k$. Moreover, if $\cV$ is finite and there is some $n \in \NN$ such that the cardinality of every structure in $\bigcup_{\TT\in\cT} \substructures_k(\TT)$ is bounded above by $n$, then this basis is finite up to isomorphism.
\end{proposition}
\begin{proof}
    Let $k \in \NN_+$ and $\cT$ be a family of topologies over a common vocabulary. Suppose $\substructures_k(\TT) \subset \cT$ for any $\TT \in \cT$. Write $\cS := \bigcup_{\TT\in\cT} \substructures_k(\TT)$. Then $\cS \subset \cT$.

    Let $\TT \in \cT$ and $\vec{a} \in \TT^k$. Let $\SS$ be the substructure of $\TT$ generated by $\vec{a}$. Note that $\SS \in \cS$ and $\vec{a} \in \SS^k$. Thus, $\cS$ is a basis for $\cT$ of rank $k$.

    Suppose $\cV$ is finite and there is some $n \in \NN$ such that the cardinality of every structure in $\bigcup_{\TT\in\cT} \substructures_k(\TT)$ is bounded above by $n$. It suffices to show that there are finitely many $k$-tuples in $\cT$ up to $\simeq$-equivalence.
    
    For every $\TT \in \cT$ and $\vec{a} \in \TT^k$, let $\chi_{\TT,\vec{a}}$ be a quantifier-free formula obtained from \Cref{lem:local-symm-fin-card} for $\TT$ and $\vec{a}$. Since $\cV$ is finite, up to logical equivalence $\equiv$, there are only finitely many quantifier-free formulas with term-height up to $n$. Consequently, the quotient $\{ \chi_{\TT,\vec{a}} \mid \TT \in \cT,\, \vec{a} \in \TT^k \} / {\equiv}$ is finite.

    For any $\TT, \TT' \in \cT$, $\vec{a} \in \TT^k$, and $\vec{b} \in {\TT'}^k$, if $\chi_{\TT,\vec{a}} \equiv \chi_{\TT',\vec{b}}$, then $\TT' \models \chi_{\TT,\vec{a}}(\vec{b})$, and therefore $\vec{a} \simeq \vec{b}$. Thus, $\{ (\TT, \vec{a}) \mid \TT \in \cT,\, \vec{a} \in \TT^k \} / {\simeq}$ is also finite.
\end{proof}

\begin{theorem}[\Cref{thm:bool-c}]
    Any safe parameterized Boolean program $\cP$ with universal structure $\TT$ has a $\TT$-proof space that proves its safety, which only depends on $\TT$ and the set of commands and their semantics. Moreover, if $\TT^2/{\simeq}$ is finite, this proof space is finitely generated.
\end{theorem}

Write $\cP = \{ \prog_i \}_{i\in\cI}$ and $\prog_i = \tp{P_i, \sem{-}_i}$. Let $\TT_i$ be the topology of $\prog_i$. We first formalize what it means that ``the (local) semantics of every indexed command is deterministic and reads and writes finitely many variables'': For every $i \in \cI$ and $a \in \TT_i$, there is a finite set $\rw_i(a) \subset \nb(a)$ and a function $\sem{-}_{\rw_i}$ that maps every $\tp{\sigma : a} \in \Sigma \times \TT_i$ to a partial function on the set of valuations $\rw_i(a) \to \DD$ such that
\begin{itemize}
    \item for any $i, j \in \cI$, $a \in \TT_i$, and $b \in \TT_j$ with local isomorphism $\beta$ from $a$ to $b$, we have $\beta(\rw_i(a)) = \rw_i(b)$ and $\sem{\sigma : a}_{\rw_i} = \beta^* \sem{\sigma : a}_{\rw_i}$ for every $\sigma \in \Sigma$; and
    \item $\sem{-}_i = \Global_{N(a)} \sem{-}_{\rw_i}$.
\end{itemize}
We further assume that each $\rw_i(a)$ is represented by a list of $\VNode$-terms. 

Let us begin the proof.
\begin{proof}
    For any $a \in \TT$, if there is some $i \in \cI$, $a' \in \TT_i$, and a local isomorphism $\beta$ from $a$ to $a'$, set $\rw(a) := \beta^{-1}(\rw_i(a'))$ and $\sem{\sigma : a}_\rw := \beta^* \sem{\sigma : a'}_\rw$ for every $\sigma \in \Sigma$.

    Let $\cH$ be the $\TT$-proof space generated by the following Hoare triples: for any $\tp{\sigma : a} \in \Sigma \times \TT$ such that $a \simeq a'$ for some $a' \in \TT_i$ and $i \in \cI$,
    \begin{enumerate}
        \item for every $b \in \rw(a)$, $x \in \DD$, and $\val \in \sem{\sigma : a}_{\rw}^{-1}(\{ \val' : \rw(a) \to \DD \mid \val'(b) = x \})$,
        \[ \textstyle \htri{\bigwedge_{c \in \rw(a)} \sval(c) = \val(c)}{\tp{\sigma : a}}{\sval(b) = x} \]
        \item for every $b \in \TT \setminus \rw(a)$,
        \[ \textstyle \htri{\sval(b) = x}{\tp{\sigma : a}}{\sval(b) = x} \]
        \item for every $\val \in \sem{\sigma : a}_{\rw}^{-1}(\varnothing)$,
        \[ \textstyle \htri{\bigwedge_{c \in \rw(a)} \sval(c) = \val(c)}{\tp{\sigma : a}}{\bot} \]
        \item if $\sem{\sigma : a}_{\rw}^{-1}(\varnothing) = \varnothing$,
        \[ \textstyle \htri{\bot}{\tp{\sigma : a}}{\bot} \]
    \end{enumerate}
    Clearly, these triples are all valid. On the other hand, one can show that for any run $\rho \in (\Sigma \times \TT_i)^*$ that is infeasible if all variables are initialized to $0$, there is a finite set $U \subset \TT_i$ such that
    \[ \textstyle \htri{\bigwedge_{c \in U} \sval(c) = 0}{\rho}{\bot} \in \cH. \]

    Due to the \textsc{S-Param} rule, for (1) and (2), it suffices to take one representative $(a, b)$ from every $\simeq$-equivalence class of $\TT^2$; for (3) and (4), it suffices to take one representative $a$ from every $\simeq$-equivalence class of $\TT$.
\end{proof}
\subsection{Proofs from Section \ref{sec:predicate-automata}}

\begin{proposition}[\Cref{prop:fraisse}]
  Let $\cT$ be a class of finite logical structures over a common vocabulary and $\cls{\cT}$ be the substructures of members of $\cT$. If $\cls{\cT}$ is a countable Fra\"{i}ss\'{e} class, then its Fra\"{i}ss\'{e} limit is a limit of $\cT$ in the sense of \Cref{def:lim-struct}.
\end{proposition}
\begin{proof}
  Suppose $\cls{\cT}$ is a countable Fra\"{i}ss\'{e} class. Let $\lmt{\TT}$ be the countable Fra\"{i}ss\'{e} limit. Then $\cls{\cT}$ can be identified with the class of finitely generated substructures of $\lmt{\TT}$. Thus, any structure in $\cT \subset \cls{\cT}$ is an embedded substructure of $\lmt{\TT}$, and any finitely generated substructure of $\lmt{\TT}$ is isomorphic to some structure in $\cls{\cT}$, i.e., some substructure of some structure in $\cT$.
\end{proof}

\begin{theorem}[\Cref{thm:limit-safety}]
  For any parameterized program $\cP$ and limit program $\lmt{\prog}$ over topology $\lmt{\TT}$ for $\cP$, we have that $\cP$ is safe if and only if $\lmt{\prog}$ is safe. Furthermore, if $\TT$ is a universal structure for $\cP$ of which $\lmt{\TT}$ can be identified with a substructure, then any $\TT$-proof space $\cH$ proves the safety of $\cP$ if and only if it proves the safety of $\lmt{\prog}$.
\end{theorem}
\begin{proof}
  Let $\cP$ be a parameterized program and $\lmt{\prog}$ be a limit program for $\cP$ over $\lmt{\TT}$. Any error run in $\cP$ is $\simeq$-equivalent to an error run in $\lmt{\prog}$, whose feasibility is the same, and vice versa. Hence, $\cP$ is safe if and only if $\lmt{\prog}$ is safe.

  Suppose $\TT$ is a universal structure for $\cP$ where $\lmt{\TT}$ is embedded. Let $\cH$ be a $\TT$-proof space. Observe that the languages $\Lang_\err(\cP) := \bigcup_{\prog \in \cP} \Lang_\err(P)$ and $\Lang_\err(\lmt{P})$ have the same $\simeq$-closure. Let $[-]^{\simeq}$ denote the $\simeq$-closure operator. Then
  \begin{align*}
    \Lang_\err(\cP) \subset \Lang(\cH) &\iff [\Lang_\err(\cP)]^{\simeq} \subset \Lang(\cH) \\
    &\iff [\Lang_\err(\lmt{P})]^{\simeq} \subset \Lang(\cH) \\
    &\iff \Lang_\err(\lmt{P}) \subset \Lang(\cH),
  \end{align*}
  where the first and last equivalence follow from \Cref{thm:psl-sym}.
\end{proof}

\begin{theorem}[\Cref{thm:PAL-closedness}]
  $\TT$-PALs are closed under intersection and complement, and so are monadic $\TT$-PALs. Any $\TT$-PAL is locally symmetric, i.e., closed under the equivalence relation $\simeq$.
\end{theorem}
\begin{proof}
  The constructions for intersection and complementation follow those for $\NN$-PAs \cite[Proposition 6.5]{FarzanKP2015}.

  Let $A := \tp{\cQ, \Sigma \times \TT, \delta, \varphi_\start, F}$ and $A' := \tp{\cQ', \Sigma \times \TT, \delta', \varphi'_\start, F'}$ be $\TT$-PAs. Assume without loss of generality that $\cQ$ and $\cQ'$ are disjoint. The automaton $\tp{\cQ \cup \cQ', \Sigma \times \TT, \delta \cup \delta', \varphi_\start \land \varphi'_\start, F \cup F'}$ recognizes the language $\Lang(A) \cap \Lang(A')$.

  Let $\overline{\cQ} := \{ \overline{q} \mid q \in \cQ \}$ be a copy of $\cQ$ with arities preserved. For any formula $\varphi \in \sF_+(\cV, \cQ)$, let $\overline{\varphi}$ be the formula obtained by syntactically negating $\varphi$, pushing $\neg$ inward, and replacing every $\neg q$ with $\overline{q}$. To recognize the language $\setcomp{\Lang(A)}$, we construct the automaton $\overline{A} := \tp{\overline{\cQ}, \Sigma \times \TT, \overline{\delta}, \overline{\varphi_\start}, \overline{F}}$, where the accepting symbols $\overline{F} := \{ \overline{Q} \mid Q \in \cQ \setminus F \}$ and the transition function is given by $\overline{\delta}(\overline{q}, \sigma) := \overline{\delta(q, \sigma)}$ for every $q \in \cQ$ and $\sigma \in \Sigma$.

  For local symmetry, observe that for any $\cC \xrightarrow{\sigma : a} \cC'$ and local isomorphism $\beta$ from $a$, we have $\beta(\cC) \xrightarrow{\sigma : \beta(a)} \beta(\cC')$, where for any configuration $\cC$, $\beta(\cC)$ is obtained from $\cC$ by replacing $a \in \TT$ with $\beta(a)$.
\end{proof}

\begin{theorem}[\Cref{thm:prog-PAL}]
  Let $\TT$ be a topology over a vocabulary $\VNode$. Let $\tp{P, \sem{-}}$ be a locally symmetric concurrent program over $\TT$. If $\TT/{\simeq}$ is finite, then $\Lang_\err(P)$ is a monadic $\TT$-PAL.
\end{theorem}
\begin{proof}
  The construction is a generalization of \cite[Proposition 6.3]{FarzanKP2015}, mirroring the (reversed) control structure of $P$, except that we also ensure $\sigma \in P(u)$ for every indexed command $\tp{\sigma : u}$ that occurs.

  We define a $\TT$-PA $A_P := \tp{\cQ, \Sigma \times \TT, \delta, \varphi_\start, F}$ as follows:
  \begin{itemize}
    \item $\cQ := \locs \cup \{ I, \id{err} \}$, where every $l \in \locs$ has arity $1$, and $I$ and $\id{err}$ are two distinguished symbols with arity $0$.
    \item The transitions are defined as follows, for any $\sigma \in \Sigma$ and $l \in \locs$:
    \begin{align*}
      \delta(I, \sigma) &:= I \land \src(\sigma)(v_0) \land \underline{v_0 \in P^{-1}(\sigma)}, \\
      \delta(l, \sigma) &:= \begin{cases}
        (v_1 = v_0 \land \src(\sigma)(v_0)) \lor (v_1 \neq v_0 \land l(v_1)), & \text{if } l = \tgt(\sigma) \\
        v_1 \neq v_0 \land l(v_1), & \text{if } l \neq \tgt(\sigma)
      \end{cases} \\
      \delta(\id{err}, \sigma) &:= \begin{cases}
        \id{err}, & \text{if } \tgt(\sigma) \neq l_\err \\
        \top. & \text{if } \tgt(\sigma) = l_\err
      \end{cases}
    \end{align*}

    Since $P^{-1}(\sigma)$ is closed under local symmetry, it is a union of $\simeq$-equivalence classes in $\TT$. Then, by the assumption that $\TT / {\simeq}$ is finite and \Cref{prop:local-symm-fin}, $P^{-1}(\sigma)$ is definable by a quantifier-free formula $\chi[v_0]$, which we abbreviate as $\underline{v_0 \in P^{-1}(\sigma)}$.

    \item $\varphi_\start := I \land \id{err}$.
    \item $F := \{ I, l_\init \}$.
  \end{itemize}
\end{proof}

\begin{theorem}[\Cref{thm:proof-PAL}]
  Let $\TT$ be a topology over a vocabulary $\VNode$. Let $\cH$ be a $\TT$-proof space generated by a finite basis in normal form. If $\TT^d/{\simeq}$ is finite for every $d \in \NN_+$, then $\Lang(\cH)$ is a $\TT$-PAL.
\end{theorem}
\begin{proof}
  Let $H$ be a finite basis in normal form for $\cH$. The construction is a generalization of \cite[Proposition 6.4]{FarzanKP2015}, mirroring the structure of $H$: assertions in $H$ become the predicates of the $\TT$-PA, and Hoare triples become transitions.

  We define a $\TT$-PA $A_H := \tp{\cQ, \Sigma \times \TT, \delta, \varphi_\start, F}$ as follows:
  \begin{itemize}
    \item $\cQ$ consists of, for every assertion $\phi[\overrightarrow{\sval(a)}]$ that occurs in $H$, a predicate symbol $[\phi]$ that is uniquely identified by $\phi$ (called the \emph{canonical name} in \cite{FarzanKP2015}), whose arity is the arity of $\phi$.

    Note that conjuncts recognized by combinatorial entailment should be separated into different assertions.

    \item Define $\delta$ as follows: for every Hoare triple
    \[ \textstyle \htri{\bigwedge_{i=1}^n \phi_i[\overrightarrow{\sval(a_i)}]}{\tp{\sigma : b}}{\psi[\overrightarrow{\sval(c)}]} \]
    in $H$, set
    \[ \delta([\psi], \sigma) := \bigwedge_{i=1}^n [\phi_i](\vec{t_i}) \land \chi(v_0, \ldots, v_{|\vec{c}|}), \]

    where $\chi$ is a quantifier-free formula that defines the $\simeq$-equivalence class of $b\vec{c}$ (\Cref{prop:local-symm-fin}), and $\vec{t_i}$ is a tuple of $\VNode$-terms $(t_{i,1}, \ldots, t_{i,m_i})$ such that $t_{i,j}(b\vec{c}) = a_{i,j}$ for every $i \in \{1,\ldots,n\}$ and $j \in \{1, \ldots, m_i \}$. If there are more than one choice for $t_{i,j}$, either is fine, as their equality is preserved by all tuples that are $\simeq$-equivalent to $b\vec{c}$ (\Cref{prop:local-symm-qf}).

    If there are multiple Hoare triples in $H$ with the same $[\psi]$ and $\sigma$, we combine their transitions disjunctively.

    \item $\varphi_\start := [\bot]$.
    \item $F := \varnothing$.
  \end{itemize}
\end{proof}

\begin{lemma}[\Cref{lem:wsts}]
  Let $\TT$ be a $\cV$-structure and $\cQ$ be a finite relational vocabulary disjoint from $\cV[\TT]$. Suppose there is a pre-order $\preceq$ on the set of configurations $2^{\sF^\TT_\atom(\cQ)}$ such that for any $\TT$-predicate automaton $\tp{\cQ, \Sigma \times \TT, \delta, \varphi_\start, F}$, we have
  \begin{enumerate}
    \item $\preceq$ is decidable.
    \item (Essentially Finite Branching) Given any configuration $\cC$, one can effectively compute a finite subset $\sC$ of the successors of $\cC$ such that for any successor $\cC'$ of $\cC$, there is some $\bar{\cC}' \in \sC$ such that $\bar{\cC}' \preceq \cC'$.
    \item (Downward Compatibility) For any configurations $\cC$ and $\bar{\cC}$ such that $\bar{\cC} \preceq \cC$, if $\cC$ is accepting, then $\bar{\cC}$ is accepting, and for any configuration $\cC'$ such that $\cC \to \cC'$, there is some configuration $\bar{\cC}'$ such that $\bar{\cC} \to \bar{\cC}'$.
  \end{enumerate}
  Then there is a semi-algorithm for $\TT$-PA non-emptiness, which becomes a decision procedure for the class of $\TT$-PA whose reachable configurations are well-quasi-ordered by $\preceq$.
\end{lemma}
\begin{proof}
The semi-algorithm is as follows (assuming a fixed $\TT$), where the procedure call $\Call{Repr-Succ}{A, \cC}$ returns a finite subset of successors of $\cC$ as described by the essentially finite branching condition:
\begin{algorithmic}[1]
\Procedure{Check-Emptiness}{$A = \tp{\cQ, \Sigma, \delta, \varphi_{\start}, F}$}
\State $\id{Closed} \gets \varnothing$
\State $\id{Open}$ $\gets \Call{Dnf}{\varphi_{\start}}$
\While{$\Call{Nonempty}{\id{Open}}$}
  \State $\mathcal{C} \gets \Call{Dequeue}{\id{Open}}$
  \If{$\not\exists \mathcal{C'} \in \id{Closed}.\, \mathcal{C'} \preceq \mathcal{C}$}
    \ForAll{$\cC' \in \Call{Repr-Succ}{A, \cC}$}
      \If{$\cC'$ is accepting}
        \State \Return False
      \ElsIf{$\cC' \notin \id{Open}$ and $\cC' \notin \id{Closed}$}
        \State $\Call{Enqueue}{\id{Open}, \mathcal{C'}}$
      \EndIf
    \EndFor
    \State $\id{Closed} \gets \id{Closed} \cup \{ \mathcal{C} \}$
  \EndIf
\EndWhile
\State \Return True
\EndProcedure
\end{algorithmic}

For termination, we follow the argument in the proof of \cite[Proposition 5.3.15]{Kincaid2016}. The search space of the algorithm forms a finitely-branching tree. By K\"{o}nig's lemma, if the algorithm does not terminate, there must be an infinite path from the root. Suppose $\preceq$ well-quasi-orders the reachable configurations. Then on the infinite path, there is a configuration $\cC$ and a descendant $\cC'$ of $\cC$ such that $\cC \preceq \cC'$. But when $\cC'$ is expanded, $\cC$ must be in the $\id{Closed}$ set, so $\cC'$ should not have been expanded. The contradiction shows that the algorithm terminates if $\preceq$ well-quasi-orders the reachable configurations.
\end{proof}

\begin{theorem}[\Cref{thm:A-PA-emp}]
  Suppose $\TT$ satisfies the following:
  \begin{enumerate}
    \item $\TT$ is homogeneous.
    \item $\TT^d/{\simeq}$ is finite for every $d \in \NN_+$, and there is an algorithm that takes $d$ and outputs a list of quantifier-free formulas that define the equivalence classes (whose existence is guaranteed by Proposition~\ref{prop:local-symm-fin}).
    \item The satisfiability of any quantifier-free $\cV[\TT]$-formula in $\TT$ is decidable.
  \end{enumerate}
  Then the covering relation satisfies conditions (1)--(3) in Lemma~\ref{lem:wsts}.
\end{theorem}
\begin{proof}
  \begin{enumerate}
    \item When $\TT$ is homogeneous, the covering relation is decidable if the local symmetry relation is. Given $\vec{a}, \vec{b} \in \TT^d$, to decide whether $\vec{a} \simeq \vec{b}$, we just compute a list of quantifier-free formulas defining the equivalence classes $\TT^d / {\simeq}$, and test whether $\vec{a}$ and $\vec{b}$ are in the same equivalence class.
    \item Below is an implementation of \textsc{Repr-Succ}, where
    \begin{itemize}
      \item $\Call{Support}{\mathcal{C}}$ returns the $\TT$ elements in $\cC$ listed in some order,
      \item $\Call{Eq-Classes}{d}$ returns a list of quantifier-free formulas defining the equivalence classes in $\TT^d / {\simeq}$ for any $d \in \NN_+$, and
      \item $\Call{Solve}{p[\vec{u}]}$ returns a satisfying assignment for any quantifier-free $\cV[\TT]$-formula $p$, or None if it is unsatisfiable.
      
      Although here we only assume the decidability of satisfiability, in \Cref{sec:predicate-automata} we have the global assumption that elements in $\TT$ are finitely representable (and therefore recursively enumerable); together, they show that \textsc{Solve} is computable.
    \end{itemize}

\begin{algorithmic}[1]
\Procedure{Repr-Succ}{$\tp{\cQ, \Sigma, \delta, \varphi_{\start}, F}, \cC$}
  \State $\sC \gets \{\}$
  \State $\vec{a} \gets \Call{Support}{\mathcal{C}}$
  \State $d \gets |\vec{a}|$
  \State $X \gets \Call{Eq-Classes}{\TT, d + 1}$
  \ForAll{$\chi[u_1, \ldots, u_{d + 1}] \in X$}
    \State $b \gets \Call{Solve}{\chi[u_1, \vec{a}]}$
    \If{$b \neq \operatorname{None}$}
      \ForAll{$\sigma \in \Sigma$}
        \ForAll{$\cC'$ s.t. $\cC \xrightarrow{\sigma : b} \cC'$}
          \State $\sC \gets \sC \cup \{ \cC' \}$
        \EndFor
      \EndFor
    \EndIf
  \EndFor
  \State \Return $\sC$
\EndProcedure
\end{algorithmic}
    \item Downward compatibility is easy to verify.
  \end{enumerate}
\end{proof}

\begin{remark}
  In a counterexample-guided abstraction refinement loop, we would also want a witness for non-emptiness. The implementation of \textsc{Repr-Succ} given above can be modified to also return the letter $\tp{\sigma : b}$ that leads to each $\cC' \in \sC$. Then, the \textsc{Check-Emptiness} procedure can be modified accordingly to return a witness instead of False.
\end{remark}

\begin{corollary}[\Cref{cor:bool-d}]
  Fix a structure $\TT$ that satisfies the conditions in Theorem \ref{thm:A-PA-emp}. Additionally, assume for any finite relational vocabulary $\cQ$ disjoint from $\cV[\TT]$ where symbols have arity at most $1$, the covering pre-order is a well-quasi-order on the set of configurations $2^{\sF^\TT_\atom(\cQ)}$. Then the safety of any parameterized Boolean program with a limit program over $\TT$ is decidable.
\end{corollary}
\begin{proof}
  Let $\cP$ be a parameterized Boolean program with a limit program $\lmt{\prog}$ over $\TT$. Let $H$ be a finite basis of the $\TT$-proof space $\cH$ constructed in \Cref{thm:bool-c}. Then $\cP$ is safe if and only if $\cH$ proves its safety, which happens if and only if $\cH$ proves the safety of $\lmt{\prog}$, the limit program of $\cP$ over $\TT$ (\Cref{thm:limit-safety}, with a slight modification to take into account the initialization of nodes in the neighbourhood of a run).

  Note that the conditions in \Cref{thm:A-PA-emp} allow the following operations:
  \begin{itemize}
    \item Enumerate the $\simeq$-equivalence classes of $\TT^d$ for any $d \in \NN_+$, each represented by a quantifier-free formula.
    \item Find a concrete tuple that satisfies a quantifier-free formula (using the \textsc{Solve} procedure in the proof of \Cref{thm:A-PA-emp}).
    \item Find a quantifier-free formula that defines the $\simeq$-equivalence class of a given tuple.
  \end{itemize}

  Our algorithm is as follows:
  \begin{enumerate}
    \item Compute the basis $H$. (\Cref{thm:bool-c})

    We enumerate the representatives for the $\simeq$-equivalence classes of $\TT$ and $\TT^2$. Since we assume there is a computable function that maps every node in the limit structure to a locally symmetric node in the topologies of $\cP$, and each $\rw_i(a)$ is represented by a list of $\VNode$-terms, the functions $\rw$ and $\sem{-}_{\rw}$ are computable.

    \item Construct the $\TT$-PA $A_H$ for $\Lang(\cH)$. (\Cref{thm:proof-PAL})
    
    Note that $H$ is already in normal form.

    We add the canonical name $[x = 0]$ to the set $F$ of accepting symbols to account for initialization.

    \item Construct the $\TT$-PA $A_{\lmt{P}}$ for $\Lang(\lmt{P})$. (\Cref{thm:prog-PAL})

    For each $\sigma \in \Sigma$, the formula defining $P^{-1}(\sigma)$ is computable: it is a disjunction of a subset of the quantifier-free formulas representing $\TT/{\simeq}$. We enumerate the representative $a$ of each equivalence class, and include the formula corresponding to $a$ if $\sigma \in P(a)$.

    \item Check whether $\Lang(A_{\lmt{P}}) \cap \setcomp{\Lang(A_H)}$ is empty. (\Cref{thm:PAL-closedness,thm:A-PA-emp}) If yes, return \textsc{Safe}; otherwise, return \textsc{Unsafe}.

    Since this language is recognized by a monadic $\TT$-PA, by assumption, the emptiness check terminates.
  \end{enumerate}
\end{proof}
}

\end{document}